\documentclass[12pt]{iopart}
\usepackage{iopams}
\usepackage{setstack}

\newcommand{\cC}{\ensuremath{\mathcal{C}}}
\newcommand{\cP}{\ensuremath{\mathcal{P}}}
\newcommand{\cT}{\ensuremath{\mathcal{T}}}
\newcommand{\half}{\mbox{$\textstyle \frac{1}{2}$}}
\newcommand{\x}{\ensuremath{\hat x}}
\newcommand{\y}{\ensuremath{\hat y}}
\newcommand{\z}{\ensuremath{\hat z}}
\newcommand{\p}{\ensuremath{\hat p}}
\newcommand{\q}{\ensuremath{\hat q}}
\newcommand{\rr}{\ensuremath{\hat r}}
\newcommand{\pslash}{\partial\raise.3ex\hbox{\kern-.5em /}}
\newcommand{\delslash}{\nabla\raise.3ex\hbox{\kern-.7em /}}
\newcommand{\ve}{\ensuremath{\varepsilon}}
\newcommand{\vf}{\ensuremath{\varphi}}

\begin{document}
\rightline{preprint LA-UR-07-1254}

\title[Making Sense of Non-Hermitian Hamiltonians]{Making Sense of Non-Hermitian
Hamiltonians}

\author[C M Bender] {Carl~M~Bender\footnote{Permanent address: Department of
Physics, Washington University, St. Louis MO 63130, USA}}

\address{Center for Nonlinear Studies, Los Alamos National Laboratory\\
Los Alamos, NM 87545, USA\\ \footnotesize\texttt{cmb@wustl.edu}}

\begin{abstract}
The Hamiltonian $H$ specifies the energy levels and time evolution of a quantum
theory. A standard axiom of quantum mechanics requires that $H$ be Hermitian
because Hermiticity guarantees that the energy spectrum is real and that time
evolution is unitary (probability-preserving). This paper describes an
alternative formulation of quantum mechanics in which the mathematical axiom of
Hermiticity (transpose + complex conjugate) is replaced by the physically
transparent condition of space-time reflection ($\cP\cT$) symmetry. If $H$ has
an unbroken $\cP\cT$ symmetry, then the spectrum is real. Examples of $\cP
\cT$-symmetric non-Hermitian quantum-mechanical Hamiltonians are $H=\p^2+i\x^3$
and $H=\p^2-\x^4$. Amazingly, the energy levels of these Hamiltonians are all
real and positive!

Does a $\cP\cT$-symmetric Hamiltonian $H$ specify a physical quantum theory in
which the norms of states are positive and time evolution is unitary? The answer
is that if $H$ has an unbroken $\cP\cT$ symmetry, then it has another symmetry
represented by a linear operator $\cC$. In terms of $\cC$, one can construct a
time-independent inner product with a positive-definite norm. Thus, $\cP
\cT$-symmetric Hamiltonians describe a new class of complex quantum theories
having positive probabilities and unitary time evolution.

The Lee Model provides an excellent example of a $\cP\cT$-symmetric Hamiltonian.
The renormalized Lee-model Hamiltonian has a negative-norm ``ghost'' state
because renormalization causes the Hamiltonian to become non-Hermitian. For the
past 50 years there have been many attempts to find a physical interpretation
for the ghost, but all such attempts failed. The correct interpretation of the
ghost is simply that the non-Hermitian Lee Model Hamiltonian is $\cP
\cT$-symmetric. The $\cC$ operator for the Lee Model is calculated exactly and
in closed form and the ghost is shown to be a physical state having a positive
norm. The ideas of $\cP\cT$ symmetry are illustrated by using many
quantum-mechanical and quantum-field-theoretic models.

(\today)
\end{abstract}

\submitto{\RPP}


\section{Introduction -- New Kinds of Quantum Theories}
\label{s1}

The theory of quantum mechanics is nearly one hundred years old and because
there have been so many experimental verifications of its theoretical
predictions, it has become an accepted component of modern science. In an
introductory course on quantum physics, one learns the fundamental axioms that
define and characterize the theory. All but one of these axioms are physical
requirements. For example, the energy spectrum is required to be real because
all measurements of the energy of a system yield real results. Another axiom
requires that the energy spectrum be bounded below so that the system has a
stable lowest-energy state. Yet another axiom requires that the time evolution
of a quantum system be {\em unitary} (probability-conserving) because the
expected result of a probability measurement of a state cannot grow or decay in
time. A quantum theory of elementary particles must also satisfy the physical
axioms of Lorentz covariance and causality. However, there is one axiom that
stands out because it is mathematical rather than physical in character, and
this is the requirement that the Hamiltonian $H$, which is the operator that
expresses the dynamics of the quantum system, be Hermitian. 

The requirement that $H$ be Hermitian dates back to the early days of quantum
mechanics. The Hermiticity of $H$ is expressed by the equation
\begin{equation}
H=H^\dag,
\label{e1}
\end{equation}
where the Dirac Hermitian conjugation symbol $\dag$ represents the combined
operations of matrix transposition and complex conjugation. The mathematical
symmetry condition (\ref{e1}) is physically obscure but very convenient because
it implies that the eigenvalues of $H$ are real and that the time-evolution
operator $e^{-iHt}$ is unitary.

Hamiltonians that are non-Hermitian have traditionally been used to describe
dissipative processes, such as the phenomenon of radioactive decay. However,
these non-Hermitian Hamiltonians are only approximate, phenomenological
descriptions of physical processes. They cannot be regarded as fundamental
because they violate the requirement of unitarity. A non-Hermitian Hamiltonian
whose purpose is to describe a particle that undergoes radioactive decay
predicts that the probability of finding the particle gradually decreases in
time. Of course, a particle cannot just disappear because this would violate the
conservation of probability; rather, the particle transforms into other
particles. Thus, a non-Hermitian Hamiltonian that describes radioactive decay
can at best be a simplified, phenomenological, and {\em nonfundamental}
description of the decay process because it ignores the precise nature of the
decay products. In his book on quantum field theory Barton gives the standard
reasons for why a non-Hermitian Hamiltonian cannot provide a fundamental
description of nature \cite{BARTON}: ``A non-Hermitian Hamiltonian is
unacceptable partly because it may lead to complex energy eigenvalues, but
chiefly because it implies a non-unitary S matrix, which fails to conserve
probability and makes a hash of the physical interpretation.''

The purpose of this paper is to describe at an elementary level the
breakthroughs that have been made in the past decade which show that while the
symmetry condition (\ref{e1}) is sufficient to guarantee that the energy
spectrum is real and that time evolution is unitary, the condition of Dirac
Hermiticity is {\em not necessary}. It is possible to describe natural processes
by means of non-Hermitian Hamiltonians. We will show that the Hermiticity
requirement (\ref{e1}) may be replaced by the analogous but physically
transparent condition of space-time reflection symmetry ($\cP\cT$ symmetry)
\begin{equation}
H=H^{\cP\cT}
\label{e2}
\end{equation}
without violating any of the physical axioms of quantum mechanics. If $H$
satisfies (\ref{e2}), it is said to be $\cP\cT$ {\em symmetric}.

The notation used in this paper is as follows: The space-reflection operator, or
{\em parity} operator, is represented by the symbol $\cP$. The effect of $\cP$
on the quantum-mechanical coordinate operator $\x$ and the momentum operator
$\p$ is to change their signs:
\begin{equation}
\cP\x\cP=-\x\quad{\rm and}\quad\cP\p\cP=-\p.
\label{e3}
\end{equation}
Note that $\cP$ is a linear operator and that it leaves invariant the
fundamental commutation relation (the Heisenberg algebra) of quantum mechanics,
\begin{equation}
\x\p-\p\x=i\hbar\,{\bf 1},
\label{e4}
\end{equation}
where ${\bf 1}$ is the identity matrix. The time-reversal operator is
represented by the symbol $\cT$. This operator leaves $\x$ invariant but changes
the sign of $\p$:
\begin{equation}
\cT\x\cT=\x\quad{\rm and}\quad\cT\p\cT=-\p.
\label{e5}
\end{equation}
Like the parity operator $\cP$, the time-reversal operator $\cT$ leaves the
commutation relation (\ref{e4}) invariant, but this requires that $\cT$ reverse
the sign of the complex number $i$:
\begin{equation}
\cT i\cT=-i.
\label{e6}
\end{equation}
Equation (\ref{e6}) demonstrates that $\cT$ is not a linear operator; $\cT$ is
said to be {\em antilinear}. Also, since $\cP$ and $\cT$ are reflection
operators, their squares are the unit operator:
\begin{equation}
\cP^2=\cT^2={\bf 1}.
\label{e7}
\end{equation}
Finally, the $\cP$ and $\cT$ operators commute:
\begin{equation}
\cP\cT-\cT\cP=0.
\label{e8}
\end{equation}
In terms of the $\cP$ and $\cT$ operators, we define the $\cP\cT$-reflected
Hamiltonian $H^{\cP\cT}$ in (\ref{e2}) as $H^{\cP\cT}\equiv(\cP\cT)H(\cP\cT)$.
Thus, if a Hamiltonian is $\cP\cT$ symmetric [that is, if it satisfies
(\ref{e2})], then the $\cP\cT$ operator commutes with $H$:
\begin{equation}
H(\cP\cT)-(\cP\cT)H=0.
\label{e9}
\end{equation}

A $\cP\cT$-symmetric Hamiltonian need not be Hermitian; that is, it need not
satisfy the Hermiticity symmetry condition (\ref{e1}). Thus, it is possible to
have a fully consistent quantum theory whose dynamics is described by a
non-Hermitian Hamiltonian. Some examples of such non-Hermitian $\cP
\cT$-symmetric Hamiltonians are
\begin{equation}
H=\p^2+i\x^3,
\label{e10}
\end{equation}
and
\begin{equation}
H=\p^2-\x^4.
\label{e11}
\end{equation}
It is amazing indeed that the eigenvalues of these strange-looking Hamiltonians
are all real and positive and that these two Hamiltonians specify a unitary time
evolution even though they are non-Hermitian.

\begin{figure}[b!]
\vspace{3.2in}
\includegraphics{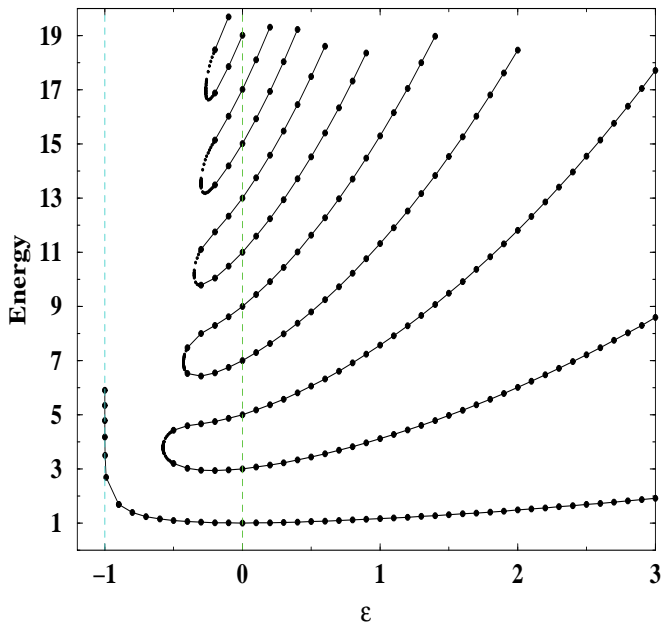}
\caption{Energy levels of the Hamiltonian $H=\p^2+\x^2(i\x)^\epsilon$ as a
function of the real parameter $\epsilon$. There are three regions: When
$\epsilon\geq0$, the spectrum is real and positive and the energy levels rise
with increasing $\epsilon$. The lower bound of this region, $\epsilon=0$,
corresponds to the harmonic oscillator, whose energy levels are $E_n=2n+1$. When
$-1<\epsilon<0$, there are a finite number of real positive eigenvalues and an
infinite number of complex conjugate pairs of eigenvalues. As $\epsilon$
decreases from $0$ to $-1$, the number of real eigenvalues decreases; when
$\epsilon\leq-0.57793$, the only real eigenvalue is the ground-state energy. As
$\epsilon$ approaches $-1^+$, the ground-state energy diverges. For $\epsilon
\leq-1$ there are no real eigenvalues. When $\epsilon\geq0$, the $\cP\cT$
symmetry is unbroken, but when $\epsilon<0$ the $\cP\cT$ symmetry is broken.}
\label{f1}
\end{figure}

The Hamiltonians in (\ref{e10}) and (\ref{e11}) are special cases of the general
parametric family of $\cP\cT$-symmetric Hamiltonians 
\begin{equation}
H=\p^2+\x^2(i\x)^\epsilon,
\label{e12}
\end{equation}
where the parameter $\epsilon$ is real. These Hamiltonians are all $\cP\cT$
symmetric because they satisfy the condition in (\ref{e2}). It was shown in 1998
that when $\epsilon\geq0$ all of the eigenvalues of these Hamiltonians are
entirely real and positive, but when $\epsilon<0$ there are complex eigenvalues
\cite{r1}. We say that $\epsilon\geq0$ is the parametric region of {\em
unbroken} $\cP\cT$ symmetry and that $\epsilon<0$ is the parametric region of
{\em broken} $\cP\cT$ symmetry. (See Fig.~\ref{f1}.)

One can think of the non-Hermitian Hamiltonians in (\ref{e12}) as complex
extensions of the harmonic oscillator Hamiltonian $H=\p^2+\x^2$. Indeed, the
quantum theories defined by $H$ are complex extensions of the conventional
quantum theory of the harmonic oscillator into the complex domain. The general
constructive principle that we are using in (\ref{e12}) is to start with a 
Hamiltonian that is {\em both} Hermitian and $\cP\cT$ symmetric. One then
introduces a real parameter $\epsilon$ in such a way that as $\epsilon$
increases from $0$ the Hamiltonian is no longer Hermitian but its $\cP\cT$
symmetry is maintained. One need not start with the harmonic oscillator. One
can, for example, begin with any of the Hermitian Hamiltonians $H=\p^2+\x^{2N}$,
where $N=1,\,2,\,3,\,\ldots$, and introduce the parameter $\epsilon$ as
follows: $H=\p^2+\x^{2N}(i\x)^\epsilon$. [The Hamiltonian in (\ref{e12}) is just
the special case $N=1$.] The properties of these Hamiltonians are discussed in
\cite{BBM}. 

We emphasize that these new kinds of Hamiltonians define valid and consistent
quantum theories in which the mathematical condition of Dirac Hermiticity $H=
H^\dagger$ has been replaced by the physical condition of $\cP\cT$ symmetry, $H=
H^{\cP\cT}$. The condition in (\ref{e2}) that the Hamiltonian is $\cP\cT$
symmetric is a physical condition because $\cP$ and $\cT$ are elements of the
homogeneous Lorentz group of spatial rotations and Lorentz boosts. The real
Lorentz group consists of four disconnected parts \cite{SW}: (i) The first part,
called the {\em proper orthochronous Lorentz group}, is a subgroup of the
Lorentz group whose elements are continuously connected to the identity. (ii)
The second part consists of all of the elements of the proper orthochronous
Lorentz group multiplied by the parity operator $\cP$. (iii) The third part
consists of all of the elements of the proper orthochronous Lorentz group
multiplied by the time-reversal operator $\cT$. (iv) The fourth part consists of
all of the elements of the proper orthochronous Lorentz group multiplied by $\cP
\cT$. Note that parts (ii) -- (iv) are not subgroups of the Lorentz group
because they do not contain the identity element. These four parts of the
Lorentz group are disconnected because there is no continuous path in group
space from one part to another.

When we say that Lorentz invariance is a physical requirement of a theory, what
we really mean is that the theory must be invariant under Lorentz
transformations belonging to the proper, orthochronous Lorentz group. We know
that the physical world is {\em not} invariant under the full homogeneous
Lorentz group because it has been demonstrated experimentally that there exist
weak processes that do not respect parity symmetry and other weak processes that
do not respect time-reversal symmetry.

One can extend the real Lorentz group to the {\em complex} Lorentz group
\cite{SW}. (To perform this extension it is necessary to make the crucial
assumption that the eigenvalues of the Hamiltonian are real and bounded below.)
The complex Lorentz group consists of {\em two} and not four disconnected parts.
In the complex Lorentz group there exists a continuous path in group space from
the elements of the real proper, orthochronous Lorentz group to the elements of
part (iv) of the real Lorentz group. There also exists a continuous path in
group space from the elements of part (ii) to the elements of part (iii) of the
real Lorentz group. Thus, while we know that the world is not invariant under
$\cP$ reflection or under $\cT$ reflection, we are proposing here to consider
the possibility suggested by complex group analysis that a fundamental discrete
symmetry of the world is $\cP\cT$ symmetry, or space-time reflection symmetry.

The most important consequence of the discovery that non-Hermitian $\cP
\cT$-symmetric Hamiltonians can define acceptable theories of quantum mechanics
is that we now can construct many new kinds of Hamiltonians that only a decade 
ago would have been rejected as being unphysical because they violate the
axiom of Hermiticity. In this paper we will examine some of these new
Hamiltonians and discuss the properties and possible physical implications of
the theories defined by these Hamiltonians. So far, there have been no
experiments that prove clearly and definitively that quantum systems defined by
non-Hermitian $\cP\cT$-symmetric Hamiltonians do exist in nature. However, one
should keep an open mind regarding the kinds of theories that one is willing to
consider. Indeed, Gell-Mann's ``totalitarian principle'' states that among all
possible physical theories ``Everything which is not forbidden is compulsory.''

\subsection{Presentation and Scope of this Paper}
\label{ss1-1}

Like many research areas in science, the study of non-Hermitian Hamiltonians
having real spectra began in a haphazard and diffuse fashion. There are numerous
early examples of isolated and disconnected discoveries of such non-Hermitian
Hamiltonians. For example, in 1980 Caliceti {\em et al.}, who were studying
Borel summation of divergent perturbation series arising from classes of
anharmonic oscillators, were astonished to find that the eigenvalues of an
oscillator having an imaginary cubic selfinteraction term are real \cite{H1}. In
the summer of 1993, when I was visiting CEN Saclay, I learned that Bessis and
Zinn-Justin had noticed on the basis of numerical work that some of the
eigenvalues of the cubic Hamiltonian in (\ref{e10}) seemed to be real, and they
wondered if the spectrum might be entirely real \cite{H2}. [Their interest in
the Hamiltonian (\ref{e10}) was piqued by the Lee-Yang edge singularity.] In
1982 Andrianov, who was doing perturbative studies of $-x^4$ potentials, found
evidence that such theories might have real eigenvalues \cite{H3}. In 1992
Hollowood \cite{H4} and Scholtz {\em et al.} \cite{H5} discovered in their own
areas of research surprising examples of non-Hermitian Hamiltonians having real
spectra.

The field of $\cP\cT$-symmetric quantum mechanics was established in 1998 with
the discovery by Bender and Boettcher that the numerical conjectures of Bessis
and Zinn-Justin were not only valid, but were just one instance of a huge class
of non-Hermitian Schr\"odinger eigenvalue problems whose spectra are entirely
real and positive \cite{r1}. Bender and Boettcher showed that the reality of
the spectra was due to a symmetry principle, namely the condition of an unbroken
space-time reflection symmetry, and they argued that this symmetry principle
could replace the usual requirement of Dirac Hermiticity.

This discovery by Bender and Boettcher relies on two essential mathematical
ingredients. First, Bender and Boettcher used the techniques of analytic
continuation of eigenvalue problems. These fundamental techniques were developed
and used heavily in the early work of Bender and Wu on divergent perturbation
series \cite{BW1,BW2} and were later used by Bender and Turbiner \cite{r4}.
These techniques are crucial because they show how to analytically continue the
boundary conditions of an eigenvalue problem as a function of a parameter in the
Hamiltonian. Second, Bender and Boettcher used the delta-expansion techniques
that had been discovered and developed by Bender {\em et al}. \cite{delta} as a
way to avoid divergent perturbation series. The delta expansion is a powerful
perturbation-theory technique in which the small perturbation parameter is a
measure of the nonlinearity of the problem. As a result, many crucial properties
of the problem are exactly preserved as this parameter varies. In the case of
$\cP\cT$-symmetric quantum mechanics, it is the reality of the eigenvalues that
is exactly maintained as the perturbation parameter $\epsilon$ in (\ref{e12}) is
varied.

Many researchers have contributed immensely to the development of $\cP
\cT$-symmetric quantum mechanics by discovering new examples and models, proving
theorems, and performing numerical and asymptotic analysis. In the past few
years a large and active research community has developed and there have been
half a dozen international conferences on the subject of $\cP\cT$ symmetry,
pseudo-Hermiticity, and non-Hermitian Hamiltonians. The proceedings of these
conferences provide a complete source of references \cite{C1,C2,C3,C4,C5,C6}.
By now, there have been so many contributions to the field that it is impossible
to describe them all in this one paper.

The purpose of this paper is to give an elementary introduction to this exciting
and active field of research. In writing this paper, my hope is that the rate of
new discoveries and the development of the field will continue at such a rapid
pace that this review will soon become obsolete.

\subsection{Organization of this Paper}
\label{ss1-2}

In Sec.~\ref{s2} we show that $\cP\cT$-symmetric Hamiltonians are complex
extensions of Hermitian Hamiltonians and we discuss the key property of $\cP
\cT$-symmetric Hamiltonians, namely, that their energy eigenvalues are real and
bounded below. We discuss techniques for calculating eigenvalues. In
Sec.~\ref{s3} we discuss classical $\cP\cT$-symmetric Hamiltonians and show how
to continue ordinary classical mechanics into the complex domain. The crucial
theoretical questions of conservation of probability and unitary time evolution
are addressed in Sec.~\ref{s4}. In Sec.~\ref{s5} we illustrate the theory of
$\cP\cT$-symmetric quantum mechanics by using a simple $2\times2$ matrix
Hamiltonian. Then in Sec.~\ref{s6} we explain how to calculate the $\cC$
operator, which is the central pillar of $\cP\cT$ symmetry and which is needed
to construct the Hilbert space for the theory. In the next two sections we
discuss the many applications of $\cP\cT$-symmetric quantum theory. We discuss
quantum-mechanical applications in Sec.~\ref{s7} and quantum-field-theoretic
applications in Sec.~\ref{s8}. Finally, in Sec.~\ref{s9} we make some brief
concluding remarks.

\section{Determining the Eigenvalues of a $\cP\cT$-Symmetric Hamiltonian}
\label{s2}

In this section we show how to calculate the eigenvalues of a $\cP\cT$-symmetric
Hamiltonian. We begin by pointing out that the Hamiltonian operator defines and
determines the physical properties of a quantum theory in three important ways:

\begin{itemize}

\item[(i)] {\em The Hamiltonian determines the energy levels of the quantum
theory}. To find these energy levels one must solve the time-independent
Schr\"odinger eigenvalue problem
\begin{equation}
H\psi=E\psi.
\label{e13}
\end{equation}
This equation usually takes the form of a differential equation that must be
solved subject to boundary conditions on the eigenfunction $\psi$. In the case 
of a $\cP\cT$-symmetric Hamiltonian it is crucial that the boundary conditions
be imposed properly. We emphasize that for a quantum theory to be physically
acceptable the energy eigenvalues of the Hamiltonian must be real and bounded
below.

\item[(ii)] {\em The Hamiltonian specifies the time evolution of the states
and operators of the quantum theory}. To determine the time evolution of a
state $\psi(t)$ in the Schr\"odinger picture we must solve the time-dependent
Schr\"odinger equation
\begin{equation}
i\frac{\partial}{\partial t}\psi(t)=H\psi(t).
\label{e14}
\end{equation}
The solution to this first-order differential equation is straightforward
because $H$ is assumed to be independent of time:
\begin{equation}
\psi(t)=e^{-iHt}\psi(0).
\label{e15}
\end{equation}
We call $e^{-iHt}$ the {\em time-evolution operator}. In conventional quantum
mechanics the time-evolution operator is unitary because the Hamiltonian $H$ is
Hermitian. As a result, the norm of the state $\psi(t)$ remains constant in
time. The constancy of the norm is an essential feature of a quantum system
because the norm of a state is a probability, and this probability must remain
constant in time. If this probability were to grow or decay in time, we would
say that the theory violates unitarity. In $\cP\cT$-symmetric quantum mechanics
$H$ is not Dirac Hermitian, but the norms of states are still time independent.

\item[(iii)] {\em The Hamiltonian incorporates the symmetries of the theory}. As
an example, suppose that the Hamiltonian $H$ commutes with the parity operator
$\cP$. We then say that the Hamiltonian is {\em parity invariant}. Since $\cP$
is a linear operator, we know that the eigenstates of the Hamiltonian [the
solutions to (\ref{e13})] will also be eigenstates of $\cP$. Thus, the
eigenstates of $H$ will have a definite parity; they will be either even or odd
under space reflection. (Of course, a general state, which is a linear
combination of the eigenstates of $H$, need not have definite parity.)

\end{itemize}

\subsection{Broken and Unbroken $\cP\cT$ Symmetry}
\label{ss2-1}

For the case of a $\cP\cT$-symmetric Hamiltonian, the $\cP\cT$ operator commutes
with the Hamiltonian $H$ [see (\ref{e9})]. However, $\cP\cT$ symmetry is more
subtle than parity symmetry because the $\cP\cT$ operator is not linear. Because
$\cP\cT$ is not linear, the eigenstates of $H$ may or may not be eigenstates of
$\cP\cT$.

Let us see what may go wrong if we assume that an eigenstate $\psi$ of the
Hamiltonian $H$ is also an eigenstate of the $\cP\cT$ operator. Call the
eigenvalue $\lambda$ and express the eigenvalue condition as
\begin{equation}
\cP\cT\psi=\lambda\psi.
\label{e16}
\end{equation}
Multiply (\ref{e16}) by $\cP\cT$ on the left and use the property that $(\cP\cT
)^2={\bf 1}$ [see (\ref{e7}) and (\ref{e8})]:
\begin{equation}
\psi=(\cP\cT)\lambda(\cP\cT)^2\psi.
\label{e17}
\end{equation}
Since $\cT$ is antilinear [see (\ref{e6})], we get
\begin{equation}
\psi=\lambda^*\lambda\psi=|\lambda|^2\psi.
\label{e18}
\end{equation}
Thus, $|\lambda|^2=1$ and the eigenvalue $\lambda$ of the $\cP\cT$ operator is a
pure phase:
\begin{equation}
\lambda=e^{i\alpha}.
\label{e19}
\end{equation}

Next, multiply the eigenvalue equation (\ref{e13}) by $\cP\cT$ on the left and
again use the property that $(\cP\cT)^2={\bf 1}$:
\begin{equation}
(\cP\cT)H\psi=(\cP\cT)E(\cP\cT)^2\psi.
\label{e20}
\end{equation}
Using the eigenvalue equation (\ref{e16}) and recalling that $\cP\cT$ commutes
with $H$, we get
\begin{equation}
H\lambda\psi=(\cP\cT)E(\cP\cT)\lambda\psi.
\label{e21}
\end{equation}
Finally, we again use the property that $\cT$ is antilinear to obtain
\begin{equation}
E\lambda\psi=E^*\lambda\psi.
\label{e22}
\end{equation}
Since $\lambda$ is nonzero [see (\ref{e19})], we conclude that the eigenvalue
$E$ is real: $E=E^*$.

In general, this conclusion is false, as Fig.~\ref{f1} clearly demonstrates.
When $\epsilon<0$, some of the eigenvalues have disappeared because they are
complex. On the other hand, for the restricted region $\epsilon\geq0$ this
conclusion is correct; all of the eigenvalues are indeed real. We are then led
to make the following definition: If every eigenfunction of a $\cP\cT$-symmetric
Hamiltonian is also an eigenfunction of the $\cP\cT$ operator, we say that the
$\cP\cT$ symmetry of $H$ is {\em unbroken}. Conversely, if some of the
eigenfunctions of a $\cP\cT$-symmetric Hamiltonian are not simultaneously
eigenfunctions of the $\cP\cT$ operator, we say that the $\cP\cT$ symmetry of
$H$ is {\em broken}.

The correct way to interpret (\ref{e22}) is that if a Hamiltonian has an
unbroken $\cP\cT$ symmetry, then all of its eigenvalues are real. Thus, to
establish that the eigenvalues of a particular $\cP\cT$-symmetric Hamiltonian
are real, it is necessary to prove that the $\cP\cT$ symmetry of $H$ is
unbroken. This is difficult to show, and it took several years after the
discovery of the family of $\cP\cT$-symmetric Hamiltonians in (\ref{e12}) before
a complete and rigorous proof was finally constructed by Dorey {\em et al.} in
2001 \cite{D1,D2}.\footnote{The proof by Dorey {\em et al.} draws from many
areas of theoretical and mathematical physics and uses spectral determinants,
the Bethe {\em ansatz}, the Baxter $T$-$Q$ relation, the monodromy group, and an
array of techniques used in conformal quantum field theory. This proof is too
technical to be described in this paper, but it was a significant advance
because it establishes a correspondence between ordinary differential equations
and integrable models. This correspondence is known as the {\em ODE-IM
correspondence} \cite{D3,D4,D5}.} Many others have contributed to the rigorous
mathematical development of the theory of $\cP\cT$ symmetry. These include Shin
\cite{SHIN}, Pham \cite{PP1}, Delabaere \cite{DD2}, Trinh \cite{TRINH}, Weigert
\cite{WWWW,WWWW1}, and Scholtz and Geyer \cite{GEY}. Mostafazadeh generalized
$\cP\cT$ symmetry to pseudo-Hermiticity (see Subsec.~\ref{ss4-5}).

\subsection{Boundary Conditions for the Schr\"odinger Eigenvalue Problem}
\label{ss2-2}

Our objective in this section is to show how to calculate the eigenvalues of the
Schr\"odinger eigenvalue problem (\ref{e13}). The most direct approach is to
write (\ref{e13}) as a differential equation in coordinate space. The principal
conceptual difficulty that we face in solving this differential equation is in
identifying and understanding the boundary conditions on the coordinate-space 
eigenfunctions.

To write (\ref{e13}) in coordinate space, we make the standard transcriptions
\begin{equation}
\x\to x\quad{\rm and}\quad\p\to-i\frac{d}{dx},
\label{e23}
\end{equation}
except that we treat the variable $x$ as complex. The Schr\"odinger eigenvalue
problem (\ref{e13}) then takes the form
\begin{equation}
-\psi''(x)+x^2(ix)^\epsilon\psi(x)=E\psi(x).
\label{e24}
\end{equation}

Although we cannot solve this equation exactly for arbitrary $\epsilon$, we can
easily find the possible asymptotic behaviors of its solutions by using the WKB
approximation. In general, for any differential equation of the form $-y''(x)+
V(x)y(x)=0$, where $V(x)$ is a function that grows as $|x|\to\infty$, we know 
that the exponential component of the asymptotic behavior of $y(x)$ for large
$|x|$ has the form
\begin{equation}
y(x)\sim\exp\left[\pm\int^x ds\sqrt{V(s)}\right].
\label{e25}
\end{equation}

To identify the appropriate boundary conditions to impose on $\psi(x)$, we
consider first the Hermitian case $\epsilon=0$ (the harmonic oscillator). From
(\ref{e25}) we can see immediately that the possible asymptotic behaviors of
solutions are $\psi(x)\sim\exp\left(\pm\half x^2\right)$. The usual requirement
that the eigenfunction be square-integrable implies that we must choose the
negative sign in the exponent, and therefore the eigenfunctions are
Gaussian-like for large $|x|$. This result extends into the complex-$x$ plane:
If the eigenfunctions vanish exponentially on the real-$x$ axis for large $|x|$,
they must also vanish in two wedges of opening angle $\half\pi$ in the complex
plane centered about the positive-real and negative-real axes. These wedges are
called {\em Stokes wedges} \cite{BO}.

What happens as $\epsilon$ increases from $0$? As soon as $\epsilon>0$
($\epsilon$ noninteger), a logarithmic branch point appears at the origin $x=0$.
Without loss of generality, we may choose the branch cut to run up the imaginary
axis from $x=0$ to $x=i\infty$. In this cut plane the solutions to the
differential equation (\ref{e24}) are single-valued. From the asymptotic
behavior of $\psi(x)$ in (\ref{e25}), we deduce that the Stokes wedges rotate
downward into the complex-$x$ plane and that the opening angles of the wedges
decrease as $\epsilon$ increases.

\begin{figure}[t!]
\vspace{2.3in}
\includegraphics{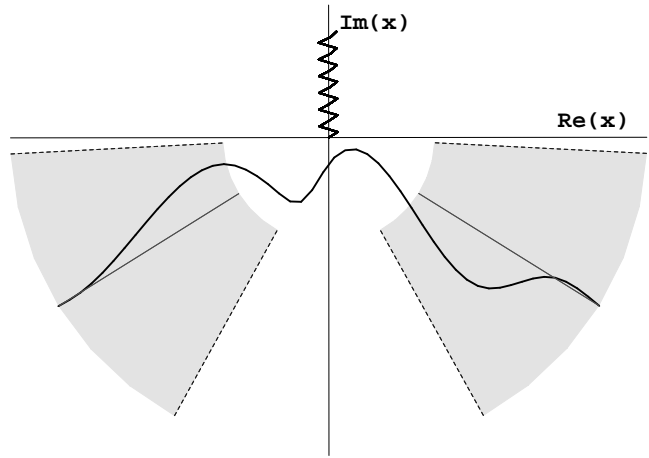}
\caption{Stokes wedges in the complex-$x$ plane containing the contour on which
the eigenvalue problem for the differential equation (\ref{e24}) for $\epsilon=
2.2$ is posed. In these wedges $\psi(x)$ vanishes exponentially as $|x|\to
\infty$. The eigenfunction $\psi(x)$ vanishes most rapidly at the centers of
the wedges.}
\label{f2}
\end{figure}

There are many wedges in which $\psi(x)\to0$ as $|x|\to\infty$. Thus, there are
many eigenvalue problems associated with the differential equation (\ref{e24}).
We choose to continue the eigenvalue differential equation (\ref{e24}) smoothly
away from the location of the harmonic oscillator wedges at $\epsilon=0$. (A
detailed description of how to extend eigenvalue equations into the complex
plane may be found in Ref.~\cite{r4}.) For $\epsilon>0$ the center lines of the
left and right wedges lie at the angles
\begin{equation}
\theta_{\rm left}=-\pi+\frac{\epsilon}{\epsilon+4}~\frac{\pi}{2}\quad{\rm and}
\quad \theta_{\rm right}=-\frac{\epsilon}{\epsilon+4}~\frac{\pi}{2}.
\label{e26}
\end{equation}
The opening angle of each of these wedges is $\Delta=2\pi/(\epsilon+4)$. The
differential equation (\ref{e24}) may be integrated along any path in the
complex-$x$ plane as long as the ends of the path approach complex infinity
inside the left wedge and the right wedge. Note that these wedges contain the
real-$x$ axis when $-1<\epsilon<2$. However, as soon as $\epsilon$ is larger
than 2, the wedges rotate below the real-$x$ axis. These wedges are shown in
Fig.~\ref{f2}.

Notice that the wedges in Fig.~\ref{f2} are mirror images of one another if they
are reflected through the imaginary-$x$ axis. This left-right symmetry is the
coordinate-space realization of $\cP\cT$ symmetry. If we choose any point $x$ in
the complex-$x$ plane and perform a parity reflection, then $x\to-x$. Time
reversal replaces $i$ by $-i$ as we saw in (\ref{e6}), and so $\cT$ replace $-x$
by its complex conjugate $-x^*$. Thus, in the coordinate representation $\cP\cT$
symmetry is left-right symmetry.

\subsection{The Flaw in Dyson's Argument}
\label{ss2-3}

The quantum theories that we are considering in this paper are obtained by
extending real quantum mechanics into the complex domain, as explained in
Subsec.~\ref{ss2-2}. The notion of analytically continuing a Hamiltonian into
the complex plane was first discussed in 1952 by Dyson, who argued heuristically
that perturbation theory for quantum electrodynamics diverges \cite{DY}. Dyson's
argument consists of rotating the electric charge $e$ into the complex-$e$
plane: $e\to ie$. Applied to the standard quantum anharmonic-oscillator
Hamiltonian $H=\p^2+g\x^4$, Dyson's argument goes as follows: Rotate the
parameter $g$ anticlockwise in the complex-$g$ plane from positive $g$ to $-g$.
Now, the potential term in the Hamiltonian is {\em no longer bounded below}, so
the resulting theory has no ground state. Hence, the ground-state energy $E_0(
g)$ has an abrupt transition at $g=0$, which implies that $E_0(g)$ must have a
singularity at $g=0$.

Following Dyson's reasoning, one would think that the spectrum of the
Hamiltonian (\ref{e11}), which is obtained by setting $\epsilon=2$ in
(\ref{e12}), would not be bounded below, and one might conclude that this
Hamiltonian is mathematically and physically unacceptable. However, this
heuristic argument is flawed. While the ground-state energy of the quantum
anharmonic oscillator does indeed have a singularity at $g=0$, the spectrum of
the Hamiltonian (\ref{e11}) that is obtained by analytically continuing a
parameter in the Hamiltonian remains ambiguous until the boundary conditions 
satisfied by the eigenfunctions are specified. The term ``bounded below'' is
inappropriate if it relies on an ordering relation applied to a complex
potential because ordering relations cannot be used for complex numbers. The
concern that the spectrum of $H$ in (\ref{e11}) is not bounded below is
unfounded because $\cP\cT$-symmetric boundary conditions on the Schr\"odinger
equation (\ref{e24}) prohibit the occurrence of negative eigenvalues.

The eigenvalues of $H$ in (\ref{e11}) depend crucially on the history of how
this negative-coupling-constant Hamiltonian constant is obtained. There are two
different ways to obtain $H$ in (\ref{e11}): First, one can substitute $g=|g|
e^{i\theta}$ into $H=\p^2+g\x^4$ and rotate from $\theta=0$ to $\theta=\pi$.
Under this rotation, the ground-state energy $E_0(g)$ becomes complex. Clearly,
$E_0(g)$ is real and positive when $g>0$ and complex when $g<
0$.\footnote{Rotating from $\theta=0$ to $\theta =-\pi$, we obtain the same
Hamiltonian as in (\ref{e11}), but the spectrum is the complex conjugate of the
spectrum obtained when we rotate from $\theta=0$ to $\theta= \pi$.} Second, one
can obtain (\ref{e11}) as a limit of the Hamiltonian $H$ in (\ref{e12}) as
$\epsilon:\,0\to2$. The spectrum of this complex Hamiltonian is real, positive,
and discrete, as is shown in Fig.~\ref{f1}.

How can the Hamiltonian (\ref{e11}) possess two such astonishingly different
spectra? The answer lies in the boundary conditions satisfied by the
eigenfunctions $\psi(x)$. In the first case, in which $\theta={\rm arg}\,g$ is
rotated in the complex-$g$ plane from $0$ to $\pi$, $\psi(x)$ vanishes in the
complex-$x$ plane as $|x|\to\infty$ inside the wedges $-\pi/3<{\rm arg}\,x<0$
and $-4\pi/3<{\rm arg}\,x<-\pi$. {\em These wedges are not $\cP\cT$-symmetric
reflections of one another}. In the second case, in which the exponent
$\epsilon$ ranges from $0$ to $2$, $\psi(x)$ vanishes in the complex-$x$ plane
as $|x|\to\infty$ inside the $\cP\cT$-symmetric pair of wedges $-\pi/3<{\rm arg}
\,x<0$ and $-\pi<{\rm arg}\,x<-2\pi/3$. We emphasize that in this second case
the boundary conditions hold in wedges that are symmetric with respect to the
imaginary axis; these boundary conditions enforce the $\cP\cT$ symmetry of $H$
and are responsible for the reality of the energy spectrum.

\medskip

\begin{itemize}
\item[~~]
\begin{footnotesize}
\noindent{\em Illustrative example:} The harmonic oscillator Hamiltonian
\begin{equation}
H=\p^2+\omega^2\x^2\quad(\omega>0)
\label{e27}
\end{equation}
is an elementary model that illustrates the dependence of the eigenvalues on the
boundary conditions imposed on the eigenfunctions. Let us assume that the
eigenfunctions of $H$ vanish on the real-$x$ axis as $x\to\pm\infty$. The
eigenfunctions have the form of a Gaussian $\exp\left(-\half\omega x^2\right)$
multiplied by a Hermite polynomial. Thus, the Stokes wedges are centered about
the positive real-$x$ axis and have angular opening $\half\pi$. The eigenvalues
$E_n$ are given exactly by the formula
\begin{equation}
E_n=\left(n+\half\right)\omega\qquad(n=0,\,1,\,2,\,3,\,\ldots).
\label{e28}
\end{equation}
Now suppose that the parameter $\omega$ is rotated by $180^\circ$ into the
complex-$\omega$ plane from the positive axis to the negative axis so that
$\omega$ is replaced by $-\omega$. This causes the Stokes wedges in the
complex-$x$ plane to rotate by $90^\circ$ so that the eigenfunctions now vanish
exponentially on the imaginary-$x$ axis rather than on the real-$x$ axis. Also,
as a consequence of this rotation, the {\em eigenvalues change sign}:
\begin{equation}
E_n=-\left(n+\half\right)\omega\qquad(n=0,\,1,\,2,\,3,\,\ldots).
\label{e29}
\end{equation}
Notice that under the rotation that replaces $\omega$ by $-\omega$ the
Hamiltonian remains invariant, and yet the signs of the eigenvalues are
reversed! This shows that the eigenspectrum depends crucially on the boundary
conditions that are imposed on the eigenfunctions.
\end{footnotesize}
\end{itemize}

\medskip
Apart from the eigenvalues, there is yet another striking difference between the
two theories corresponding to $H$ in (\ref{e11}). The expectation value of the
operator $\x$ in the ground-state eigenfunction $\psi_0(x)$ is given by
\begin{equation}
\frac{\langle0|x|0\rangle}{\langle0|0\rangle}\equiv\frac{\int_C dx\,x
\psi_0^2(x)}{\int_C dx\,\psi_0^2(x)},
\label{e30}
\end{equation}
where $C$ is a complex contour that lies in the asymptotic wedges described
above. The value of $\langle0|x|0\rangle/\langle0|0\rangle$ for $H$ in
(\ref{e11}) depends on the limiting process by which we obtain $H$. If we
substitute $g=g_0e^{i\theta}$ into the Hamiltonian $H=\p^2+g\x^4$ and rotate
$g$ from $\theta=0$ to $\theta=\pi$, we find by an elementary symmetry argument
that this expectation value vanishes for all $g$ on the semicircle in the
complex-$g$ plane. The expectation value vanishes because this rotation in the
complex-$g$ plane preserves parity symmetry ($x\to-x$). However, if we define
$H$ in (\ref{e11}) by using the Hamiltonian in (\ref{e12}) and by allowing
$\epsilon$ to range from $0$ to $2$, we find that this expectation value is
nonzero. In fact, this expectation value is nonvanishing for all $\epsilon>0$.
On this alternate path $\cP\cT$ symmetry (reflection about the imaginary axis,
$x\to-x^*$) is preserved, but parity symmetry is permanently broken. (We suggest
in Subsec.~\ref{ss8-4} that, as a consequence of broken parity symmetry, one
might be able to describe the dynamics of the Higgs sector by using a $\cP
\cT$-symmetric $-g\vf^4$ quantum field theory.)

\subsection{Using WKB Phase-Integral Techniques to Calculate Eigenvalues}
\label{ss2-4}

Now that we have identified the boundary conditions to be imposed on the
eigenfunctions of the $\cP\cT$-symmetric Hamiltonian (\ref{e12}), we can use a
variety of techniques to calculate the eigenvalues of this Hamiltonian. Not
surprisingly, it is impossible to solve the differential-equation eigenvalue
problem (\ref{e24}) analytically and in closed form except in two special cases,
namely, for $\epsilon=0$ (the harmonic oscillator) and for $\epsilon\to\infty$
(the $\cP\cT$-symmetric version of the square-well potential, whose solution is
given in Ref.~\cite{SQ}). Thus, it is necessary to use approximate analytic or
numerical methods.

The simplest analytic approach uses WKB theory, which gives an excellent
approximation to the eigenvalues when $\epsilon>0$. The WKB calculation is
interesting because it must be performed in the complex plane rather than on the
real-$x$ axis. The turning points $x_{\pm}$ are those roots of $E=x^2(ix
)^\epsilon$ that {\em analytically continue} off the real axis as $\epsilon$
increases from $0$. These turning points,
\begin{equation}
x_-=E^\frac{1}{\epsilon+2}e^{i\pi\left(\frac{3}{2}-\frac{1}{\epsilon+2}\right)},
\quad x_+=E^\frac{1}{\epsilon+2}e^{-i\pi\left(\frac{1}{2}-\frac{1}{\epsilon+2}
\right)},
\label{e31}
\end{equation}
lie in the lower-half (upper-half) $x$ plane in Fig.~\ref{f2} when $\epsilon>0$
($\epsilon<0$).

The leading-order WKB phase-integral quantization condition is given by
\begin{equation}
(n+1/2)\pi=\int_{x_-}^{x_+}dx\,\sqrt{E-x^2(ix)^\epsilon}.
\label{e32}
\end{equation}
When $\epsilon>0$ this path lies entirely in the lower-half $x$ plane, and when
$\epsilon=0$ (the case of the harmonic oscillator) the path lies on the real
axis. However, when $\epsilon<0$ the path lies in the upper-half $x$ plane and
crosses the cut on the positive imaginary-$x$ axis. In this case there is no
{\em continuous path joining the turning points.} Hence, WKB fails when
$\epsilon<0$.

When $\epsilon\geq0$, we deform the phase-integral contour so that it follows
the rays from $x_-$ to $0$ and from $0$ to $x_+$:
\begin{equation}
\left(n+\half\right)\pi=2\sin\left(\frac{\pi}{\epsilon+2}\right)
E^{\frac{1}{\epsilon+2}+\half}\int_0^1 ds\,\sqrt{1-s^{\epsilon+2}}.
\label{e33}
\end{equation}
We then solve for $E_n$:
\begin{equation}
E_n\sim\left[\frac{\Gamma\left(\frac{3}{2}+\frac{1}{\epsilon+2}\right)\sqrt{\pi}
\left(n+\half\right)}{\sin\left(\frac{\pi}{\epsilon+2}\right)\Gamma\left(1+\frac
{1}{\epsilon+2}\right)}\right]^\frac{2\epsilon+4}{\epsilon+4}\quad(n\to\infty).
\label{e34}
\end{equation}
This formula gives a very accurate approximation to the eigenvalues plotted in
Fig.~\ref{f1} and it shows, at least in the WKB approximation, that the energy
eigenvalues of $H$ in (\ref{e12}) are real and positive (see Table \ref{t1}). We
can, in addition, perform a higher-order WKB calculation by replacing the phase
integral by a {\em closed contour} that encircles the path joining the turning
points (see Refs.~\cite{BBM} and \cite{BO}).

\begin{table}
\caption[t1]{Comparison of the exact eigenvalues (obtained with Runge-Kutta)
and the WKB result in (\ref{e34}).}
\begin{tabular}{llrrllrr}
$\epsilon$ & $n$ & $E_{\rm exact}$ & $E_{\rm WKB}$ & $\epsilon$ & $n$ &
$E_{\rm exact}$ & $E_{\rm WKB}$ \\ \hline
1 & 0 & $1.156\,267\,072$ & 1.0943& 2 & 0 & $1.477\,149\,753$  &  1.3765 \\
&1 & $4.109\,228\,752$ & 4.0895&      & 1 & $6.003\,386\,082$  &  5.9558 \\
&2 & $7.562\,273\,854$ & 7.5489&      & 2 & $11.802\,433\,593$ & 11.7690 \\
&3 & $11.314\,421\,818$ & 11.3043&    & 3 & $18.458\,818\,694$ & 18.4321 \\
&4 & $15.291\,553\,748$ & 15.2832&    & 4 & $25.791\,792\,423$ & 25.7692 \\
&5 & $19.451\,529\,125$ & 19.4444&    & 5 & $33.694\,279\,298$ & 33.6746 \\
&6 & $23.766\,740\,439$ & 23.7606&    & 6 & $42.093\,814\,569$ & 42.0761 \\
&7 & $28.217\,524\,934$ & 28.2120&    & 7 & $50.937\,278\,826$ & 50.9214 \\
&8 & $32.789\,082\,922$ & 32.7841&    & 8 & $60.185\,767\,651$ & 60.1696 \\
&9 & $37.469\,824\,697$ & 37.4653&    & 9 & $69.795\,703\,031$ & 69.7884 \\
\end{tabular}
\label{t1}
\end{table}

\begin{figure}[t!]
\vspace{2.6in}
\includegraphics{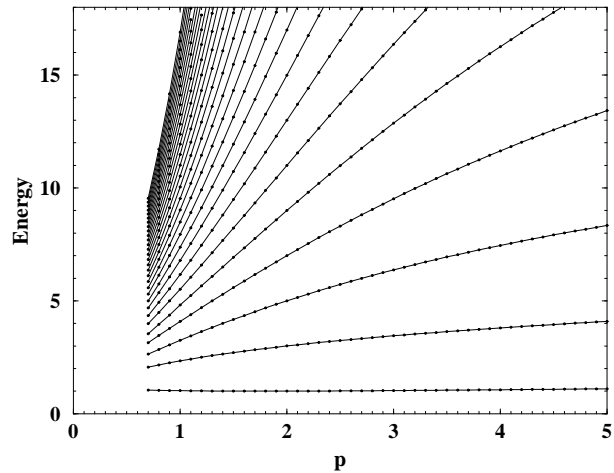}
\caption{Energy levels of the Hamiltonian $H=p^2+|x|^P$ as a function of the
real parameter $P$. This figure is similar to Fig.~\ref{f1}, but the eigenvalues
do not pinch off and go into the complex plane because the Hamiltonian is
Hermitian. (The spectrum becomes dense at $P=0$.)}
\label{f3}
\end{figure}

It is interesting that the spectrum of the real $|x|^{\epsilon+2}$ potential
strongly resembles that of the $x^2(ix)^\epsilon$ potential. The leading-order
WKB quantization condition (accurate for $\epsilon>-2$) is like that in
(\ref{e34}) except that the factor of $\sin\left(\frac{\pi}{\epsilon+2}\right)$
is absent. However, as $\epsilon\to\infty$, the spectrum of $|x|^{\epsilon+2}$
approaches that of the square-well potential [$E_n=(n+1)^2\pi^2/4$], while the
energies of the complex $x^2(ix)^\epsilon$ potential diverge, as Fig.~\ref{f1}
indicates. The energies of the $|x|^P$ potential are shown in Fig.~\ref{f3}.

\subsection{Numerical Calculation of Eigenvalues}
\label{ss2-5}

There are several highly accurate numerical techniques for computing the
energy spectrum that is displayed in Fig.~\ref{f1}. The simplest and most direct
method is to integrate the Schr\"odinger differential equation (\ref{e24}) using
a Runga-Kutta approach. To do so, we convert this complex differential equation
to a system of coupled, real, second-order equations. The convergence is most
rapid when we integrate along paths located at the centers of the Stokes wedges
and follow these paths out to $\infty$. We then patch the two solutions in each
Stokes wedge together at the origin. This procedure, which is described in
detail in Ref.~\cite{r1}, gives highly accurate numerical results.

An alternative is to perform a variational calculation in which we determine
the energy levels by finding the stationary points of the functional
\begin{equation}
\langle H\rangle(a,b,c)\equiv\frac{\int_C dx\,\psi(x)H\psi(x)}{\int_C dx\,\psi^2
(x)},
\label{e35}
\end{equation}
where
\begin{equation}
\psi(x)=(ix)^c\exp\left[a(ix)^b\right]
\label{e36}
\end{equation}
is a three-parameter class of $\cP\cT$-invariant trial wave functions
\cite{VAN}. The integration contour $C$ used to define $\langle H\rangle(a,b,c)$
must lie inside the wedges in the complex-$x$ plane in which the wave function
falls off exponentially at infinity (see Fig.~\ref{f2}). Rather than having a
local minimum, the functional has a saddle point in $(a,b,c)$-space. At this
saddle point the numerical prediction for the ground-state energy is extremely
accurate for a wide range of $\epsilon$. This method also determines approximate
eigenfunctions and eigenvalues of the excited states of these non-Hermitian
Hamiltonians. Handy used the numerical technique of solving the coupled moment
problem \cite{HANDY}. This technique, which produces accurate results for the
eigenvalues, is the exact quantum-mechanical analog of solving the
Schwinger-Dyson equations in quantum field theory.

\subsection{The Remarkable Case of a $\cP\cT$-Symmetric $-x^4$ Potential}
\label{ss2-6}

The $\cP\cT$-symmetric $-x^4$ Hamiltonian in (\ref{e11}), which is obtained by
setting $\epsilon=2$ in (\ref{e12}), is particularly interesting because it is
possible to obtain the energy spectrum by using real analysis alone; that is,
one can focus on real $x$ only and avoid having to perform analysis in the
complex-$x$ plane. One way to proceed is to deform the integration contour shown
in Fig.~\ref{f2} to the upper edges of the wedges so that it lies entirely on
the real-$x$ axis. If this is done carefully, the exact eigenvalues for this
potential can be obtained by solving the Schr\"odinger equation (\ref{e24})
subject to the boundary conditions that the potential be {\em reflectionless}
\cite{BERRY}. That is, an incoming incident wave from the left gives rise to an
outgoing transmitted wave on the right, but no reflected wave on the left. (This
observation may have consequences in cosmological models, as explained in
Subsec.~\ref{ss8-7}.)

Another way to proceed is to show that the eigenvalues of the non-Hermitian $-
x^4$ Hamiltonian are identical with the eigenvalues of a conventional Hermitian
Hamiltonian having a positive $x^4$ potential. A number of authors have observed
and discussed this equivalence \cite{H3,BG,JM}. Here, we use elementary
differential-equation methods \cite{BBCJM} to prove that the spectrum of the
non-Hermitian $\cP\cT$-symmetric Hamiltonian
\begin{equation}
H=\frac{1}{2m}\p^2-g\x^4\quad(g>0)
\label{e37}
\end{equation}
is identical to the spectrum of the Hermitian Hamiltonian
\begin{equation}
\tilde H=\frac{1}{2m}\p^2+4g\x^4-\hbar\sqrt{\frac{2g}{m}}\,\x\quad(g>0).
\label{e38}
\end{equation}
(We have included the dimensional constants $m$, $g$, and $\hbar$ because they
help to elucidate the physical significance of the spectral equivalence of these
two very different Hamiltonians.) To show that $H$ in (\ref{e37}) and $\tilde H$
in (\ref{e38}) are equivalent, we examine the corresponding Schr\"odinger
eigenvalue equations
\begin{equation}
-\frac{\hbar^2}{2m}\psi''(x)-gx^4\psi(x)=E\psi(x)
\label{e39}
\end{equation}
and
\begin{equation}
-\frac{\hbar^2}{2m}\Phi''(x)+\left(-\hbar\sqrt{\frac{2g}{m}}\,x+4gx^4\right)
\Phi(x)=E\Phi(x).
\label{e40}
\end{equation}

\begin{figure*}[t!]
\vspace{1.9in}
\includegraphics{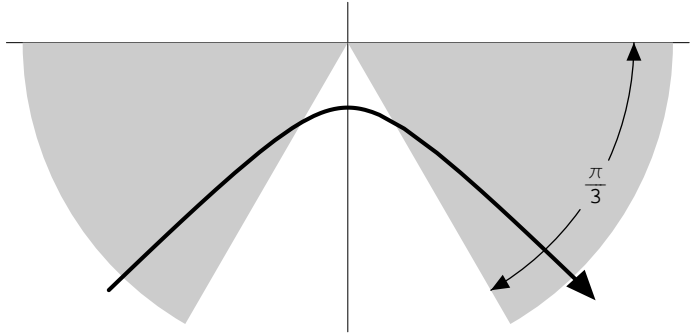}
\caption{Stokes wedges in the lower-half complex-$x$ plane for the
Schr\"odinger equation (\ref{e39}) arising from the Hamiltonian $H$ in
(\ref{e37}). The eigenfunctions of $H$ decay exponentially as $|x|\to\infty$
inside these wedges. Also shown is the contour in (\ref{e41}).}
\label{f4}
\end{figure*}

We begin by moving the complex integration contour for the Schr\"odinger
equation (\ref{e39}) to the real axis. To do so, we parameterize the integration
contour using
\begin{equation}
x=-2iL\sqrt{1+iy/L},
\label{e41}
\end{equation}
where
\begin{equation}
L=\lambda\left[\hbar^2/(mg)\right]^{1/6}
\label{e42}
\end{equation}
and $y$ is a real parameter that ranges from $-\infty$ to $\infty$. A graph of
the contour in (\ref{e41}) is shown in Fig.~\ref{f4}. The transformed
differential equation then reads
\begin{equation}
-\frac{\hbar^2}{2m}\left(1+\frac{iy}{L}\right)\phi''(y)
-\frac{i\hbar^2}{4Lm}\phi'(y)-16gL^4\left(1+\frac{iy}{L}
\right)^2\phi(y)=E\phi(y).
\label{e43}
\end{equation}

Next, we perform a Fourier transform defined by
\begin{equation}
\tilde f(p)\equiv\int_{-\infty}^\infty dy\,e^{-iyp/\hbar}f(y).
\label{e44}
\end{equation}
By this definition the Fourier transforms of a derivative and a product are
given by
\begin{equation}
f'(y)\to ip\tilde f(p)/\hbar\quad{\rm and}\quad yf(y)\to i\hbar\tilde f'(p).
\label{e45}
\end{equation}
Thus, the transformed version of (\ref{e43}) reads
\begin{eqnarray}
&&\frac{1}{2m}\left(1-\frac{\hbar}{L}\frac{d}{dp}\right)p^2\tilde\phi(p)
+\frac{\hbar}{4Lm}p\tilde\phi(p)\nonumber\\
&&\qquad -16gL^4\left(1-\frac{\hbar}{L}\frac{d}{dp}
\right)^2\tilde\phi(p)=E\tilde\phi(p).
\label{e46}
\end{eqnarray}
We expand and simplify the differential equation in (\ref{e46}) and get
\begin{eqnarray}
&&-16gL^2\hbar^2\tilde\phi''(p)+\left(-\frac{\hbar p^2}{2mL}+32gL^3\hbar\right)
\tilde\phi'(p)\nonumber\\
&&\qquad+\left(\frac{p^2}{2m}-\frac{3p\hbar}{4mL}-16gL^4\right)\tilde\phi(p)
=E\tilde\phi(p).
\label{e47}
\end{eqnarray}

Next, we eliminate the one-derivative term in the differential equation
(\ref{e47}) to convert it to the form of a Schr\"odinger equation. To do so,
we substitute
\begin{equation}
\tilde\phi(p)=e^{F(p)}\Phi(p),
\label{e48}
\end{equation}
where
\begin{equation}
F(p)=\frac{L}{\hbar}p-\frac{1}{192gmL^3\hbar}p^3.
\label{e49}
\end{equation}
The resulting equation is
\begin{equation}
-16gL^2\hbar^2\Phi''(p)+\left(\frac{p^4}{256gm^2 L^4}
-\frac{\hbar p}{4mL}\right)\Phi(p)=E\Phi(p).
\label{e50}
\end{equation}

Last, we rescale (\ref{e50}) by substituting
\begin{equation}
p=xL\sqrt{32mg}
\label{e51}
\end{equation}
and we obtain the Schr\"odinger differential equation (\ref{e40}). The completes
the proof and verifies that the eigenvalues of the two Hamiltonians (\ref{e37})
and (\ref{e38}) have identical eigenvalues. This demonstration of equivalence is
exact; no approximations were made in this argument.\footnote{A simple and exact
transformation like the one presented in this subsection for mapping a $\cP
\cT$-symmetric non-Hermitian Hamiltonian to a Hermitian Hamiltonian has been
found only for the isolated case $\epsilon=2$ in (\ref{e12}). [A more
complicated spectral equivalence exists for the special case $\epsilon=4$ (see
Ref.~\cite{D1,D2}).]}

\subsection{Parity Anomaly}
\label{ss2-7}

The proof in Subsec.~\ref{ss2-6} that the Hamiltonians in (\ref{e37}) and
(\ref{e38}) are equivalent helps to clarify some of the physical content of the
non-Hermitian $\cP\cT$-symmetric Hamiltonian (\ref{e37}). The interpretation of
the result in (\ref{e40}) is that the linear term in the potential of the
equivalent quartic Hermitian Hamiltonian in (\ref{e38}) is a parity {\em
anomaly}. In general, an {\em anomaly} is a purely quantum (nonclassical) effect
that vanishes in the classical limit $\hbar\to0$. There is no classical analog
of an anomaly because Planck's constant $\hbar$ does not appear in classical
mechanics.

We refer to the linear term in (\ref{e38}) as a {\em parity} anomaly for the
following reason: Even though the $\cP\cT$-symmetric Hamiltonian (\ref{e37})
is symmetric under the parity reflections defined in (\ref{e3}), $H$ does not
respect parity symmetry. The violation of parity symmetry is subtle because it
is contained in the boundary conditions that the eigenfunctions of the
associated Schr\"odinger equation must satisfy. Since these boundary conditions
are given at $|x|=\infty$, the violation of parity symmetry is not detectable
in any finite domain in the complex-$x$ plane. Classical motion is a local
phenomenon. Therefore, a classical particle that is moving under the influence
of this Hamiltonian (see Sec.~\ref{s3}) will act as if it is subject to
parity-symmetric forces; the classical particle cannot feel the influences of
quantum boundary conditions imposed at $|x|=\infty$. In contrast, a quantum
wave function is inherently nonlocal because it must obey boundary conditions
that are imposed at $|x|=\infty$. Thus, only a quantum particle ``knows'' about
the violation of parity symmetry. To establish the equivalence between the
Hamiltonians in (\ref{e37}) and (\ref{e38}) it was necessary to perform a
Fourier transform [see (\ref{e46})]. This transformation maps the point at $x=
\infty$ to the point at $p=0$, and this explains the presence of the linear
parity-violating term in the potential of $\tilde H$ in (\ref{e38}). The
violation of parity is now a visible local effect in the Hamiltonian $\tilde H$.
However, this violation of parity is proportional to $\hbar$ and evaporates in
the classical limit $\hbar\to0$.

The Hamiltonian (\ref{e38}) is Hermitian in the usual Dirac sense and its energy
spectrum is bounded below. This Hamiltonian is also $\cP\cT$-symmetric because
at every stage in the sequence of differential-equation transformations in
Subsec.~\ref{ss2-6}, $\cP\cT$ symmetry is preserved. However, the variable
$x$ that gives rise to the parity anomaly in (\ref{e40}) is {\em not} a
coordinate variable. Its behavior is that of a momentum variable because $x$
changes sign under time reversal.

The violation of parity symmetry at the quantum level has important physical
implications. It is the lack of parity symmetry that implies that the one-point
Green's function in the corresponding quantum field theory does not vanish. A
possible consequence of this is that the elusive Higgs particle, which is a
fundamental ingredient in the standard model of particle physics, is a quantum
anomaly. In the next subsection, we show that the parity anomaly has a major
impact on the spectrum of bound states in a quantum theory.

\subsection{Physical Consequence of the Parity Anomaly: Appearance of Bound
States in a $\cP\cT$-Symmetric Quartic Potential}
\label{ss2-8}

A direct physical consequence of the parity anomaly is the appearance of bound
states. To elucidate the connection between the parity anomaly and bound states,
we generalize the Hamiltonian (\ref{e37}) to include a harmonic ($\x^2$) term in
the potential:
\begin{eqnarray}
H=\frac{1}{2m}\p^2+\frac{\mu^2}{2}\x^2-g\x^4.
\label{e52}
\end{eqnarray}
The same differential-equation analysis used in Subsec.~\ref{ss2-6}
straightforwardly yields the equivalent Hermitian Hamiltonian \cite{BG,JM}
\begin{eqnarray}
\tilde H=\frac{\p^2}{2m}-\hbar\sqrt{\frac{2g}{m}}\,\x+4g\left(\x^2-\frac{\mu^2}
{8g}\right)^2.
\label{e53}
\end{eqnarray}
Note that for these more general Hamiltonians the linear anomaly term remains
unchanged from that in (\ref{e38}).

It was shown in an earlier paper \cite{BOUND} that the Hamiltonian (\ref{e52})
exhibits bound states. In particle physics a {\em bound state} is a state having
a negative binding energy. Bound states in the context of a quantum mechanics
are defined as follows: Let the energy levels of the Hamiltonian be $E_n$ ($n=0,
1,2,\ldots$). The {\em renormalized mass} is the mass gap; that is, $M=E_1-E_0$.
Higher excitations must be measured relative to the vacuum energy: $E_n-E_0$
($n=2,3,4,\ldots$). We say that the $n$th higher excitation is a bound state if
the binding energy
\begin{equation}
B_n\equiv E_n-E_0-nM
\label{e54}
\end{equation}
is negative. If $B_n$ is positive, then we regard the state as {\em unbound}
because this state can decay into $n$ 1-particle states of mass $M$ in the
presence of an external field.

In Ref.~\cite{BOUND} it was shown numerically that for small positive values of
$g$ the first few states of $H$ in (\ref{e52}) are bound. As $g$ increases, the
number of bound states decreases until, when $g/\mu^3$ is larger than the
critical value $0.0465$, there are no bound states at all.\footnote{In Ref.~
\cite{BOUND} a heuristic argument was given to explain why there is such a
critical value. This argument is heuristic because the non-Hermitian Hamiltonian
is evaluated for $x$ in the complex plane. As is explained in
Subsec.~\ref{ss2-3}, when $x$ is complex, one cannot use
order relationships such as $>$ or $<$, which only apply to real numbers.}

Because $H$ in (\ref{e52}) has the same spectrum as the Hermitian Hamiltonian in
(\ref{e53}), it is easy to explain the appearance of bound states and to show
that the bound states are a direct consequence of the linear anomaly term. To
probe the influence of the anomaly, we generalize (\ref{e53}) by inserting a
dimensionless parameter $\epsilon$ that measures the strength of the anomaly
term:
\begin{eqnarray}
\tilde H=\half\p^2-\epsilon\sqrt{2g}\,\x+4g\left(\x^2-\frac{1}{8g}
\right)^2,
\label{e55}
\end{eqnarray}
where for simplicity we have set $m=\mu=\hbar=1$.

\begin{figure}[t!]
\vspace{2.05in}
\includegraphics{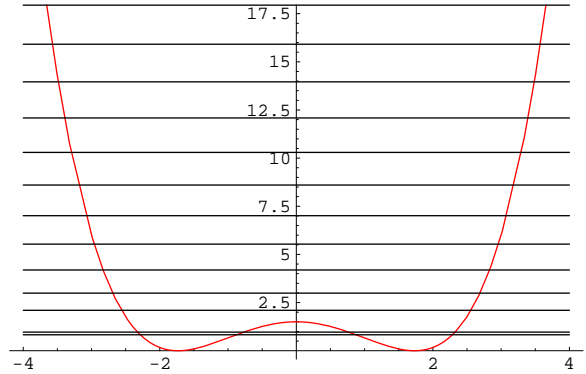}
\caption{Potential of the Hermitian Hamiltonian (\ref{e55}) plotted as a
function of the real variable $x$ for the case $\epsilon=0$ and $g=0.046$. The
energy levels are indicated by horizontal lines. Because $\epsilon=0$, there is
no anomaly and the double-well potential is symmetric. Therefore, the mass gap
is very small and thus there are no bound states.}
\label{f5}
\end{figure}

If we set $\epsilon=0$, there is no anomaly term and the potential is a {\em
symmetric} double well. The mass gap for a double well is exponentially small
because it is a result of the tunneling between the wells and thus the
renormalized mass $M$ is very small. Therefore, $B_n$ in (\ref{e54}) is positive
and there are no bound states. In Fig.~\ref{f5} we display the double-well
potential and the first several states of the system for the case $g=0.046$ and
$\epsilon=0$. There is a very small splitting between the lowest two states.

\begin{figure}[b!]
\vspace{1.95in}
\includegraphics{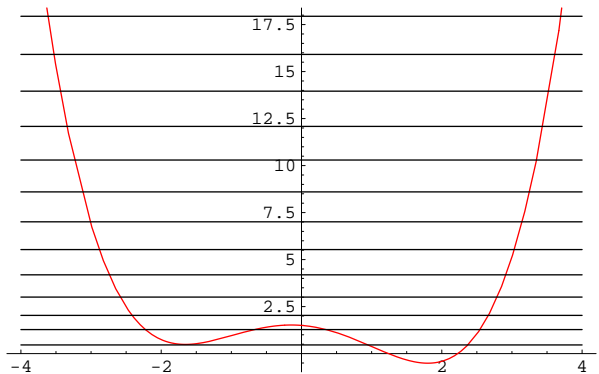}
\caption{Asymmetric potential well plotted as a function of the real variable
$x$ for the Hermitian Hamiltonian (\ref{e55}) with $\epsilon=1$ and $g=0.046$.
The energy levels are indicated by horizontal lines. There is one bound state.
The occurrence of bound states is due to the anomaly.}
\label{f6}
\end{figure}

If $\epsilon=1$, the double-well potential is asymmetric and the two lowest
states are not approximately degenerate. Therefore, bound states can occur near
the bottom of the potential well. Higher-energy states eventually become unbound
because, as we know from WKB theory, in a quartic well the $n$th energy level
grows like $n^{4/3}$ for large $n$. As $g$ becomes large, the number of bound
states becomes smaller because the depth of the double well decreases. For large
enough $g$ there are no bound states. In Fig.~\ref{f6} we display the potential
for $\epsilon=1$ and for $g=0.046$; for this $g$ there is one bound state.

\begin{figure}[t!]
\vspace{2.1in}
\includegraphics{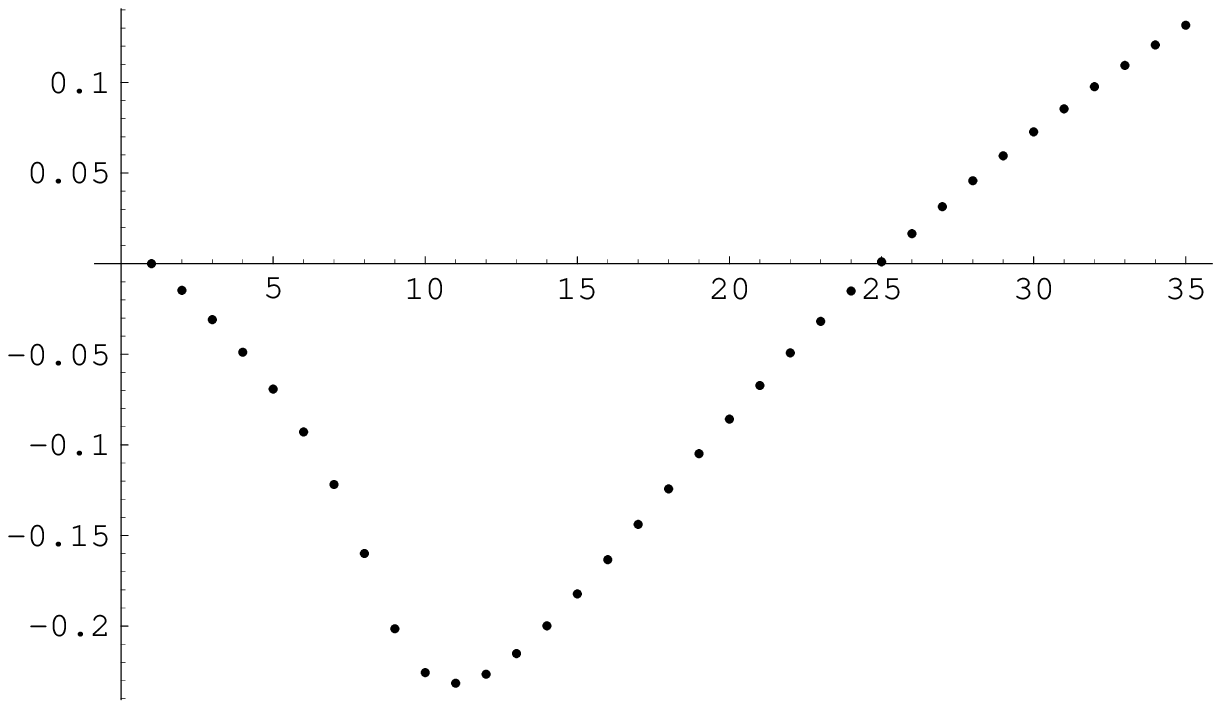}
\caption{Binding energies $B_n=E_n-E_0-nM$ plotted as a function of $n$ for $g=
0.008333$ and $\epsilon=1$. A negative value of $B_n$ indicates a bound state.
Observe that there are 23 bound states for these parameter values. Note that
$B_n$ is a smooth function of $n$.}
\label{f7}
\end{figure}

Another way to display the bound states is to plot the value of the binding 
energy $B_n$ as a function of $n$. For example, in Fig.~\ref{f7} we display the
bound states for $\epsilon=1$ and $g=0.008333$. Note that for these values there
are 23 bound states. Observe also that the binding energy $B_n$ is a smooth
function of $n$.

\begin{figure}[b!]
\vspace{1.95in}
\includegraphics{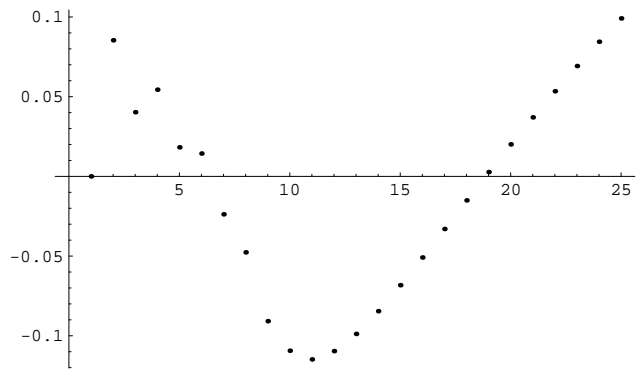}
\caption{Binding energies $B_n$ plotted as a function of $n$ for $g=0.008333$
and $\epsilon=0.9$. The first five states have now become unbound and $B_n$ is
not a smooth function of $n$ for $n\leq6$. The next 12 states are bound, and in
this region $B_n$ is a smooth function of $n$. Comparison of this figure with
Fig.~\ref{f7} shows that the bound-state spectrum is exquisitely sensitive to
the strength of the linear anomaly term.}
\label{f8}
\end{figure}

\begin{figure}[t!]
\vspace{2.15in}
\includegraphics{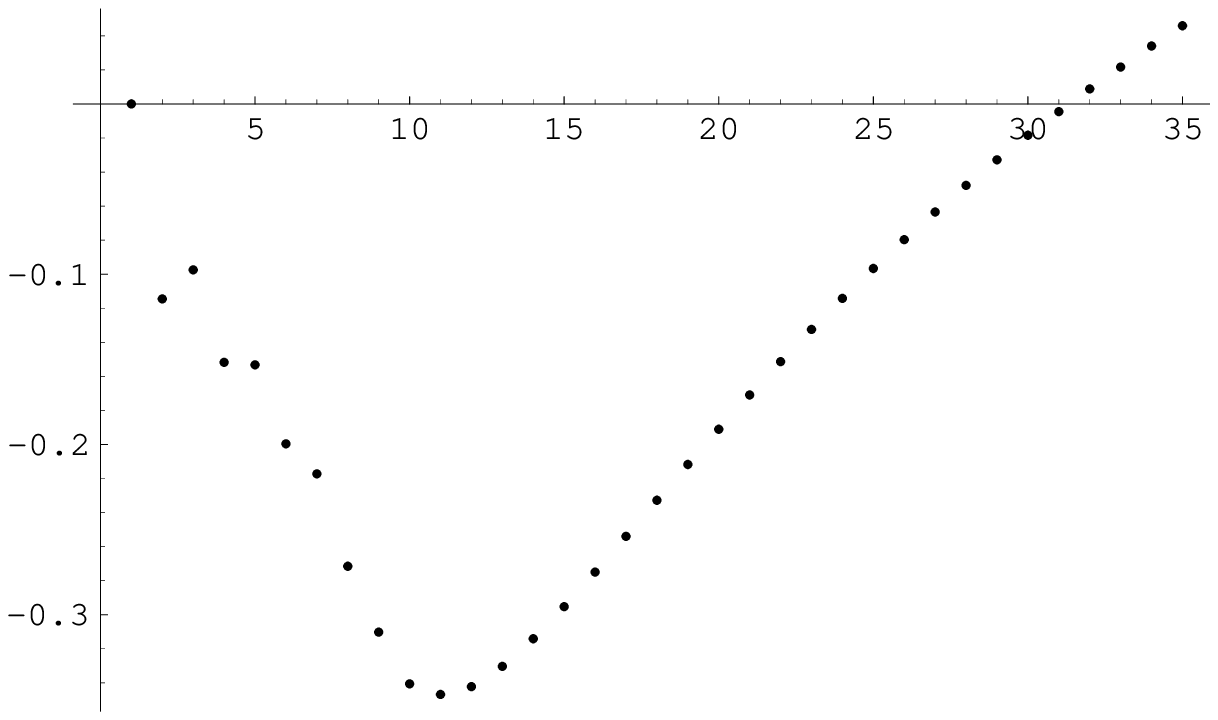}
\caption{Binding energies $B_n$ plotted as a function of $n$ for $g=0.008333$
and $\epsilon=1.1$. Note that there are 30 bound states and that $B_n$ is not a
smooth function of $n$ when $n$ is small.}
\label{f9}
\end{figure}

It is noteworthy that the bound-state spectrum depends so sensitively on the
strength of the anomaly term in the Hamiltonian (\ref{e55}). If $\epsilon$ is
slightly less than $1$, the first few states become unbound, as shown in
Fig.~\ref{f8}. In this figure $g=0.008333$ and $\epsilon=0.9$. If $\epsilon$ is
slightly greater than $1$, the binding energy $B_n$ is not a smooth function of
$n$ for small $n$. In Fig.~\ref{f9} we plot $B_n$ as a function of $n$ for $g=0.
008333$ and $\epsilon=1.1$. Note that for these values of the parameters there
are 30 bound states. Figures~\ref{f7}, \ref{f8}, and \ref{f9} are strikingly
different, which demonstrates the extreme sensitivity of the bound-state
spectrum to the anomaly term.

\section{$\cP\cT$-Symmetric Classical Mechanics --- The Strange Dynamics of a
Classical Particle Subject to Complex Forces}
\label{s3}

In this section we describe the properties of the $\cP\cT$-symmetric
classical-mechanical theory that underlies the quantum-mechanical theory
described by the Hamiltonian (\ref{e12}). We describe the motion of a particle
that feels complex forces and responds by moving about in the complex plane.
Several papers have been published in this area \cite{BBM,CL1,CL2,CL3,CL4,CL5}
and we summarize here some of the surprising discoveries.

One objective here is to explain heuristically how an upside-down potential like
that in (\ref{e11}) can have {\em positive}-energy quantum-mechanical
eigenstates. One might think (incorrectly!) that since a classical particle
would slide down the positive real axis to infinity, the corresponding quantum
states would be unstable and the spectrum of the quantum system would be 
unbounded below. In fact, when $\epsilon\geq0$ (the region of unbroken $\cP\cT$
symmetry), all but a set of measure zero of the possible classical paths are
confined and periodic, and thus the classical particle does not slide off to
infinity. When $\epsilon<0$ (the region of broken $\cP\cT$ symmetry), the
classical trajectories do indeed run off to infinity, and we can begin to
understand why the energy levels of the corresponding quantum system are
complex.

The equation of motion of a classical particle described by $H$ in (\ref{e12})
follows from Hamilton's equations:
\begin{equation}
{dx\over dt}={\partial H\over\partial p}=2p,\qquad
{dp\over dt}=-{\partial H\over\partial x}=i(2+\epsilon)(ix)^{1+\epsilon}.
\label{e56}
\end{equation}
Combining these two equations gives
\begin{equation}
{d^2x\over dt^2}=2i(2+\epsilon)(ix)^{1+\epsilon},
\label{e57}
\end{equation}
which is the {\em complex} version of Newton's second law, $F=ma$.

We can integrate (\ref{e57}) to give
\begin{equation}
{1\over2}{dx\over dt}=\pm\sqrt{E+(ix)^{2+\epsilon}},
\label{e58}
\end{equation}
where $E$ is the energy of the classical particle (the time-independent value of
$H$). We treat time $t$ as a real variable that parameterizes the complex path
$x(t)$ of this particle. Equation (\ref{e58}) is a complex generalization of
the concept that the velocity is the time derivative of the position ($v=\frac{d
x}{dt}$). Here, $t$ is real, but $v$ and $x$ are complex.

We now describe and classify the solutions to Eq.~(\ref{e58}). Because the
corresponding quantum theory possesses ${\cP\cT}$ invariance, we restrict our
attention to real values of $E$. Given this restriction, we can rescale $x$ and
$t$ by real numbers so that without loss of generality Eq.~(\ref{e58}) reduces
to
\begin{equation}
{dx\over dt}=\pm\sqrt{1+(ix)^{2+\epsilon}}.
\label{e59}
\end{equation}

\subsection{The Case $\epsilon=0$}
\label{ss3-1}

The classical solutions to Eq.~(\ref{e59}) have elaborate topologies, so we
begin by considering some special values of $\epsilon$. For the simplest case,
$\epsilon=0$, there are two turning points and these lie on the real axis at
$\pm1$. To solve Eq.~(\ref{e59}) we must specify the initial condition $x(0)$.
An obvious choice for $x(0)$ is a turning point. If the path begins at $\pm1$,
there is a unique trajectory in the complex-$x$ plane that solves (\ref{e59}).
This trajectory lies on the real axis and oscillates between the turning points.
This is the usual sinusoidal harmonic motion.

Choosing the energy determines the locations of the turning points, and choosing
the initial position of the particle determines the initial velocity (up to a
plus or minus sign) as well. So if the path of the particle begins anywhere on
the real axis between the turning points, the initial velocity is fixed up to a
sign and the trajectory of the particle still oscillates between the turning
points.

\begin{figure*}[b!]
\vspace{2.2in}
\includegraphics{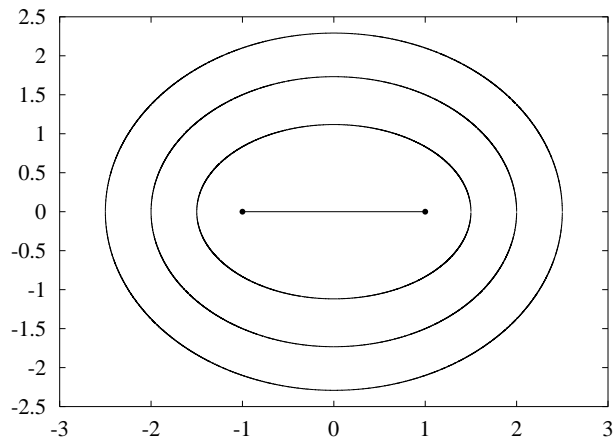}
\caption{Classical trajectories in the complex-$x$ plane for the
harmonic-oscillator Hamiltonian $H=p^2+x^2$. These trajectories are the
complex paths of a particle whose energy is $E=1$. The trajectories are nested
ellipses with foci located at the turning points at $x=\pm1$. The real line
segment (degenerate ellipse) connecting the turning points is the conventional
real periodic classical solution to the harmonic oscillator. All paths are
closed orbits having the same period $2\pi$.}
\label{f10}
\end{figure*}

In conventional classical mechanics the only possible initial positions for the
particle are on the real-$x$ axis between the turning points because the
velocity is real; all other points on the real axis belong to the so-called
classically forbidden region. However, because we are analytically continuing
classical mechanics into the complex plane, we can choose any point $x(0)$ in
the complex plane as an initial position. For all complex initial positions
outside of the conventional classically allowed region the classical trajectory
is an ellipse whose foci are the turning points. The ellipses are nested because
no trajectories may cross. These ellipses are shown in Fig.~\ref{f10}.

The exact harmonic-oscillator ($\epsilon=0$) solution to (\ref{e59}) is
\begin{equation}
x(t)=\cos[{\rm arccos}\,x(0)\pm t],
\label{e60}
\end{equation}
where the sign of $t$ determines the direction (clockwise or anticlockwise) in
which the particle traces out the ellipse. For {\em any} ellipse the period is
$2\pi$. The period is the same for all trajectories because we can join the
square-root branch points by a single finite branch cut lying along the real
axis from $x=-1$ to $x=1$. The complex path integral that determines the period
can then be shrunk (by Cauchy's theorem) to the usual real integral joining the
turning points.

Note that all of the elliptical orbits in Fig.~\ref{f10} are symmetric with
respect to parity $\cP$ (reflections through the origin) and time reversal $\cT$
(reflections about the real axis) as well as to $\cP\cT$ (reflections about the
imaginary axis). Furthermore, $\cP$ and $\cT$ individually preserve the
directions in which the ellipses are traversed.

\subsection{The Case $\epsilon=1$}
\label{ss3-2}

When $\epsilon=1$, there are three turning points. These turning points solve
the equation $ix^3=1$. Two lie below the real axis and are symmetric with
respect to the imaginary axis:
\begin{equation}
x_-=e^{-5i\pi/6}\quad{\rm and}\quad x_+=e^{-i\pi/6}.
\label{e61}
\end{equation}
Under $\cP\cT$ reflection $x_-$ and $x_+$ are interchanged. The third turning
point lies on the imaginary axis at $x_0=i$.

\begin{figure*}[t!]
\vspace{2.4in}
\includegraphics{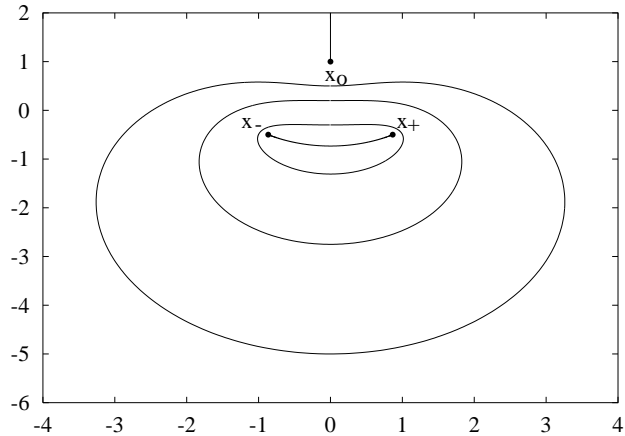}
\caption{Classical trajectories in the complex-$x$ plane for a particle of
energy $E=1$ described by the Hamiltonian $H=p^2+ix^3$. An oscillatory
trajectory connects the turning points $x_{\pm}$. This trajectory is enclosed by
a set of closed, nested paths that fill the finite complex-$x$ plane except for
points on the imaginary axis at or above the turning point $x_0=i$. Trajectories
that originate on the imaginary axis above $x=i$ either move off to $i\infty$ or
else approach $x_0$, stop, turn around, and then move up the imaginary axis to
$i\infty$.}
\label{f11}
\end{figure*}

Like the case $\epsilon=0$, the trajectory of a particle that begins at the
turning point $x_-$ follows a path in the complex-$x$ plane to the turning point
at $x_+$. Then, the particle retraces its path back to the turning point at
$x_-$, and it continues to oscillate between these two turning points. This path
is shown on Fig.~\ref{f11}. The period of this motion is $2\sqrt{3\pi}\Gamma
\left(\frac{4}{3}\right)/\Gamma\left(\frac{5}{6}\right)$. A particle beginning
at the third turning point $x_0$ exhibits a completely distinct motion: It
travels up the imaginary axis and reaches $i\infty$ in a finite time $\sqrt{\pi}
\Gamma\left(\frac{4}{3}\right)/\Gamma\left(\frac{5}{6}\right)$. This motion is
not periodic.

Paths originating from all other points in the finite complex-$x$ plane follow
closed periodic orbits. No two orbits may intersect; rather they are all nested,
like the ellipses for the case $\epsilon=0$. All of these orbits encircle the
turning points $x_\pm$ and, by virtue of Cauchy's theorem, have the same period
$2\sqrt{3\pi}\Gamma\left(\frac{4}{3}\right)/\Gamma\left(\frac{5}{6}\right)$ as
the oscillatory path connecting $x_\pm$. Because these orbits must avoid
crossing the trajectory that runs up the positive imaginary axis from the
turning point at $x_0=i$, they are pinched in the region just below $x_0$, as 
shown on Fig.~\ref{f11}.

\subsection{The Case $\epsilon=2$}
\label{ss3-3}

\begin{figure*}[t!]
\vspace{2.4in}
\includegraphics{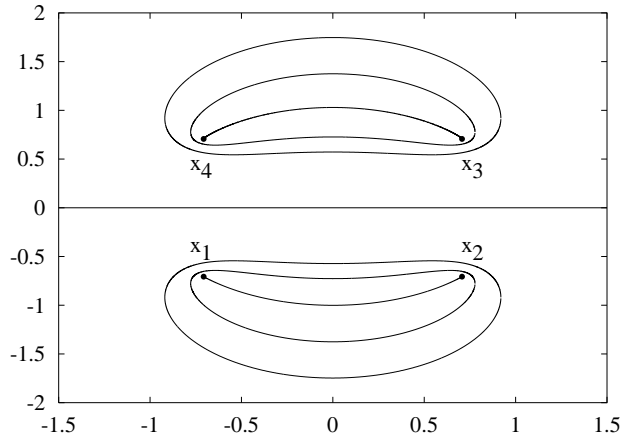}
\caption{Classical trajectories in the complex-$x$ plane for a particle
described by the Hamiltonian $H=p^2-x^4$ and having energy $E=1$. There are two
oscillatory trajectories connecting the pairs of turning points $x_1$ and $x_2$
in the lower-half $x$-plane and $x_3$ and $x_4$ in the upper-half $x$-plane. [A
trajectory joining any other pair of turning points is forbidden because it
would violate $\cP\cT$ (left-right) symmetry.] The oscillatory trajectories are
surrounded by closed orbits of the same period. In contrast to these periodic
orbits, there is a special class of trajectories having unbounded path length
and running along the real-$x$ axis.}
\label{f12}
\end{figure*}

When $\epsilon=2$, there are four turning points, two below the real axis and
symmetric with respect to the imaginary axis, $x_1=e^{-3i\pi/4}$ and $x_2=e^{-i
\pi/4}$, and two more located above the real axis and symmetric with respect to
the imaginary axis, $x_3=e^{i\pi/4}$ and $x_4=e^{3i\pi/4}$. These turning points
are solutions to the equation $-x^4=1$. Classical trajectories that oscillate
between the pair $x_1$ and $x_2$ and the pair $x_3$ and $x_4$ are shown on
Fig.~\ref{f12}. The period of these oscillations is $2\sqrt{2\pi}\Gamma\left(
\frac{5}{4}\right)/\Gamma\left(\frac{3}{4}\right)$. Trajectories that begin
elsewhere in the complex-$x$ plane are also shown on Fig.~\ref{f10}. By virtue
of Cauchy's theorem all these nested nonintersecting trajectories have the same
period. All classical motion is periodic except for the special trajectories
that begin on the real axis. A particle that begins on the real-$x$ axis runs
off to $\pm\infty$; its trajectory is nonperiodic.

\subsection{Broken and Unbroken Classical $\cP\cT$ Symmetry}
\label{ss3-4}

We can now understand heuristically why the energies of the corresponding $\cP
\cT$-symmetric quantum systems are real. In each of Figs.~\ref{f10}, \ref{f11},
and \ref{f12} we can see that all of the orbits are localized and periodic. We
may regard the pictured classical motions as representing particles confined to
and orbiting around {\em complex atoms}! We can then use Bohr-Sommerfeld
quantization to determine the discrete energies of the system:
\begin{equation}
\oint_C dx\,p=\oint_C dx\,\sqrt{E-x^2(ix)^\epsilon}=\left(n+\half\right)\pi,
\label{e62}
\end{equation}
where $C$ represents the orbit of a classical particle in the complex-$x$ plane.
By Cauchy's theorem, any closed orbit leads to the same result for the energy 
$E_n$, and because of $\cP\cT$ symmetry the integral above gives a {\em real}
value for the energy.

\begin{figure*}[t!]
\vspace{2.6in}
\includegraphics{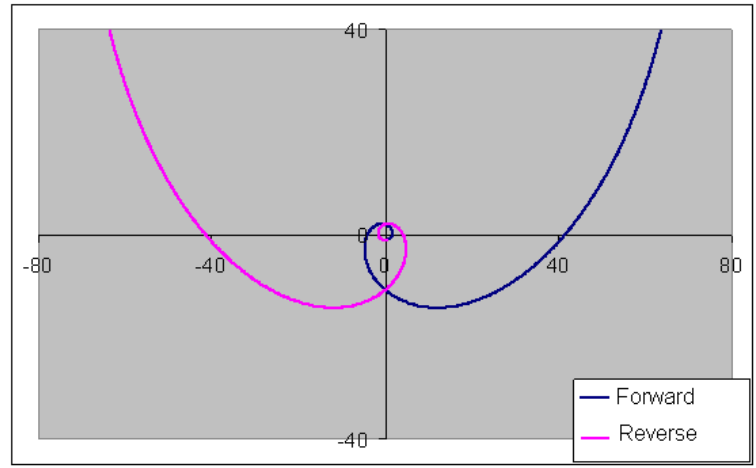}
\caption{Classical trajectories in the complex-$x$ plane for the Hamiltonian in
(\ref{e21}) with $\epsilon=-0.2$. These trajectories begin on the negative
imaginary axis very close to the origin. One trajectory evolves forward in time
and the other goes backward in time. The trajectories are open orbits and show
the particle spiraling off to infinity. The trajectories begin on the principal
sheet of the Riemann surface; as they cross the branch cut on the positive
imaginary axis, they visit the higher and lower sheets of the surface. The
trajectories do not cross because they lie on different Riemann sheets.}
\label{f13}
\end{figure*}

The key difference between classical paths for $\epsilon>0$ and for $\epsilon<0$
is that in the former case the paths (except for isolated examples) are closed
orbits and in the latter case the paths are open orbits. In Fig.~\ref{f13} we
consider the case $\epsilon=-0.2$ and display two paths that begin on the
negative imaginary axis. Because $\epsilon$ is noninteger, there is a branch cut
and the classical particle travels on a Riemann surface rather than on a single
sheet of the complex plane. In this figure one path evolves forward in time and
the other path evolves backward in time. Each path spirals outward and 
eventually moves off to infinity. Note that the pair of paths forms a $\cP
\cT$-symmetric structure. We remark that the paths do not cross because they are
on different sheets of the Riemann surface. The function $(ix)^{-0.2}$ requires
a branch cut, and we take this branch cut to lie along the positive imaginary
axis. The forward-evolving path leaves the principal sheet (sheet 0) of the
Riemann surface and crosses the branch cut in the positive sense and continues
on sheet 1. The reverse path crosses the branch cut in the negative sense and
continues on sheet $-1$. Figure \ref{f13} shows the projection of the classical
orbit onto the principal sheet.

Figure \ref{f13} shows why the energies of the quantum system in the broken
$\cP\cT$-symmetric region $\epsilon<0$ are not real. The answer is simply that
the trajectories are not closed orbits; the trajectories are {\em open} orbits,
and all classical particles drift off to $x=\infty$. As explained after
(\ref{e32}), if we attempt to quantize the system for the case $\epsilon<0$
using the Bohr-Sommerfeld integral in (\ref{e62}), the integral does not exist
because the integration contour is not closed.

\subsection{Noninteger Values of $\epsilon$}
\label{ss3-5}

\begin{figure}[t!]
\vspace{4.3in}
\includegraphics{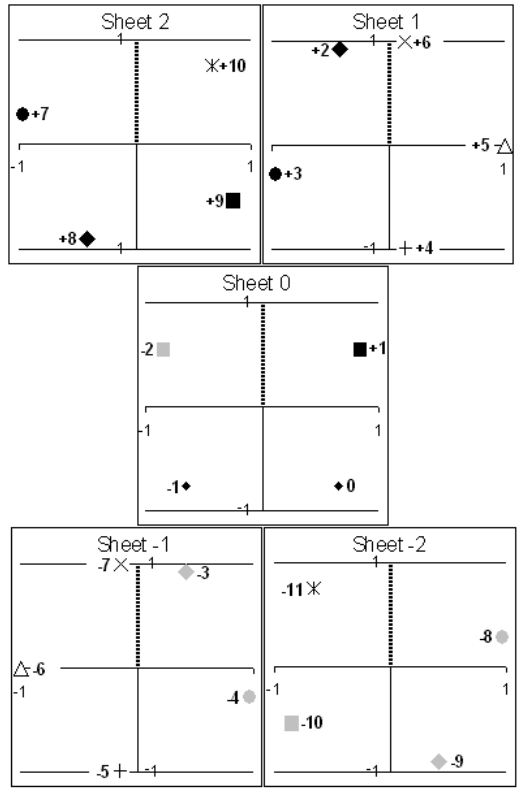}
\caption{Locations of the turning points for $\epsilon=\frac{12}{5}$. There are
11 $\cP\cT$-symmetric pairs of turning points, with each pair being mirror
images under reflection through the imaginary-$x$ axis on the principal sheet.
All 22 turning points lie on the unit circle on a five-sheeted Riemann surface,
where the sheets are joined by cuts on the positive-imaginary axis.}
\label{f14}
\end{figure}

As $\epsilon$ increases from 0, the turning points at $x=1$ (and at $x=-1$), as
shown in Fig.~\ref{f10} rotate downward and clockwise (anticlockwise) into the
complex-$x$ plane. These turning points are solutions to the equation $1+(ix)^{2
+\epsilon}=0$. When $\epsilon$ is noninteger, this equation has many solutions
that all lie on the unit circle and have the form
\begin{equation}
x=\exp\left(i\pi\frac{4N-\epsilon}{4+2\epsilon}\right)\quad(N~{\rm integer}).
\label{e63}
\end{equation}
These turning points occur in $\cP\cT$-symmetric pairs (pairs that are symmetric
when reflected through the imaginary axis) corresponding to the $N$ values
$(N=-1,~N=0)$, $(N=-2,~N=1)$, $(N=-3,~N=2)$, $(N=-4,~N=3)$, and so on. We label
these pairs by the integer $K$ ($K=0,~1,~2,~3,~\ldots$) so that the $K$th pair
corresponds to $(N=-K-1,~N=K)$. The pair of turning points on the real-$x$ axis
for $\epsilon=0$ deforms continuously into the $K=0$ pair of turning points when
$\epsilon>0$. When $\epsilon$ is rational, there are a finite number of turning
points in the complex-$x$ Riemann surface. For example, when $\epsilon=\frac{12}
{5}$, there are 5 sheets in the Riemann surface and 11 pairs of turning points.
The $K=0$ pair of turning points are labeled $N=-1$ and $N=0$, the $K=1$ pair
are labeled $N=-2$ and $N=1$, and so on. The last ($K=10$) pair of turning
points are labeled $N=-11$ and $N=10$. These turning points are shown in
Fig.~\ref{f14}.

As $\epsilon$ increases from 0, the elliptical complex trajectories in
Fig.~\ref{f10} for the harmonic oscillator begin to distort but the trajectories
remain closed and periodic except for special singular trajectories that run off
to complex infinity, as we see in Fig.~\ref{f11}. These singular trajectories
only occur when $\epsilon$ is an integer. Nearly all of the orbits that one
finds are $\cP\cT$ symmetric (left-right symmetric), and until very recently it
was thought that all closed periodic orbits are $\cP\cT$ symmetric. This is, in
fact, not so \cite{CL4}. Closed non-$\cP\cT$-symmetric orbits exist, and these
orbits are crucial for understanding the behavior of the periods of the complex
orbits as $\epsilon$ varies.

\begin{figure*}[t!]
\vspace{2.9in}
\includegraphics{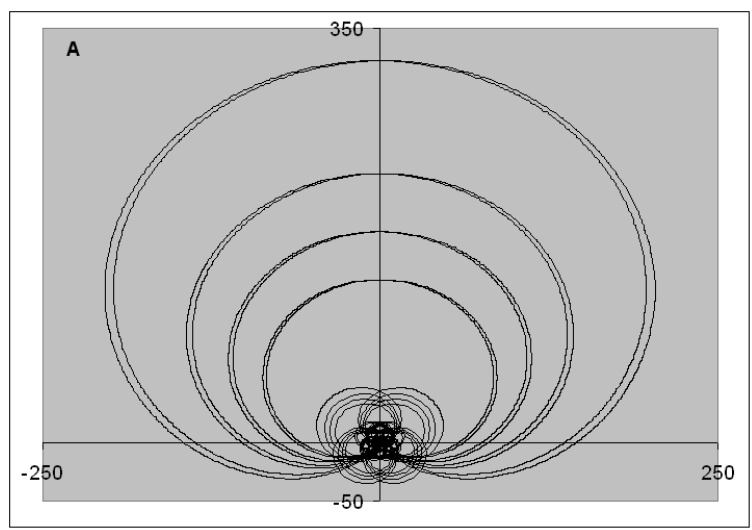}
\caption{A classical trajectory in the complex-$x$ plane for the complex
Hamiltonian $H=p^2x^2(ix)^{\pi-2}$. This complicated trajectory begins at $x(0)=
-7.1i$ and visits 11 sheets of the Riemann surface. Its period is approximately
$T=255.3$. This figure displays the projection of the trajectory onto the
principal sheet of the Riemann surface. This trajectory does not cross itself.}
\label{f15}
\end{figure*}

\begin{figure*}[t!]
\vspace{2.4in}
\includegraphics{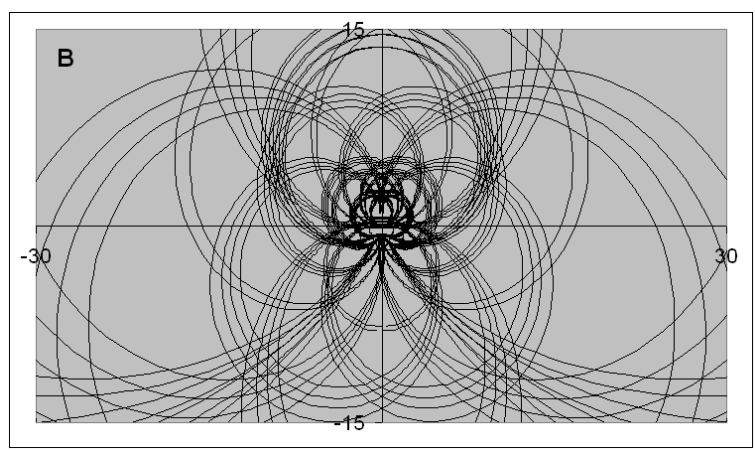}
\caption{An enlargement of the classical trajectory $x(t)$ in Fig.~\ref{f15}
showing the detail near the origin in the complex-$x$ plane. We emphasize that
this classical path never crosses itself; the apparent self-intersections are
paths that lie on different sheets of the Riemann surface.}
\label{f16}
\end{figure*}

In Fig.~\ref{f15} we display a $\cP\cT$-symmetric orbit of immense topological
intricacy that visits many sheets of the Riemann surface. This figure shows a
classical trajectory corresponding to $\epsilon=\pi-2$. The trajectory starts at
$x(0)=-7.1i$ and visits $11$ sheets of the Riemann surface. Its period is
$T=255.3$. The structure of this orbit near the origin is so complicated that we
provide a magnified version in Fig.~\ref{f16}.

The period of any classical orbit depends on the specific pairs of turning
points that are enclosed by the orbit and on the number of times that the orbit
encircles each pair. As shown in Ref.~\cite{CL2}, any given orbit can be
deformed to a simpler orbit of exactly the same period. This simpler orbit
connects two turning points and oscillates between them rather than encircling
them. For the elementary case of orbits that enclose only the $K=0$ pair of
turning points, the formula for the period of the closed orbit is
\begin{eqnarray}
T_0(\epsilon)=2\sqrt{\pi}\frac{\Gamma\left(\frac{3+\epsilon}{2+\epsilon}\right)}
{\Gamma\left(\frac{4+\epsilon}{r4+2\epsilon}\right)}\cos\left(\frac{\epsilon\pi}
{4+2\epsilon}\right)\qquad(\epsilon\geq0).
\label{e64}
\end{eqnarray}
The derivation of (\ref{e64}) goes as follows: The period $T_0$ is given by a
closed contour integral along the trajectory in the complex-$x$ plane. This
trajectory encloses the square-root branch cut that joins the $K=0$ pair of
turning points. This contour can be deformed into a pair of rays that run from
one turning point to the origin and then from the origin to the other turning
point. The integral along each ray is easily evaluated as a beta function, which
is then written in terms of gamma functions.

When the classical orbit encloses more than just the $K=0$ pair of turning
points, the formula for the period of the orbit becomes more complicated
\cite{CL2}. In general, there are contributions to the period integral from many
enclosed pairs of turning points. We label each such pair by the integer $j$.
The formula for the period of the topological class of classical orbits whose
central orbit terminates on the $K$th pair of turning points is
\begin{eqnarray}
T_K(\epsilon)=2\sqrt{\pi}\frac{\Gamma\left(\frac{3+\epsilon}{2+\epsilon}\right)}
{\Gamma\left(\frac{4+\epsilon}{4+2\epsilon}\right)}\sum_{j=0}^{\infty}a_j(K,
\epsilon)\left|\cos\left(\frac{(2j+1)\epsilon\pi}{4+2\epsilon}\right)\right|.
\label{e65}
\end{eqnarray}
In this formula the cosines originate from the angular positions of the turning
points in (\ref{e63}). The coefficients $a_j(K,\epsilon)$ are all nonnegative
integers. The $j$th coefficient is nonzero only if the classical path encloses
the $j$th pair of turning points. Each coefficient is an {\em even} integer
except for the $j=K$ coefficient, which is an odd integer. The coefficients $a_j
(K,\epsilon)$ satisfy
\begin{eqnarray}
\sum_{j=0}^{\infty}a_j(K,\epsilon)=k,
\label{e66}
\end{eqnarray}
where $k$ is the number of times that the central classical path crosses the
imaginary axis. Equation (\ref{e66}) truncates the summation in (\ref{e65}) to
a finite number of terms.

The period $T_0$ in (\ref{e64}) of orbits connecting the $K=0$ turning points is
a smoothly decreasing function of $\epsilon$. However, for classical orbits
connecting the $K$th ($K>0$) pair of turning points, the classical orbits
exhibit fine structure that is exquisitely sensitive to the value of $\epsilon$.
Small variations in $\epsilon$ can cause huge changes in the topology and in the
periods of the closed orbits. Depending on $\epsilon$, there are orbits having
short periods as well as orbits having long and possibly arbitrarily long
periods.

\subsection{Classical Orbits Having Spontaneously Broken $\cP\cT$ Symmetry}
\label{ss3-6}

\begin{figure*}[t!]
\vspace{4.45in}
\includegraphics{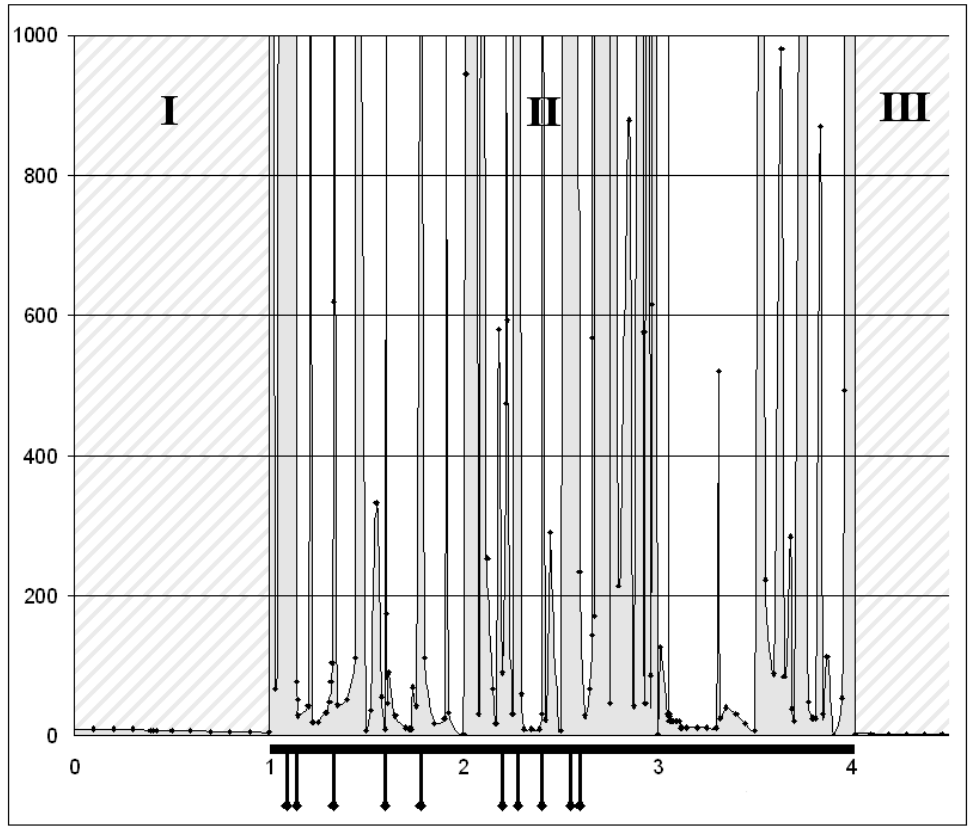}
\caption{Period of a classical trajectory beginning at the $N=1$ turning point
in the complex-$x$ plane. The period is plotted as a function of $\epsilon$. The
period decreases smoothly for $0\leq\epsilon<1$ (Region I). However, when $1\leq
\epsilon\leq4$ (Region II), the period becomes a rapidly varying and noisy
function of $\epsilon$. For $\epsilon>4$ (Region III) the period is once again a
smoothly decaying function of $\epsilon$. Region II contains short subintervals
where the period is a small and smoothly varying function of $\epsilon$. At the
edges of these subintervals the period suddenly becomes extremely long. Detailed
numerical analysis shows that the edges of the subintervals lie at special
rational values of $\epsilon$. Some of these special rational values of
$\epsilon$ are indicated by vertical line segments that cross the horizontal
axis. At these rational values the orbit does not reach the $N=-2$ turning point
and the $\cP\cT$ symmetry of the classical orbit is spontaneously broken.}
\label{f17}
\end{figure*}

\begin{figure*}[t!]
\vspace{4.45in}
\includegraphics{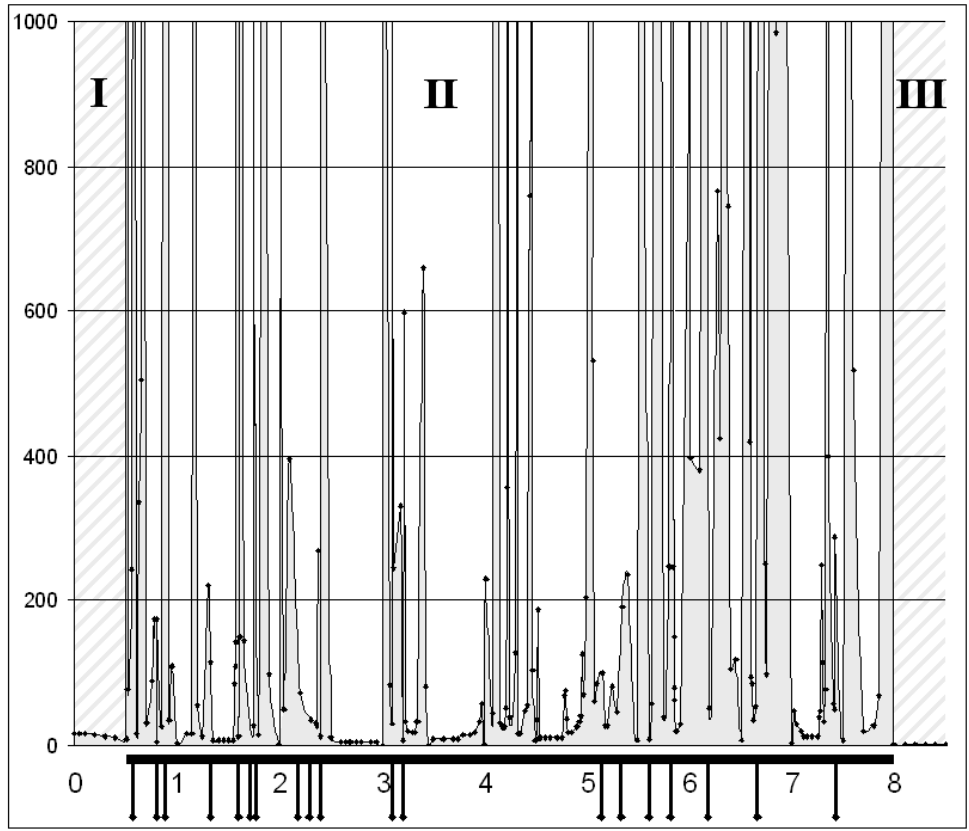}
\caption{Period of a classical trajectory joining (except when $\cP\cT$ symmetry
is broken) the $K=2$ pair of turning points. The period is plotted as a function
of $\epsilon$. As in the $K=1$ case shown in Fig.~\ref{f17}, there are three
regions. When $0\leq\epsilon\leq\half$ (Region I), the period is a smooth
decreasing function of $\epsilon$; when $\half<\epsilon\leq8$ (Region II), the
period is a rapidly varying and choppy function of $\epsilon$; when $8<\epsilon$
(Region III), the period is again a smooth and decreasing function of
$\epsilon$.}
\label{f18}
\end{figure*}

There is a general pattern that holds for all $K$. For classical orbits that
oscillate between the $K$th pair of turning points, there are three regions of
$\epsilon$. The domain of Region I is $0\leq\epsilon\leq\frac{1}{K}$, the domain
of Region II is $\frac{1}{K}<\epsilon<4K$, and the domain of Region III is $4K<
\epsilon$. In Regions I and III the period is a small and smoothly decreasing
function of $\epsilon$. However, in Region II the period is a rapidly varying
and noisy function of $\epsilon$. We illustrate this behavior for the case $K=1$
and $K=2$ in Figs.~\ref{f17} and \ref{f18}.

The abrupt changes in the topology and the periods of the orbits for $\epsilon$
in Region II are associated with the appearance of orbits having {\em
spontaneously broken} $\cP\cT$ symmetry. In Region II there are short patches
where the period is relatively small and is a slowly varying function of
$\epsilon$. These patches are bounded by special values of $\epsilon$ for which
the period of the orbit suddenly becomes extremely long. Numerical studies of
the orbits connecting the $K$th pair of turning points indicate that these
special values of $\epsilon$ are always {\em rational} \cite{CL5}. Furthermore,
at these special rational values of $\epsilon$, the closed orbits are {\em not}
$\cP\cT$-symmetric (left-right symmetric). Such orbits exhibit {\em
spontaneously broken} $\cP\cT$ symmetry. Some special values of $\epsilon$ at
which spontaneously broken $\cP\cT$-symmetric orbits occur are indicated in
Figs.~\ref{f17} and \ref{f18} by short vertical lines below the horizontal axis.
These special values of $\epsilon$ have the form $\frac{p}{q}$, where $p$
is a multiple of 4 and $q$ is odd.

\begin{figure*}[t!]
\vspace{2.60in}
\includegraphics{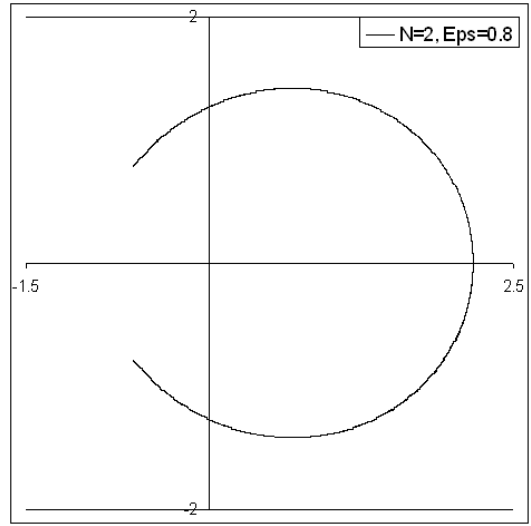}
\caption{A horseshoe-shaped non-$\cP\cT$-symmetric orbit. This orbit is not
symmetric with respect to the imaginary axis, but it is symmetric with respect
to the real axis. The orbit terminates at a complex-conjugate pair of turning
points. For this orbit $\epsilon=\frac{4}{5}$.}
\label{f19}
\end{figure*}

A broken-$\cP\cT$-symmetric orbit is a failed $\cP\cT$-symmetric orbit. Figure
\ref{f19} displays a spontaneously-broken-$\cP\cT$-symmetric orbit for $\epsilon
=\frac{4}{5}$. The orbit starts at the $N=2$ turning point, but it never reaches
the $\cP\cT$-symmetric turning point $N=-3$. Rather, the orbit terminates when
it runs into and is reflected back from the complex conjugate $N=4$ turning
point [see (\ref{e63})]. The period of the orbit is short ($T=4.63$). While this
orbit is not $\cP\cT$ (left-right) symmetric, it does possess complex-conjugate
(up-down) symmetry. In general, for a non-$\cP\cT$-symmetric orbit to exist, it
must join or encircle a pair of complex-conjugate turning points.

\begin{figure*}[ht!]
\vspace{2.95in}
\includegraphics{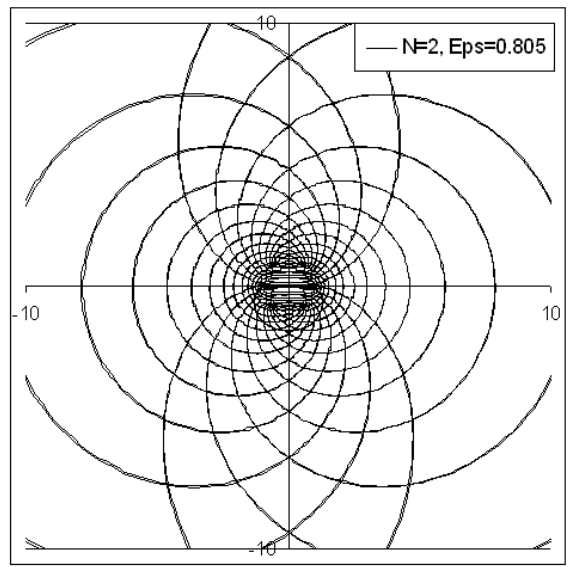}
\caption{$\cP\cT$-symmetric orbit for $\epsilon=0.805$. This orbit connects
the $K=2$ pair of turning points.}
\label{f20}
\end{figure*}

If we change $\epsilon$ slightly, $\cP\cT$ symmetry is restored and one can only
find orbits that are $\cP\cT$ symmetric. For example, if we take $\epsilon=
0.805$, we obtain the complicated orbit in Fig.~\ref{f20}. The period of this
orbit is large ($T=173.36$).

\begin{figure*}[t!]
\vspace{3.65in}
\includegraphics{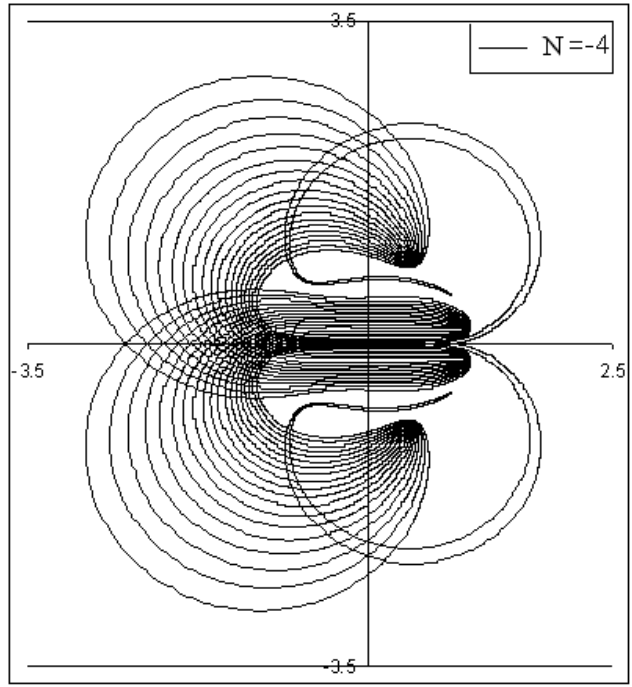}
\caption{Non-$\cP\cT$-symmetric orbit for $\epsilon=\frac{16}{9}$. This
topologically complicated orbit originates at the $N=-4$ turning point but
does not reach the $\cP\cT$-symmetric $N=3$ turning point. Instead, it is
reflected back at the complex-conjugate $N=-14$ turning point. The period of
this orbit is $T=186.14$.}
\label{f21}
\end{figure*}

Broken-$\cP\cT$-symmetric orbits need not be simple looking, like the orbit
shown in Fig.~\ref{f19}. Indeed, they can have an elaborate topology. As an
example we plot in Fig.~\ref{f21} the complicated orbit that arises when
$\epsilon=\frac{16}{9}$. This orbit is a failed $K=3$ $\cP\cT$-symmetric orbit
that originates at the $N=-4$ turning point, but never reaches the $\cP
\cT$-symmetric $N=3$ turning point. Instead, it is reflected back by the
complex-conjugate $N=-14$ turning point. 

This study of classical orbits provides a heuristic explanation of the quantum
transition from a broken to an unbroken $\cP\cT$ symmetry as $\epsilon$
increases past $0$. The quantum transition corresponds to a change from open to
closed classical orbits. Furthermore, we can now see why the quantum theory in
the unbroken region is unitary. At the classical level, particles are bound in a
complex atom and cannot escape to infinity; at the quantum level the probability
is conserved and does not leak away as time evolves. Quantum mechanics is
obtained by summing over all possible classical trajectories, and in the case of
$\cP\cT$-symmetric classical mechanics we have seen some bizarre classical
trajectories. To understand how summing over such trajectories produces $\cP
\cT$-symmetric quantum mechanics will require much more research.

\section{$\cP\cT$-Symmetric Quantum Mechanics}
\label{s4}

Establishing that the eigenvalues of many $\cP\cT$-symmetric Hamiltonians are
real and positive raises an obvious question: Does a non-Hermitian Hamiltonian
such as $H$ in (\ref{e12}) define a physical theory of quantum mechanics or is
the reality and positivity of the spectrum merely an intriguing mathematical
curiosity exhibited by some special classes of complex eigenvalue problems?
Recall that a {\em physical} quantum theory must (i) have an energy spectrum
that is bounded below; (ii) possess a Hilbert space of state vectors that is
endowed with an inner product having a positive norm; (iii) have unitary time
evolution. The simplest condition on the Hamiltonian $H$ that guarantees that
the quantum theory satisfies these three requirements is that $H$ be real and
symmetric. However, this condition is overly restrictive. One can allow $H$ to
be complex as long as it is Dirac Hermitian: $H^\dagger=H$. In this section we
explain why we can replace the condition of Hermiticity by the condition that
$H$ have an unbroken $\cP\cT$ symmetry and still satisfy the above requirements
for a physical quantum theory.\footnote{All of the $\cP\cT$-symmetric
Hamiltonians considered in this paper are symmetric under matrix transposition.
This matrix symmetry condition is not necessary, but it has the simplifying
advantage that the we do not need to have a biorthogonal set of basis states. We
can consider $\cP\cT$-symmetric Hamiltonians that are not symmetric under matrix
transposition, but only at the cost of introducing a biorthogonal basis
\cite{WWWW1,BIOX}.}

\subsection{Recipe for a Quantum-Mechanical Theory Defined by a Hermitian
Hamiltonian}
\label{ss4-1}

For purposes of comparison, we summarize in this subsection the standard
textbook procedure that one follows in analyzing a theory defined by a
conventional Hermitian quantum-mechanical Hamiltonian. In the next subsection
we repeat these procedures for a non-Hermitian Hamiltonian.

\begin{itemize}
\item[(a)] {\em Eigenfunctions and eigenvalues of $H$}. Given the Hamiltonian
$H$ one can write down the time-independent Schr\"odinger equation associated
with $H$ and calculate the eigenvalues $E_n$ and eigenfunctions $\psi_n(x)$.
Usually, this calculation is difficult to perform analytically, so it must be
done numerically.

\item[(b)] {\em Orthogonality of eigenfunctions.} Because $H$ is Hermitian, the
eigenfunctions of $H$ will be orthogonal with respect to the standard Hermitian
inner product:
\begin{equation}
(\psi,\phi)\equiv\int dx\,[\psi(x)]^*\phi(x)
\label{e67}
\end{equation}
{\em Orthogonality} means that the inner product of two eigenfunctions $\psi_m(
x)$ and $\psi_n(x)$ associated with different eigenvalues $E_m\neq E_n$
vanishes:
\begin{equation}
(\psi_m,\phi_n)=0.
\label{e68}
\end{equation}
(We do not discuss here the technical problems associated with degenerate 
spectra.)

\item[(c)] {\em Orthonormality of eigenfunctions}. Since the Hamiltonian is
Hermitian, the norm of any vector is guaranteed to be positive. This means that
we can normalize the eigenfunctions of $H$ so that the norm of every
eigenfunction is unity:
\begin{equation}
(\psi_n,\psi_n)=1.
\label{e69}
\end{equation}

\item[(d)] {\em Completeness of eigenfunctions}. It is a deep theorem of the
theory of linear operators on Hilbert spaces that the eigenfunctions of a
Hermitian Hamiltonian are {\em complete}. This means that any (finite-norm)
vector $\chi$ in the Hilbert space can be expressed as a linear combination of
the eigenfunctions of $H$:
\begin{equation}
\chi=\sum_{n=0}^\infty a_n\psi_n.
\label{e70}
\end{equation}
The formal statement of completeness in coordinate space is the reconstruction
of the unit operator (the delta function) as a sum over the eigenfunctions:
\begin{equation}
\sum_{n=0}^\infty[\psi_n(x)]^*\psi_n(y)=\delta(x-y).
\label{e71}
\end{equation}

\item[(e)] {\em Reconstruction of the Hamiltonian $H$ and the Green's function
$G$, and calculation of the spectral Zeta function}. The Hamiltonian matrix in
coordinate space has the form
\begin{equation}
\sum_{n=0}^\infty [\psi_n(x)]^*\psi_n(y)E_n=H(x,y)
\label{e72}
\end{equation}
and the Green's function is given by
\begin{equation}
\sum_{n=0}^\infty [\psi_n(x)]^*\psi_n(y)\frac{1}{E_n}=G(x,y).
\label{e73}
\end{equation}
The Green's function is the matrix inverse of the Hamiltonian in the sense that
\begin{equation}
\int dy\,H(x,y)G(y,z)=\delta(x-z).
\label{e74}
\end{equation}
The formula for the Green's function in (\ref{e73}) allows us to calculate the
sum of the reciprocals of the energy eigenvalues. We simply set $x=y$ in
(\ref{e73}), integrate with respect to $x$, and use the normalization condition
in (\ref{e69}) to obtain the result that
\begin{equation}
\int dx\,G(x,x)=\sum_{n=0}^\infty\frac{1}{E_n}.
\label{e75}
\end{equation}
The summation on the right side of (\ref{e75}) is called the {\em spectral zeta
function}. This sum is convergent if the energy levels $E_n$ rise faster than
linearly with $n$. Thus, the spectral zeta function for the harmonic oscillator
is divergent, but it exists for the $|x|^{\epsilon+2}$ and $x^2(ix)^\epsilon$
potentials if $\epsilon>0$.

\item[(f)] {\em Time evolution and unitarity}. For a Hermitian Hamiltonian the
time-evolution operator $e^{-iHt}$ [see (\ref{e15})] is unitary, and it
automatically preserves the inner product:
\begin{equation}
\Big(\chi(t),\chi(t)\Big)=\Big(\chi(0)e^{iHt},e^{-iHt}\chi(t)\Big)=
\Big(\chi(0),\chi(0)\Big).
\label{e76}
\end{equation}

\item[(g)] {\em Observables}. An observable is represented by a linear Hermitian
operator. The outcome of a measurement is one of the {\em real} eigenvalues of
this operator.

\item[(h)] {\em Miscellany}. One can study a number of additional topics, such
as the classical and semiclassical limits of the quantum theory, probability
density and currents, perturbative and nonperturbative calculations, and so on.
We do not address these issues in depth in this paper.

\end{itemize}

\subsection{Recipe for $\cP\cT$-Symmetric Quantum Mechanics}
\label{ss4-2}

Let us follow the recipe outlined in Subsec.~\ref{ss4-1} for the case of a
non-Hermitian $\cP\cT$-symmetric Hamiltonian having an unbroken $\cP\cT$
symmetry. For definiteness, we will imagine that the non-Hermitian Hamiltonian
has the form in (\ref{e12}). The novelty here is that we do not know {\em a
priori} the definition of the inner product, as we do in the case of ordinary
Hermitian quantum mechanics. We will have to discover the correct inner product
in the course of our analysis. The inner product is determined by the
Hamiltonian itself, so $\cP\cT$-symmetric quantum mechanics is a kind of
``bootstrap'' theory. The Hamiltonian operator chooses its own Hilbert space
(and associated inner product) in which it prefers to live!

\begin{itemize}
\item[(a)] {\em Eigenfunctions and eigenvalues of $H$}. In Sec.~\ref{s2} we
discussed various techniques for determining the coordinate-space eigenfunctions
and eigenvalues of a non-Hermitian Hamiltonian. We assume here that we have
found the eigenvalues $E_n$ by using either analytical or numerical methods and
that these eigenvalues are all real. [This is equivalent to assuming that the
$\cP\cT$ symmetry of $H$ is unbroken; that is, all eigenfunctions $\psi_n(x)$ of
$H$ are also eigenfunctions of $\cP\cT$.].

\item[(b)] {\em Orthogonality of eigenfunctions}. To test the orthogonality of
the eigenfunctions, we must specify an inner product. (A pair of vectors can be
orthogonal with respect to one inner product and not orthogonal with respect to
another inner product.) Since we do not yet know what inner product to use, one
might try to guess an inner product. Arguing by analogy, one might think that
since the inner product in (\ref{e67}) is appropriate for Hermitian Hamiltonians($H=H^\dag$), a good choice for an inner product associated with a $\cP
\cT$-symmetric Hamiltonian ($H=H^{\cP\cT}$) might be
\begin{equation}
(\psi,\phi)\equiv\int_C dx\,[\psi(x)]^{\cP\cT}\phi(x)=\int_C dx\,[\psi(-x)]^*
\phi(x),
\label{e77}
\end{equation}
where $C$ is a contour in the Stokes wedges shown in Fig.~\ref{f2}. With this
inner-product definition one can show by a trivial integration-by-parts argument
using the time-independent Schr\"odinger equation (\ref{e24}) that pairs of
eigenfunctions of $H$ associated with different eigenvalues are orthogonal.
However, this guess for an inner product is not acceptable for formulating a
valid quantum theory because the norm of a state is not necessarily positive.

\item[(c)] {\em The $\cC\cP\cT$ inner product}. To construct an inner product
with a positive norm for a complex non-Hermitian Hamiltonian having an {\em
unbroken} $\cP\cT$ symmetry, we will construct a new linear operator $\cC$ that
commutes with both $H$ and $\cP\cT$. Because $\cC$ commutes with the
Hamiltonian, it represents a {\em symmetry} of $H$. We use the symbol $\cC$ to
represent this symmetry because, as we will see, the properties of $\cC$ are
similar to those of the charge conjugation operator in particle physics. The
inner product with respect to $\cC\cP\cT$ conjugation is defined as
\begin{equation}
\langle\psi|\chi\rangle^{\cC\cP\cT}=\int dx\,\psi^{\cC\cP\cT}(x)\chi(x),
\label{e78}
\end{equation}
where $\psi^{\cC\cP\cT}(x)=\int dy\,{\cC}(x,y)\psi^*(-y)$. We will show that
this inner product satisfies the requirements for the quantum theory defined by
$H$ to have a Hilbert space with a positive norm and to be a unitary theory of
quantum mechanics. We will represent the $\cC$ operator as a sum over the
eigenfunctions of $H$, but before doing so we must first show how to normalize
these eigenfunctions.

\item[(d)] {\em $\cP\cT$-symmetric normalization of the eigenfunctions and the
strange statement of completeness}. We showed in (\ref{e19}) that the
eigenfunctions $\psi_n(x)$ of $H$ are also eigenfunctions of the $\cP\cT$
operator with eigenvalue $\lambda=e^{i\alpha}$, where $\lambda$ and $\alpha$
depend on $n$. Thus, we can construct $\cP\cT$-normalized eigenfunctions
$\phi_n(x)$ defined by
\begin{equation}
\phi_n(x)\equiv e^{-i\alpha/2}\psi_n(x).
\label{e79}
\end{equation}
By this construction, $\phi_n(x)$ is still an eigenfunction of $H$ and it is
also an eigenfunction of $\cP\cT$ with eigenvalue $1$. One can also
show both numerically and analytically that the algebraic sign of the $\cP\cT$
norm in (\ref{e77}) of $\phi_n(x)$ is $(-1)^n$ for all $n$ and for all values
of $\epsilon>0$ \cite{BBJ1}. Thus, we {\em define} the eigenfunctions so that
their $\cP\cT$ norms are exactly $(-1)^n$:
\begin{equation}
\int_C dx\,[\phi_n(x)]^{\cP\cT}\phi_n(x)=\int_C dx\,[\phi_n(-x)]^*\phi_n(x)=
(-1)^n,
\label{e80}
\end{equation}
where the contour $C$ lies in the Stokes wedges shown in Fig.~\ref{f2}. In terms
of these $\cP\cT$-normalized eigenfunctions there is a simple but unusual
statement of completeness:
\begin{equation}
\sum_{n=0}^\infty(-1)^n\phi_n(x)\phi_n(y)=\delta(x-y).
\label{e81}
\end{equation}
This unusual statement of completeness has been verified both numerically and
analytically to great precision for all $\epsilon>0$ \cite{rr2,rr3} and a
mathematical proof has been given \cite{WWWW1}. It is easy to verify using
(\ref{e80}) that the left side of (\ref{e81}) satisfies the integration rule for
delta functions: $\int dy\,\delta(x-y)\delta(y-z)=\delta(x-z)$.
\end{itemize}

\begin{itemize}
\begin{footnotesize}
\item[~~] {\em Example: $\cP\cT$-symmetric normalization of harmonic-oscillator
eigenfunctions.} For the harmonic-oscillator Hamiltonian $H=\p^2+\x^2$, the
eigenfunctions are Gaussians multiplied by Hermite polynomials: $\psi_0(x)=\exp
\left(-\half x^2\right)$, $\psi_1(x)=x\exp\left(-\half x^2\right)$, $\psi_2(x)=
(2x^2-1)\exp\left(-\half x^2\right)$, $\psi_3(x)=(2x^3-3x)\exp\left(-\half x^2
\right)$, and so on. To normalize these eigenfunctions so that they are
also eigenfunctions of the $\cP\cT$ operator with eigenvalue 1, we choose
\begin{eqnarray}
\phi_0(x)&=&a_0\exp\left(-\half x^2\right),\nonumber\\
\phi_1(x)&=&a_1ix\exp\left(-\half x^2\right),\nonumber\\
\psi_2(x)&=&a_2(2x^2-1)\exp\left(-\half x^2\right),\nonumber\\
\psi_3(x)&=&a_3i(2x^3-3x)\exp\left(-\half x^2\right),
\label{e82}
\end{eqnarray}
and so on, where the real numbers $a_n$ are chosen so that the integral in
(\ref{e80}) evaluates to $(-1)^n$ for all $n$. It is easy to verify that if the
eigenfunctions $\phi_n(x)$ are substituted into (\ref{e81}) and the summation
is performed, then the result is the Dirac delta function on the right side of
(\ref{e81}).
\end{footnotesize}
\end{itemize}

\begin{itemize}
\item[(e)] {\em Coordinate-space representation of $H$ and $G$ and the spectral
Zeta function}. From the statement of completeness in (\ref{e81}) we can
construct coordinate-space representations of the linear operators. For example,
since the coordinate-space representation of the parity operator is $\cP(x,y)=
\delta(x+y)$, we have
\begin{equation}
\cP(x,y)=\sum_{n=0}^\infty(-1)^n\phi_n(x)\phi_n(-y).
\label{e83}
\end{equation}
We can also construct the coordinate-space representations of the Hamiltonian
and the Green's function,
\begin{eqnarray}
H(x,y)&=&\sum_{n=0}^\infty(-1)^nE_n\phi_n(x)\phi_n(y),\nonumber\\
G(x,y)&=&\sum_{n=0}^\infty(-1)^n\frac{1}{E_n}\phi_n(x)\phi_n(y),
\label{e84}
\end{eqnarray}
and using (\ref{e80}) it is straightforward to show that $G$ is the functional
inverse of $H$: $\int dy\,H(x,y)G(y,z)=\delta(x-z)$. For the class of $\cP
\cT$-symmetric Hamiltonians in (\ref{e12}) this equation takes the form of a
differential equation satisfied by $G(x,y)$:
\begin{equation}
\left(-\frac{d^2}{dx^2}+x^2(ix)^\epsilon\right)G(x,y)=\delta(x-y).
\label{e85}
\end{equation}
Equation (\ref{e85}) can be solved in terms of associated Bessel functions in
each of two regions, $x>y$ and $x<y$. The solutions can then be patched together
at $x=y$ to obtain a closed-form expression for $G(x,y)$ \cite{rr2,rr3}. One
then uses (\ref{e75}) to find an exact formula for the spectral zeta function
for all values of $\epsilon$:
\begin{equation}
\sum_n\frac{1}{E_n}=\left[1+\frac{\cos\left(\frac{3\epsilon\pi}{2\epsilon+8}
\right)\sin\left(\frac{\pi}{4+\epsilon}\right)}{\cos\left(\frac{\epsilon\pi}
{4+2\epsilon}\right)\sin\left(\frac{3\pi}{4+\epsilon}\right)}\right]
\frac{\Gamma\left(\frac{1}{4+\epsilon}\right)\Gamma\left(\frac{2}{4+\epsilon}
\right)\Gamma\left(\frac{\epsilon}{4+\epsilon}\right)}{(4+\epsilon)^\frac{4+
2\epsilon}{4+\epsilon}\Gamma\left(\frac{1+\epsilon}{4+\epsilon}\right)
\Gamma\left(\frac{2+\epsilon}{4+\epsilon}\right)}.
\label{e86}
\end{equation}

\item[(f)] {\em Construction of the $\cC$ operator}. The situation here in which
half of the energy eigenstates have positive norm and half have negative norm
[see (\ref{e80})] is analogous to the problem that Dirac encountered in
formulating the spinor wave equation in relativistic quantum theory \cite{rf17}.
Following Dirac, we attack the problem of an indefinite norm by finding an
interpretation of the negative-norm states. For any Hamiltonian $H$ having an
unbroken $\cP\cT$ symmetry there exists an additional symmetry of $H$ connected
with the fact that there are equal numbers of positive- and negative-norm
states. The linear operator $\cC$ that embodies this symmetry can be represented
in coordinate space as a sum over the $\cP\cT$-normalized eigenfunctions of the
$\cP\cT$-symmetric Hamiltonian in (\ref{e12}):
\begin{equation}
\cC(x,y)=\sum_{n=0}^\infty\phi_n(x)\phi_n(y).
\label{e87}
\end{equation}
Notice that this equation is identical to the statement of completeness in 
(\ref{e81}) except that the factor of $(-1)^n$ is absent. We can use (\ref{e80})
and (\ref{e81}) to verify that the square of $\cC$ is unity ($\cC^2={\bf 1}$):
\begin{equation}
\int dy\,\cC(x,y)\cC(y,z)=\delta(x-z).
\label{e88}
\end{equation}
Thus, the eigenvalues of $\cC$ are $\pm1$. Also, $\cC$ commutes with $H$.
Therefore, since $\cC$ is linear, the eigenstates of $H$ have definite values of
$\cC$. Specifically,
\begin{eqnarray}
\cC\phi_n(x)&=&\int dy\,\cC(x,y)\phi_n(y)\nonumber\\
&=&\sum_{m=0}^\infty\phi_m(x)\int dy\,\phi_m(y)\phi_n(y)=(-1)^n\phi_n(x).
\label{e89}
\end{eqnarray}
This new operator $\cC$ resembles the charge-conjugation operator in quantum
field theory. However, the precise meaning of $\cC$ is that it represents the
measurement of the sign of the $\cP\cT$ norm in (\ref{e80}) of an eigenstate.

The operators $\cP$ and $\cC$ are distinct square roots of the unity operator
$\delta(x-y)$. That is, $\cP^2=\cC^2={\bf 1}$, but $\cP\neq\cC$ because $\cP$ is
real, while $\cC$ is complex. The parity operator in coordinate space is
explicitly real ($\cP(x,y)=\delta(x+y)$), while the operator $\cC(x,y)$ is
complex because it is a sum of products of complex functions. The two operators
$\cP$ and $\cC$ do not commute. However, $\cC$ {\em does} commute with $\cP\cT$.

\item[(g)] {\em Positive norm and unitarity in $\cP\cT$-symmetric quantum
mechanics}. Having constructed the operator $\cC$, we can now use the new $\cC
\cP\cT$ inner-product defined in (\ref{e78}). Like the $\cP\cT$ inner product,
this new inner product is phase independent. Also, because the time-evolution
operator (as in ordinary quantum mechanics) is $e^{-iHt}$ and because $H$
commutes with $\cP\cT$ and with $\cC\cP\cT$, both the $\cP\cT$ inner product and
the $\cC\cP\cT$ inner product remain time independent as the states evolve.
However, unlike the $\cP\cT$ inner product, the $\cC\cP\cT$ inner product is
{\em positive definite} because $\cC$ contributes a factor of $-1$ when it acts
on states with negative $\cP\cT$ norm. In terms of the $\cC\cP\cT$ conjugate,
the completeness condition reads
\begin{equation}
\sum_{n=0}^\infty\phi_n(x)[\cC\cP\cT\phi_n(y)]=\delta(x-y).
\label{e90}
\end{equation}
\end{itemize}

\subsection{Comparison of Hermitian and $\cP\cT$-Symmetric Quantum Theories}
\label{ss4-3}

We have shown in Subsecs.~\ref{ss4-1} and \ref{ss4-2} how to construct quantum
theories based on Hermitian and on non-Hermitian Hamiltonians. In the
formulation of a conventional quantum theory defined by a Hermitian Hamiltonian,
the Hilbert space of physical states is specified even before the Hamiltonian is
known. The inner product (\ref{e67}) in this vector space is defined with
respect to Dirac Hermitian conjugation (complex conjugate and transpose). The
Hamiltonian is then chosen and the eigenvectors and eigenvalues of the
Hamiltonian are determined. In contrast, the inner product for a quantum theory
defined by a non-Hermitian $\cP\cT$-symmetric Hamiltonian depends on the
Hamiltonian itself and thus is determined {\em dynamically}. One must solve for
the eigenstates of $H$ before knowing the Hilbert space and the associated inner
product of the theory because the $\cC$ operator is defined and constructed in
terms of the eigenstates of the Hamiltonian. The Hilbert space, which consists
of all complex linear combinations of the eigenstates of $H$, and the $\cC\cP
\cT$ inner product are determined by these eigenstates.

The operator $\cC$ does not exist as a distinct entity in ordinary Hermitian
quantum mechanics. Indeed, if we allow the parameter $\epsilon$ in (\ref{e12})
to tend to 0, the operator $\cC$ in this limit becomes identical to $\cP$ and
the $\cC\cP\cT$ operator becomes $\cT$, which performs complex conjugation.
Hence, the inner product defined with respect to $\cC\cP\cT$ conjugation reduces
to the inner product of conventional quantum mechanics and (\ref{e81}) reduces
to the usual statement of completeness $\sum_n\phi_n(x)\phi_n^*(y)=\delta(x-y)$.

The $\cC\cP\cT$ inner product is independent of the choice of integration
contour $C$ as long as $C$ lies inside the asymptotic wedges associated with the
boundary conditions for the eigenvalue problem. In ordinary quantum mechanics,
where the positive-definite inner product has the form $\int dx\,f^*(x)g(x)$,
the integral must be taken along the real axis and the path of integration
cannot be deformed into the complex plane because the integrand is not analytic.
The $\cP\cT$ inner product shares with the $\cC\cP\cT$ inner-product the
advantage of analyticity and path independence, but it suffers from
nonpositivity. It is surprising that we can construct a positive-definite metric
by using $\cC\cP\cT$ conjugation without disturbing the path independence of the
inner-product integral.

Time evolution is expressed by the operator $e^{-iHt}$ whether the theory is
determined by a $\cP\cT$-symmetric Hamiltonian or just an ordinary Hermitian
Hamiltonian. To establish unitarity we must show that as a state vector evolves,
its norm does not change in time. If $\psi_0(x)$ is any given initial vector
belonging to the Hilbert space spanned by the energy eigenstates, then it
evolves into the state $\psi_t(x)$ at time $t$ according to $\psi_t(x)=e^{-iHt}
\psi_0(x)$. With respect to the $\cC\cP\cT$ inner product the norm of $\psi_t
(x)$ does not change in time because $H$ commutes with $\cC\cP\cT$.

\subsection{Observables}
\label{ss4-4}

How do we represent an observable in $\cP\cT$-symmetric quantum mechanics?
Recall that in ordinary quantum mechanics the condition for a linear operator
$A$ to be an observable is that $A=A^\dagger$. This condition guarantees that
the expectation value of $A$ in a state is real. Operators in the Heisenberg
picture evolve in time according to $A(t)=e^{iHt}A(0)e^{-iHt}$, so this
Hermiticity condition is maintained in time. In $\cP\cT$-symmetric quantum
mechanics the equivalent condition is that at time $t=0$ the operator $A$ must
obey the condition $A^{\rm T}=\cC\cP\cT A\,\cC\cP\cT$, where $A^{\rm T}$ is the
{\em transpose} of $A$ \cite{BBJ1}. If this condition holds at $t=0$, then it
will continue to hold for all time because we have assumed that $H$ is symmetric
($H=H^{\rm T}$). This condition also guarantees that the expectation value of
$A$ in any state is real.\footnote{The requirement given here for $A$ to be an
observable involves matrix transposition. This condition is more restrictive
than is necessary and it has been generalized by Mostafazadeh. See
Refs.~\cite{Z4,P7x,Jobs}. Note that if the matrix transpose symmetry condition
on the Hamiltonian is removed we must introduce a biorthogonal basis. See
Refs.~\cite{WWWW1} and \cite{BIOX}.}

The operator $\cC$ itself satisfies this requirement, so it is an observable.
The Hamiltonian is also an observable. However, the $\x$ and $\p$ operators are
not observables. Indeed, the expectation value of $\x$ in the ground state is a
negative imaginary number. Thus, there is no position operator in $\cP
\cT$-symmetric quantum mechanics. In this sense $\cP\cT$-symmetric quantum
mechanics is similar to fermionic quantum field theories. In such theories the
fermion field corresponds to the $\x$ operator. The fermion field is complex and
does not have a classical limit. One cannot measure the position of an electron;
one can only measure the position of the {\em charge} or the {\em energy} of the
electron!

One can see why the expectation of the $x$ operator in $\cP\cT$-symmetric
quantum mechanics is a negative imaginary number by examining a classical
trajectory like that shown in Fig.~\ref{f15}. Note that this classical
trajectory has left-right ($\cP\cT$) symmetry, but not up-down symmetry. Also,
the classical paths favor (spend more time in) the lower-half complex-$x$ plane.
Thus, the average classical position is a negative imaginary number. Just as the
classical particle moves about in the complex plane, the quantum probability
current flows about in the complex plane. It may be that the correct
interpretation is to view $\cP\cT$-symmetric quantum mechanics as describing the
interaction of extended, rather than pointlike objects.

\subsection{Pseudo-Hermiticity and $\cP\cT$ Symmetry}
\label{ss4-5}

The thesis of this paper -- replacing the mathematical condition of Hermiticity
by the more physical condition of $\cP\cT$ symmetry -- can be placed in a more
general mathematical context known as pseudo-Hermiticity. A linear operator $A$
is {\em pseudo-Hermitian} if there is a Hermitian operator $\eta$ such that
\begin{equation}
A^\dag=\eta A\eta.
\label{e91}
\end{equation}
The operator $\eta$ is often called an {\em intertwining} operator. The
condition in (\ref{e91}) obviously reduces to ordinary Hermiticity when the
intertwining operator $\eta$ is the identity ${\bf 1}$. The concept of
pseudo-Hermiticity was introduced in the 1940s by Dirac and Pauli, and later
discussed by Lee, Wick, and Sudarshan, who were trying to resolve the problems
that arise in quantizing electrodynamics and other quantum field theories in
which negative-norm states states appear as a consequence of renormalization
\cite{P1,P2,P3,P4,P5}. These problems are illustrated very clearly by the Lee
model, which is discussed in Subsec.~\ref{ss8-3}.

Mostafazadeh observed that because the parity operator $\cP$ is Hermitian, it
may be used as an intertwining operator. The class of Hamiltonians $H$ in
(\ref{e12}) is pseudo-Hermitian because the parity operator $\cP$ changes the
sign of $\x$ while Dirac Hermitian conjugation changes the sign of $i$
\cite{M1,P6,P7,P7x}:
\begin{equation}
H^\dag=\cP H\cP.
\label{e92}
\end{equation}
For some other references on the generalization of $\cP\cT$ symmetry to
pseudo-Hermiticity see \cite{P8}.

\section{Illustrative $2\times2$ Matrix Example of a $\cP\cT$-Symmetric
Hamiltonian}
\label{s5}

It is always useful to study exactly solvable models when one is trying to
understand a formal procedure like that discussed in Sec.~\ref{s4}. One
exactly solvable model, which has been used in many papers on $\cP\cT$ symmetry,
is due to Swanson \cite{Swan}. This model is exactly solvable because it is
quadratic in $\x$ and $\p$. In this section we use an even simpler model to
illustrate the construction of a quantum theory described by a $\cP
\cT$-symmetric Hamiltonian. We consider the elementary $2\times2$ Hamiltonian
matrix
\begin{equation}
H=\left(\begin{array}{cc} re^{i\theta} & s \cr s & re^{-i\theta}
\end{array}\right),
\label{e93}
\end{equation}
where the three parameters $r$, $s$, and $\theta$ are real \cite{AJP}. The
Hamiltonian in (\ref{e93}) is not Hermitian, but it is easy to see that it is
$\cP\cT$ symmetric, where we define the parity operator as
\begin{equation}
\cP=\left(\begin{array}{cc} 0 & 1 \cr 1 & 0\end{array}\right)
\label{e94}
\end{equation}
and we define the operator $\cT$ to perform complex conjugation.

As a first step in analyzing the Hamiltonian (\ref{e93}), we calculate its two
eigenvalues:
\begin{equation}
E_\pm=r\cos\theta\pm(s^2-r^2\sin^2\theta)^{1/2}.
\label{e95}
\end{equation}
There are clearly two parametric regions to consider, one for which the square
root in (\ref{e95}) is real and the other for which it is imaginary. When $s^2<
r^2\sin^2\theta$, the energy eigenvalues form a complex conjugate pair. This is
the region of broken $\cP\cT$ symmetry. On the other hand, when $s^2\geq r^2
\sin^2\theta$, the eigenvalues $\varepsilon_\pm=r\cos\theta\pm(s^2-r^2\sin^2
\theta)^{1/2}$ are real. This is the region of unbroken $\cP\cT$ symmetry. In
the unbroken region the simultaneous eigenstates of the operators $H$ and
$\cP\cT$ are
\begin{equation}
|E_+\rangle=\frac{1}{\sqrt{2\cos\alpha}}
\left(\begin{array}{c} e^{i\alpha/2}\cr e^{-i\alpha/2}\end{array}\right)\quad
{\rm and}\quad|E_-\rangle=\frac{i}{\sqrt{2\cos\alpha}}\left(\begin{array}{c}
e^{-i\alpha/2}\cr-e^{i\alpha/2}\end{array}\right),
\label{e96}
\end{equation}
where
\begin{equation}
\sin\alpha =\frac{r}{s}\,\sin\theta.
\label{e97}
\end{equation}
The $\cP\cT$ inner product gives
\begin{equation}
(E_{\pm},E_{\pm})=\pm1\quad{\rm and}\quad (E_{\pm},E_{\mp})=0,
\label{e98}
\end{equation}
where $(u,v)=(\cP\cT u)\cdot v$. With respect to the $\cP\cT$ inner product, the
vector space spanned by the energy eigenstates has a metric of signature $(+,-
)$. If the condition $s^2>r^2\sin^2\theta$ for an unbroken $\cP\cT$ symmetry is
violated, the states (\ref{e96}) are no longer eigenstates of $\cP\cT$ because
$\alpha$ becomes imaginary. When $\cP\cT$ symmetry is broken, the $\cP\cT$ norm
of the energy eigenstate vanishes.

Next, we construct the operator $\cC$ using (\ref{e87}):
\begin{equation}
\cC=\frac{1}{\cos\alpha}\left(\begin{array}{cc} i\sin\alpha & 1 \cr 1 & -i\sin
\alpha\end{array}\right).
\label{e99}
\end{equation}
Note that $\cC$ is distinct from $H$ and $\cP$ and it has the key property that
\begin{equation}
\cC|E_{\pm}\rangle=\pm|E_{\pm}\rangle.
\label{e100}
\end{equation}
The operator $\cC$ commutes with $H$ and satisfies $\cC^2=1$. The eigenvalues of
$\cC$ are precisely the signs of the $\cP\cT$ norms of the corresponding
eigenstates. Using the operator $\cC$ we construct the new inner product
structure $\langle u|v\rangle= (\cC\cP\cT u)\cdot v$. This inner product is
positive definite because $\langle E_{\pm}|E_{\pm}\rangle=1$. Thus, the
two-dimensional Hilbert space spanned by $|E_\pm\rangle$, with inner product
$\langle\cdot|\cdot\rangle$, has signature $(+,+)$.

Finally, we prove that the $\cC\cP\cT$ norm of any vector is positive. For the
arbitrary vector $\psi=\left({a\atop b}\right)$, where 
$a$ and $b$ are any complex numbers, we see that
\begin{eqnarray}
&&\cT\psi=\left(\begin{array}{c}a^*\cr b^*\end{array}\right),\qquad
\cP\cT\psi=\left(\begin{array}{c} b^*\cr a^*\end{array}\right),\nonumber\\
&&\cC\cP\cT\psi=\frac{1}{\cos\alpha}\,\left(\begin{array}{c}a^*+ib^*\sin\alpha
\cr b^*-ia^*\sin\alpha\end{array}\right).
\label{e101}
\end{eqnarray}
Thus, $\langle\psi|\psi\rangle=(\cC\cP\cT\psi)\cdot\psi=\frac{1}{\cos\alpha}
[a^*a+b^*b+i(b^*b-a^*a)\sin\alpha]$. Now let $a=x+iy$ and $b=u+iv$, where $x$,
$y$, $u$, and $v$ are real. Then
\begin{equation}
\langle\psi|\psi\rangle=\frac{1}{\cos\alpha}\left(x^2+v^2+2xv\sin\alpha
+y^2+u^2-2yu\sin\alpha\right),
\label{e102}
\end{equation}
which is explicitly positive and vanishes only if $x=y=u=v=0$.

Since $\langle u|$ denotes the $\cC\cP\cT$-conjugate of $|u\rangle$, the
completeness condition reads
\begin{equation}
|E_+\rangle\langle E_+|+|E_-\rangle\langle E_-|=\left(\begin{array}{cc} 1 & 0
\cr 0 & 1\end{array}\right).
\label{e103}
\end{equation}
Furthermore, using the $\cC\cP\cT$ conjugate $\langle E_\pm|$, we can represent
$\cC$ as
\begin{equation}
\cC=|E_+\rangle\langle E_+|-|E_-\rangle\langle E_-|.
\label{e104}
\end{equation}

In the limit $\theta\to0$, the Hamiltonian (\ref{e93}) for this two-state system
becomes Hermitian and $\cC$ reduces to the parity operator $\cP$. Thus, $\cC\cP
\cT$ invariance reduces to the standard condition of Hermiticity for a symmetric
matrix; namely, $H=H^*$.

\section{Calculation of the $\cC$ Operator in Quantum Mechanics}
\label{s6}

The distinguishing feature of $\cP\cT$-symmetric quantum mechanics is the $\cC$
operator. In ordinary Hermitian quantum mechanics there is no such operator.
Only a non-Hermitian $\cP\cT$-symmetric Hamiltonian possesses a $\cC$ operator
distinct from the parity operator $\cP$. Indeed, if we were to sum the series in
(\ref{e87}) for a $\cP\cT$-symmetric Hermitian Hamiltonian, the result would be
$\cP$, which in coordinate space is $\delta(x+y)$. [See (\ref{e83}).]

While the $\cC$ operator is represented formally in (\ref{e87}) as an infinite
series, it is not easy to evaluate the sum of this series. Calculating $\cC$ by
direct brute-force evaluation of the sum in (\ref{e87}) is not easy in quantum
mechanics because it is necessary to find all the eigenfunctions $\phi_n(x)$ of
$H$. Furthermore, such a procedure cannot be used in quantum field theory
because in field theory there is no simple analog of the Schr\"odinger
eigenvalue differential equation and its associated coordinate-space
eigenfunctions.

The first attempt to calculate $\cC$ relied on a perturbative approach
\cite{BMW}. In this paper the $\cP\cT$-symmetric Hamiltonian
\begin{equation}
H=\half\p^2+\half\x^2+i\epsilon\x^3
\label{e105}
\end{equation}
is considered, where $\epsilon$ is treated as a small real parameter. When
$\epsilon=0$, the Hamiltonian reduces to the Hamiltonian for the quantum
harmonic oscillator, all of whose eigenfunctions can be calculated exactly.
Thus, it is possible to express the eigenfunctions of $H$ in (\ref{e105}) in the
form of perturbation series in powers of $\epsilon$. For each of the
eigenfunctions the first few terms in the perturbation series were calculated.
These perturbation series were then substituted into (\ref{e87}) and the
summation over the $n$th eigenfunction was performed. The result is a
perturbation-series expansion of the $\cC$ operator. This calculation is long
and tedious and the final result is quite complicated. However, the calculation
turned out to be of great value because while the final answer is complicated,
it was discovered that the answer in coordinate space simplified dramatically if
the $\cC$ operator is written as the exponential of a derivative operator $Q$
multiplying the parity operator $\cP$:
\begin{equation}
\cC(x,y)=\exp\left[Q\left(x,-i\frac{d}{dx}\right)\right]\delta(x+y).
\label{e106}
\end{equation}
The simplification that occurs when $\cC$ is written in the form $\cC=e^Q\cP$ is
that while the expression for $\cC$ is a series in all positive integer powers
of $\epsilon$, $Q$ is a series in {\em odd} powers of $\epsilon$ only. Since $Q$
is a series in odd powers of $\epsilon$, in the limit $\epsilon\to0$ the
function $Q$ vanishes. Thus, in this limit the $\cC$ operator tends to the
parity operator $\cP$.

The expression in (\ref{e106}) need not be limited to coordinate space. A more
general way to represent the $\cC$ operator is to express it generically in
terms of the fundamental dynamical operators $\x$ and $\p$:
\begin{equation}
\cC=e^{Q(\x,\p)}\cP.
\label{e107}
\end{equation}
Written in this form, $Q$ is a {\em real} function of its two variables. By
seeking the $\cC$ operator in the form (\ref{e107}) we will be able to devise
powerful analytic tools for calculating it.

We illustrate the representation $\cC=e^Q\cP$ by using two elementary
Hamiltonians. First, consider the shifted harmonic oscillator $H=\half\p^2+\half
\x^2+i\epsilon\x$. This Hamiltonian has an unbroken $\cP\cT$ symmetry for all
real $\epsilon$. Its eigenvalues $E_n=n+\half+\half\epsilon^2$ are all real. The
exact formula for $\cC$ for this theory is given exactly by $\cC=e^Q\cP$, where
\begin{equation}
Q=-\epsilon\p.
\label{e108}
\end{equation}
Note that in the limit $\epsilon\to0$, where the Hamiltonian becomes Hermitian,
$\cC$ becomes identical with $\cP$.

As a second example, consider the non-Hermitian $2\times2$ matrix Hamiltonian
(\ref{e93}). The associated $\cC$ operator in (\ref{e99}) can be rewritten in
the form $\cC=e^Q\cP$, where
\begin{equation}
Q=\half\sigma_2\ln\left(\frac{1-\sin\alpha}{1+\sin\alpha}\right)
\label{e109}
\end{equation}
and
\begin{equation}
\sigma_2=\left(\begin{array}{cc}0 & -i\cr i & 0\end{array}\right)
\label{e110}
\end{equation}
is the Pauli sigma matrix. Again, observe that in the limit $\theta\to0$, where
the Hamiltonian becomes Hermitian, the $\cC$ operator becomes identical with
$\cP$.

\subsection{Algebraic Equations Satisfied by the $\cC$ Operator}
\label{ss6-1}

Fortunately, there is a relatively easy algebraic way to calculate the $\cC$
operator, and the procedure circumvents the difficult problem of evaluating the
sum in (\ref{e87}). As a result, the technique readily generalizes from quantum
mechanics to quantum field theory. In this subsection we show how to use this
technique to calculate $\cC$ for the $\cP\cT$-symmetric Hamiltonian in
(\ref{e105}). We explain how to calculate $\cC$ perturbatively to high order in
powers of $\epsilon$ for this cubic Hamiltonian. Calculating $\cC$ for other
kinds of interactions is more difficult and may require the use of semiclassical
approximations \cite{rf20.5}.

To calculate $\cC$ we make use of its three crucial algebraic properties. First,
$\cC$ commutes with the space-time reflection operator $\cP\cT$,
\begin{eqnarray}
[\cC,\cP\cT]=0,
\label{e111}
\end{eqnarray}
although $\cC$ does not, in general, commute with $\cP$ or $\cT$ separately.
Second, the square of $\cC$ is the identity,
\begin{eqnarray}
\cC^2={\bf 1},
\label{e112}
\end{eqnarray}
which allows us to interpret $\cC$ as a reflection operator. Third, $\cC$
commutes with $H$,
\begin{eqnarray}
[\cC,H]=0,
\label{e113}
\end{eqnarray}
and thus is time independent. To summarize, $\cC$ is a time-independent $\cP
\cT$-symmetric reflection operator.

The procedure for calculating $\cC$ is simply to substitute the generic operator
representation in (\ref{e107}) into the three algebraic equations (\ref{e111})
-- (\ref{e113}) in turn and to solve the resulting equations for the function
$Q$. First, we substitute (\ref{e107}) into the condition (\ref{e111}) to obtain
\begin{equation}
e^{Q(\x,\p)}=\cP\cT e^{Q(\x,\p)}\cP\cT=e^{Q(-\x,\p)},
\label{e114}
\end{equation}
from which we conclude that $Q(\x,\p)$ is an {\em even} function of $\x$.

Second, we substitute (\ref{e107}) into the condition (\ref{e112}) and find that
\begin{equation}
e^{Q(\x,\p)}\cP e^{Q(\x,\p)}\cP=e^{Q(\x,\p)}e^{Q(-\x,-\p)}=1,
\label{e115}
\end{equation}
which implies that $Q(\x,\p)=-Q(-\x,-\p)$. Since we already know that $Q(\x,\p)$
is an even function of $\x$, we conclude that it is also an {\em odd} function
of $\p$.

The remaining condition (\ref{e113}) to be imposed is that the operator $\cC$
commutes with $H$. While the first two conditions are, in effect, {\em
kinematic} conditions on $Q$ that are generally true for any Hamiltonian,
condition (\ref{e113}) is equivalent to imposing the specific dynamics of the
particular Hamiltonian that defines the quantum theory. Substituting $\cC=e^{Q(
\x,\p)}\cP$ into (\ref{e113}), we get
\begin{equation}
e^{Q(\x,\p)}[\cP,H]+[e^{Q(\x,\p)},H]\cP=0.
\label{e116}
\end{equation}
This equation is difficult to solve in general, and to do so we must use
perturbative methods, as we explain in the next subsection.

\subsection{Perturbative Calculation of $\cC$.}
\label{ss6-2}

To solve (\ref{e116}) for the Hamiltonian in (\ref{e105}), we express this
Hamiltonian in the form $H=H_0+\epsilon H_1$. Here, $H_0$ is the
harmonic-oscillator Hamiltonian $H_0=\half\p^2+\half\x^2$, which commutes with
the parity operator $\cP$, and $H_1=i\x^3$, which {\em anticommutes} with $\cP$.
Thus, the condition (\ref{e116}) becomes
\begin{eqnarray}
2\epsilon e^{Q(\x,\p)}H_1=[e^{Q(\x,\p)},H].
\label{e117}
\end{eqnarray}

We expand the operator $Q(\x,\p)$ as a perturbation series in odd powers of
$\epsilon$:
\begin{eqnarray}
Q(\x,\p)=\epsilon Q_1(\x,\p)+\epsilon^3Q_3(\x,\p)+\epsilon^5Q_5(\x,\p)+\cdots\,.
\label{e118}
\end{eqnarray}
Substituting the expansion in (\ref{e118}) into the exponential $e^{Q(\x,\p)}$,
we get after some algebra a sequence of equations that can be solved 
systematically for the operator-valued functions $Q_n(\x,\p)$ $(n=1,3,5,\ldots)$
subject to the symmetry constraints that ensure the conditions (\ref{e111}) and 
(\ref{e112}). The first three of these equations are
\begin{eqnarray}
\left[H_0,Q_1\right] &=& -2H_1,\nonumber\\
\left[H_0,Q_3\right] &=& -{\textstyle\frac{1}{6}}[Q_1,[Q_1,H_1]],\nonumber\\
\left[H_0,Q_5\right] &=& {\textstyle\frac{1}{360}}[Q_1,[Q_1,[Q_1,[Q_1,H_1]]]]
-{\textstyle\frac{1}{6}}[Q_1,[Q_3,H_1]]\nonumber\\
&&\qquad+{\textstyle\frac{1}{6}}[Q_3,[Q_1,H_1]].
\label{e119}
\end{eqnarray}

Let us solve these equations for the Hamiltonian in (\ref{e105}), for which
$H_0=\half\p^2+\half\x^2$ and $H_1=i\x^3$. The procedure is to substitute the
most general polynomial form for $Q_n$ using arbitrary coefficients and then to
solve for these coefficients. For example, to solve $\left[H_0,Q_1\right]=-2i
\x^3$, the first of the equations in (\ref{e119}), we take as an {\em ansatz}
for $Q_1$ the most general Hermitian cubic polynomial that is even in $\x$ and
odd in $p$:
\begin{eqnarray}
Q_1(\x,\p)=M\p^3+N\x\p\x,
\label{e120}
\end{eqnarray}
where $M$ and $N$ are numerical coefficients to be determined. The operator
equation for $Q_1$ is satisfied if $M=-\frac{4}{3}$ and $N=-2$.

It is algebraically tedious but completely straightforward to continue this
process. In order to present the solutions for $Q_n(\x,\p)$ ($n>1$), it is
convenient to introduce the following notation: Let $S_{m,n}$ represent the {\em
totally symmetrized} sum over all terms containing $m$ factors of $\p$ and $n$
factors of $\x$. For example,
\begin{equation}
S_{0,0}=1,~S_{0,3}=\x^3,~S_{1,1}=\half\left(\x\p+\p\x\right),~
S_{1,2}={\textstyle\frac{1}{3}}\left(\x^2\p+\x\p\x+\p\x^2\right),
\label{e121}
\end{equation}
and so on. (The properties of the operators $S_{m,n}$ are summarized in
Ref.~\cite{rf23}.)

In terms of the symmetrized operators $S_{m,n}$ the first three functions
$Q_{2n+1}$ are
\begin{eqnarray}
Q_1 &=& -{\textstyle\frac{4}{3}}\p^3-2S_{1,2},\nonumber\\
Q_3 &=& {\textstyle\frac{128}{15}}\p^5+{\textstyle\frac{40}{3}}S_{3,2}+8S_{1,4}
-12\p,\nonumber\\
Q_5 &=& -{\textstyle\frac{320}{3}}\p^7-{\textstyle\frac{544}{3}}S_{5,2}-
{\textstyle\frac{512}{3}}S_{3,4}-64S_{1,6}+{\textstyle\frac{24\,736}{45}}\p^3
+{\textstyle\frac{6\,368}{15}}S_{1,2}.
\label{e122}
\end{eqnarray}
Together, (\ref{e107}), (\ref{e118}), and (\ref{e122}) represent an explicit
perturbative expansion of $\cC$ in terms of the operators $\x$ and $\p$, correct
to order $\epsilon^6$.

To summarize, using the {\em ansatz} (\ref{e107}) we can calculate $\cC$ to high
order in perturbation theory. We are able to perform this calculation because
this {\em ansatz} obviates the necessity of calculating the $\cP\cT$-normalized
wave functions $\phi_n(x)$. We show how use these same techniques for quantum
field theory in Sec.~\ref{s8}.

\subsection{Perturbative Calculation of $\cC$ for Other Quantum-Mechanical
Hamiltonians}
\label{ss6-3}

The $\cC$ operator has been calculated perturbatively for a variety of
quantum-mechanical models. For example, let us consider first the case of the
Hamiltonian
\begin{equation}
H=\half\left(\p^2+\q^2\right)+\half\left(\x^2+\y^2\right)+i\epsilon \x^2\y,
\label{e123}
\end{equation}
which has two degrees of freedom. The energy levels of this complex
H\'enon-Heiles theory were studied in Ref.~\cite{BDMS}. The $\cC$ operator for 
this Hamiltonian was calculated in \cite{BBRR,BBJ2,BBJ3}. The perturbative
result for the $Q=Q_1\epsilon+Q_3\epsilon^3+\ldots$ operator is
\begin{eqnarray}
Q_1(\x,\y,\p,\q) &=& -{\textstyle\frac{4}{3}}\p^2\q-{\textstyle\frac{1}{3}}
S_{1,1}y-{\textstyle\frac{2}{3}}\x^2\q,\nonumber\\
Q_3(\x,\y,\p,\q) &=& {\textstyle\frac{512}{405}}\p^2\q^3+{\textstyle\frac{512}
{405}}\p^4\q+{\textstyle\frac{1088}{405}}S_{1,1}T_{2,1}\nonumber\\
&& -{\textstyle\frac{256}{405}}\p^2T_{1,2}+{\textstyle\frac{512}{405}}S_{3,1}\y
+{\textstyle\frac{288}{405}}S_{2,2}\q -{\textstyle\frac{32}{405}}\x^2\q^3
\nonumber\\
&&+{\textstyle\frac{736}{405}}\x^2T_{1,2}-{\textstyle\frac{256}{405}}S_{1,1}\y^3
+{\textstyle\frac{608}{405}}S_{1,3}\y-{\textstyle\frac{128}{405}}\x^4\q
-{\textstyle\frac{8}{9}}q\,,
\label{e124}
\end{eqnarray}
where $T_{m,n}$ represents a totally symmetric product of $m$ factors of $\q$
and $n$ factors of $\y$.

For the Hamiltonian
\begin{eqnarray}
H=\half\left(\p^2+\q^2+\rr^2\right)+\half\left(\x^2+\y^2+\z^2\right)+i
\epsilon\x\y\z,
\label{e125}
\end{eqnarray}
which has three degrees of freedom, we have \cite{BBRR,BBJ2,BBJ3}
\begin{eqnarray}
Q_1(\x,\y,\z,\p,\q,\rr) &=& -{\textstyle\frac{2}{3}}(\y\z\p+\x\z\q+\x\y\rr)
-{\textstyle\frac{4}{3}}\p\q\rr,\nonumber\\
Q_3(\x,\y,\z,\p,\q,\rr) &=& {\textstyle\frac{128}{405}}\left(\p^3\q\rr+\q^3\p\rr
+\rr^3\q\p\right)\nonumber\\
&&+{\textstyle\frac{136}{405}}[\p\x\p(\y\rr+\z\q)+\q\y\q(\x\rr+\z\p)+\rr\z\rr
(\x\q+\y\p)]\nonumber\\
&&-{\textstyle\frac{64}{405}}(\x\p\x\q\rr+\y\q\y\p\rr+\z\rr\z\p\q)+
{\textstyle\frac{184}{405}}(\x\p\x\y\z+\y\q\y\x\z\nonumber\\
&&+\z\rr\z\x\y)-{\textstyle\frac{32}{405}}\big[\x^3(\y\rr+\z\q)+\y^3(\x\rr+\z\p)
\nonumber\\
&&+\z^3(\x\q+\y\p)\big]-{\textstyle\frac{8}{405}}\left(\p^3\y\z+\q^3\x\z+\rr^3
\x\y\right).
\label{e126}
\end{eqnarray}

An extremely interesting and deceptively simple quantum-mechanical model is the 
$\cP\cT$-symmetric square well, whose Hamiltonian on the domain $0<x<\pi$ is
given by
\begin{equation}
H=\p^2+V(\x),
\label{e127}
\end{equation}
where in the coordinate representation $V(x)=\infty$ for $x<0$ and $x>\pi$ and
\begin{equation} V(x) = \left\{ \begin{array}{cl}
i\epsilon \quad & \mbox{for $\frac{\pi}{2}<x<\pi,$} \\
-i\epsilon \quad & \mbox{for $0<x<\frac{\pi}{2}.$} \\
\end{array}\right.
\label{e128}
\end{equation}
The $\cP\cT$-symmetric square-well Hamiltonian was invented by Znojil \cite{Z1}
and it has been heavily studied by many researchers \cite{Z2,Z3,Z4,Z5}. This
Hamiltonian reduces to the conventional Hermitian square well in the limit as
$\epsilon\to0$. For $H$ in (\ref{e127}) the parity operator $\cP$ performs a
reflection about $x=\frac{\pi}{2}$: $\cP:x\to\pi-x$.

In all the examples discussed so far the coordinate-space representation of the
$\cC$ operator is a combination of integer powers of $x$ and integer numbers of
derivatives multiplying the parity operator $\cP$. Hence, the $Q$ operator is a
polynomial in the operators $\x$ and $\p=-i\frac{d}{dx}$. The novelty of the
$\cP\cT$-symmetric square-well Hamiltonian (\ref{e127}) and (\ref{e128}) is that
$\cC$ contains {\em integrals} of $\cP$. Thus, the $Q$ operator, while it is a
simple function, is {\em not} a polynomial in $\x$ and $\p$ and therefore it
cannot be found easily by the algebraic perturbative methods introduced in
Subsec.~\ref{ss6-1}.

Instead, in Ref.~\cite{BT} the $\cC$ operator for the square-well Hamiltonian
was calculated by using the brute-force approach of calculating the $\cP
\cT$-normalized eigenfunctions $\phi_n(x)$ of $H$ and summing over these
eigenfunctions. The eigenfunctions $\phi_n(x)$ were obtained as perturbation
series to second order in powers of $\epsilon$. The eigenfunctions were then
normalized according to (\ref{e80}), substituted into the formula (\ref{e87}),
and the sum was evaluated directly to obtain the $\cC$ operator accurate to
order $\epsilon^2$. The advantage of the domain $0<x<\pi$ is that this sum
reduces to a set of Fourier sine and cosine series that can be evaluated in
closed form. After evaluating the sum, the result was translated to the
symmetric region $-\frac{\pi}{2}<x<\frac{\pi}{2}$. On this domain the parity
operator in coordinate space is $\cP(x,y)=\delta(x+y)$.

The last step in the calculation is to show that the $\cC$ operator to order
$\epsilon^2$ has the form $e^Q\cP$ and then to evaluate the function $Q$ to
order $\epsilon^2$. The final result for $Q(x,y)$ on the domain $-\frac{\pi}
{2}<x<\frac{\pi}{2}$ has the relatively simple structure
\begin{equation}
Q(x,y)=\textstyle{\frac{1}{4}}i\epsilon[x-y+\varepsilon(x-y)\,(|\,x+y\,|-\pi)]+
\mathcal{O}(\epsilon^3),
\label{e129}
\end{equation}
where $\varepsilon(x)$ is the standard step function
\begin{equation}
\varepsilon(x)=\left\{\begin{array}{cl}1\quad&\mbox{($x>0$),}\\
0\quad&\mbox{($x=0$),}\\ -1\quad&\mbox{($x<0$).}\\ \end{array}\right.
\label{e130}
\end{equation}

\begin{figure}[th]\vspace{2.65in}
\includegraphics{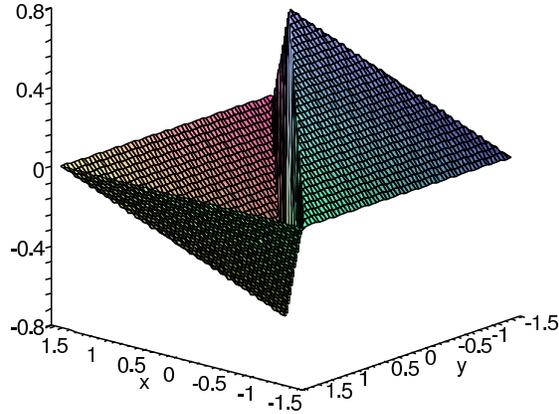}
\caption{Three-dimensional plot of the imaginary part of $\cC^{(1)}(x,y)$, the
first-order perturbative contribution in (\ref{e131}) to the $\cC$ operator in
coordinate space. The plot is on the symmetric square domain $-\frac{\pi}{2}<(x,
y)<\frac{\pi}{2}$. Note that $\cC^{(1)}(x,y)$ vanishes on the boundary of this
square domain because the eigenfunctions $\phi_n(x)$ are required to vanish at
$x=0$ and $x=\pi$.}
\label{f22}
\end{figure}

\begin{figure}[th]\vspace{2.65in}
\includegraphics{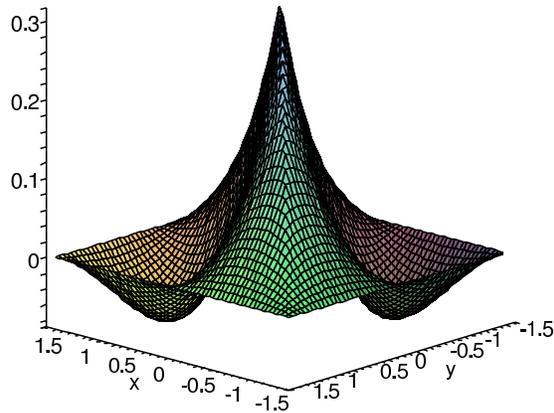}
\caption{Three-dimensional plot of $\cC^{(2)}(x,y)$ in (\ref{e131}) on the
symmetric square domain $-\frac{\pi}{2}<(x,y)<\frac{\pi}{2}$. The function
$\cC^{(2)}(x,y)$ vanishes on the boundary of this square domain because the
eigenfunctions $\phi_n(x)$ from which it was constructed vanish at the
boundaries of the square well.}
\label{f23}
\end{figure}

When we say that the formula for $Q(x,y)$ has a relatively simple structure, we
mean that this structure is simple in comparison with the formula for the
$C$ operator expressed as an expansion $C(x,y)=C^{(0)}+\epsilon C^{(1)}+
\epsilon^2C^{(2)}+\mathcal{O}(\epsilon^3)$. The formulas for the first three
coefficients in this series for $\cC$ are
\begin{eqnarray}
\cC^{(0)}(x,y)&=&\delta(x+y),\nonumber\\
\cC^{(1)}(x,y)&=&{\textstyle\frac{1}{4}}i[x+y+\varepsilon(x+y)\,(|x-y|-\pi)],
\nonumber\\
\cC^{(2)}(x,y)&=&{\textstyle\frac{1}{96}}\pi^3-{\textstyle\frac{1}{24}}(x^3+y^3)
\,\varepsilon(x+y)-{\textstyle\frac{1}{24}}(y^3-x^3)\,\varepsilon(y-x)
\nonumber\\
&&+{\textstyle\frac{1}{8}}xy\pi-{\textstyle\frac{1}{16}}\pi^2(x+y)\,\varepsilon(
x+y)+{\textstyle\frac{1}{8}}\pi(x|x|+y|y|)\,\varepsilon(x+y)\nonumber\\
&&-{\textstyle\frac{1}{4}}xy\{|x|[\theta(x-y)\,\theta(-x-y)+\theta(y-x)\,\theta
(x+y)]\nonumber\\
&&+|y|[\theta(y-x)\,\theta(-x-y)+\theta(x-y)\,\theta(x+y)]\}.
\label{e131}
\end{eqnarray}
We plot the imaginary part of $\cC^{(1)}(x,y)$ in Fig.~\ref{f22}, and we plot
$\cC^{(2)}(x,y)$ in Fig.~\ref{f23}. These three-dimensional plots show
$\cC^{(1)}(x,y)$ and $\cC^{(2)}(x,y)$ on the symmetric domain $-\frac{\pi}{2}<
(x,y)<\frac{\pi}{2}$.

The most noteworthy property of the $\cC$ operator for the square-well model is
that the associated operator $Q$ is a nonpolynomial function, and this kind of
structure had not been seen in previous studies of $\cC$. It was originally
believed that for such a simple $\cP\cT$-symmetric Hamiltonian it would be
possible to calculate the $\cC$ operator in closed form. It is a surprise that
for this elementary model the $\cC$ operator is so nontrivial.

\subsection{Mapping from a $\cP\cT$-Symmetric Hamiltonian to a Hermitian
Hamiltonian}
\label{ss6-4}

Mostafazadeh observed that the square root of the positive operator $e^Q$ can be
used to construct a similarity transformation that maps a non-Hermitian $\cP
\cT$-symmetric Hamiltonian $H$ to an equivalent Hermitian Hamiltonian $h$
\cite{M1,M2}:
\begin{equation}
h=e^{-Q/2}He^{Q/2}.
\label{e132}
\end{equation}
He noted that $h$ is equivalent to $H$ because it has the same eigenvalues as
$H$.

To understand why (\ref{e132}) is valid, recall from (\ref{e107}) that the $\cC$
operator has the general form $\cC=e^Q\cP$, where $Q=Q(\x,\p)$ is a Hermitian
function of the fundamental dynamical operator variables of the quantum theory.
Multiplying $\cC$ on the right by $\cP$ gives an expression for $e^Q$:
\begin{equation}
e^Q=\cC\cP,
\label{e133}
\end{equation}
which indicates that $\cC\cP$ is a positive and invertible operator.

To verify that $h$ in (\ref{e132}) is Hermitian, take the Hermitian conjugate of
(\ref{e132})
\begin{equation}
h^\dag=e^{Q/2}H^\dag e^{-Q/2},
\label{e134}
\end{equation}
which can be rewritten as
\begin{equation}
h^\dag=e^{-Q/2}e^QH^\dag e^{-Q}e^{Q/2}.
\label{e135}
\end{equation}
Use (\ref{e133}) to replace $e^Q$ by $\cC\cP$ and $e^{-Q}$ by $\cP\cC$
\begin{equation}
h^\dag=e^{-Q/2}\cC\cP H^\dag\cP\cC e^{Q/2}
\label{e136}
\end{equation}
and recall from (\ref{e92}) that $H^\dag$ can be replaced by $\cP H\cP$. This
gives
\begin{equation}
h^\dag=e^{-Q/2}\cC\cP\cP H\cP\cP\cC e^{Q/2}=e^{-Q/2}\cC H\cC e^{Q/2}.
\label{e137}
\end{equation}
Finally, one use the fact that $\cC$ commutes with $H$ [see (\ref{e113})] and
that the square of $\cC$ is unity [see (\ref{e112})] to reduce the right side 
of (\ref{e136}) to the right side of (\ref{e132}). This verifies that $h$ is
Hermitian in the Dirac sense.

We conclude from this calculation that for every non-Hermitian $\cP
\cT$-symmetric Hamiltonian $H$ whose $\cP\cT$ symmetry is unbroken, it is,
in principle, possible to construct via (\ref{e132}) a Hermitian Hamiltonian $h$
that has exactly the same eigenvalues as $H$. Note that it is crucial that $H$
have an unbroken $\cP\cT$ symmetry because this allows us to construct $\cC$,
which in turn allows us to construct the similarity operator $\e^{Q/2}$.

This construction poses the following question: Are $\cP\cT$-symmetric
Hamiltonians physically new and distinct from ordinary Hermitian Hamiltonians,
or do they describe exactly the same physical processes that ordinary
Hermitian Hamiltonians describe?

There are two answers to this question, the first being technical and practical
and the second being an answer in principle. First, while it is theoretically
possible to construct the Hermitian Hamiltonian $h$ whose spectrum is identical
to that of $H$, it cannot in general be done except at the perturbative level
because the transformation is so horribly complicated. (See, for example, the
discussion of the square well in Subsec.~\ref{ss6-3}.) Furthermore, while the
$\cP\cT$-symmetric Hamiltonian $H$ is simple in structure and easy to use in
calculations because the interaction term is local, it is shown in
Ref.~\cite{k1} that $h$ is {\em nonlocal} (its interaction term has arbitrarily
high powers of the variables $\x$ and $\p$.) Thus, it is not just difficult to
calculate using $h$; it is almost impossible because the usual regulation
schemes are hopelessly inadequate.

There is only one known nontrivial example for which there is actually a
closed-form expression for both $H$ and $h$, and this is the case of the quartic
Hamiltonian discussed in Subecs.~\ref{ss2-6} -- \ref{ss2-8}. Even this case, it
is not possible to construct the $\cC$ operator in closed form because this
operator is nonlocal (it contains a Fourier transform) and it performs a
transformation in the complex plane. This is the only known example for which it
is practical to calculate with both $H$ and $h$. Hence, while the mapping from
$H$ to $h$ is of great theoretical interest, it does not have much practical
value.

The second answer is of greater importance because it leads immediately to
physical considerations. The transformation from $H$ to $h$ in (\ref{e132}) is
a {\em similarity} and not a unitary transformation. Thus, while the eigenvalues
of $H$ and $h$ are the same, relationships between vectors are changed; pairs of
vectors that are orthogonal are mapped into pairs of vectors that are not
orthogonal. Thus, experiments that measure vectorial relationships can
distinguish between $H$ and $h$. One plausible experiment, which involves the
speed of unitary time evolution, is described in detail in Subsec.~\ref{ss7-3}.

\section{Applications of $\cP\cT$-Symmetric Hamiltonians in Quantum Mechanics}
\label{s7}

It is not yet known whether non-Hermitian $\mathcal{PT}$-symmetric Hamiltonians
describe phenomena that can be observed experimentally. However, non-Hermitian
$\cP\cT$-symmetric Hamiltonians have {\em already} appeared in the literature
very often and their remarkable properties have been noticed and used by many 
authors. For example, in 1959 Wu showed that the ground state of a Bose system
of hard spheres is described by a non-Hermitian Hamiltonian \cite{A1}. Wu found
that the ground-state energy of this system is real and he conjectured that all
the energy levels were real. Hollowood showed that the non-Hermitian Hamiltonian
of a complex Toda lattice has real energy levels \cite{H4}. Cubic non-Hermitian
Hamiltonians of the form $H=\p^2+i\x^3$ (and also cubic quantum field theories)
arise in studies of the Lee-Yang edge singularity \cite{A3,A4,A5,A6} and in
various Reggeon field theory models \cite{A7,A8}. In all of these cases a
non-Hermitian Hamiltonian having a real spectrum appeared mysterious at the
time, but now the explanation is simple: In every case the non-Hermitian
Hamiltonian is $\cP\cT$ symmetric. In each case the Hamiltonian is constructed
so that the position operator $\x$ or the field operator $\phi$ is always
multiplied by $i$. Hamiltonians having $\cP\cT$ symmetry have also been used to
describe magnetohydrodynamic systems \cite{G1,G2} and to study nondissipative
time-dependent systems interacting with electromagnetic fields \cite{Frig}.

In this section we describe briefly four areas of quantum mechanics in which
non-Hermitian $\cP\cT$-Hamiltonians play a crucial and significant role. 

\subsection{New $\cP\cT$-Symmetric Quasi-Exactly Solvable Hamiltonians}
\label{ss7-1}

A quantum-mechanical Hamiltonian is said to be {\em quasi-exactly solvable}
(QES) if a finite portion of its energy spectrum and associated eigenfunctions
can be found exactly and in closed form \cite{A9}. QES potentials depend on an
integer parameter $J$; for positive values of $J$ one can find exactly the first
$J$ eigenvalues and eigenfunctions, typically of a given parity. QES systems are
classified using an algebraic approach in which the Hamiltonian is expressed in
terms of the generators of a Lie algebra \cite{A10}. This approach generalizes
the dynamical-symmetry analysis of {\em exactly solvable} quantum-mechanical
systems whose {\em entire} spectrum may be found in closed form by algebraic
means.

Prior to the discovery of non-Hermitian $\cP\cT$-symmetric Hamiltonians the
lowest-degree one-dimensional QES polynomial potential that was known was a
sextic potential having one continuous parameter as well as a discrete parameter
$J$. A simple case of such a potential is \cite{BD}
\begin{equation}
V(x)=x^6-(4J-1)x^2.
\label{e138}
\end{equation}
For this potential, the Schr\"odinger equation $-\psi''(x)+[V(x)-E]\psi(x)=0$
has
$J$ even-parity solutions of the form
\begin{equation}
\psi(x)=e^{-x^4/4}\sum_{k=0}^{J-1}c_k x^{2k}.
\label{e139}
\end{equation}
The coefficients $c_k$ for $0\leq k\leq J-1$ satisfy the recursion relation
\begin{equation}
4(J-k)c_{k-1}+Ec_k+2(k+1)(2k+1)c_{k+1}=0,
\label{e140}
\end{equation}
where $c_{-1}=c_{J}=0$. The linear equations (\ref{e140}) have a nontrivial
solution for $c_0,\,c_1,\,...,\,c_{J-1}$ if the determinant of the coefficients
vanishes. For each integer $J$ this determinant is a polynomial of degree $J$ in
the variable $E$. The roots of this polynomial are all real and they are the $J$
quasi-exact energy eigenvalues of the potential (\ref{e138}).

The discovery of $\cP\cT$ symmetry allows us to introduce an entirely new class
of QES {\em quartic} polynomial potentials having {\em two} continuous
parameters in addition to the discrete parameter $J$. The Hamiltonian has the
form \cite{A11}
\begin{equation}
H=\p^2-\x^4+2ia\x^3+(a^2-2b)\x^2+2i(ab-J)\x,
\label{e141}
\end{equation}
where $a$ and $b$ are real and $J$ is a positive integer. The spectra of this
family of Hamiltonians are real, discrete, and bounded below. Like the
eigenvalues of the potential (\ref{e138}), the lowest $J$ eigenvalues of $H$ are
the roots of a polynomial of degree $J$.

The eigenfunction $\psi(x)$ satisfies $\cP\cT$-symmetric boundary conditions;
it vanishes in the Stokes wedges shown in Fig.~\ref{f4}. The eigenfunction
satisfies
\begin{equation}
-\psi''(x)+\left[-x^4+2iax^3+(a^2-2b)x^2+2i(ab-J)x\right]\psi(x)=E\psi(x).
\label{e142}
\end{equation}

We obtain the QES portion of the spectrum of $H$ in (\ref{e141}) by making
the {\em ansatz} $\psi(x)=\exp\left(-\frac{1}{3}ix^3-\half ax^2-ibx\right)
P_{J-1}(x)$, where 
\begin{equation}
P_{J-1}(x)=x^{J-1}+\sum_{k=0}^{J-2}c_k x^k
\label{e143}
\end{equation}
is a polynomial in $x$ of degree $J-1$. Substituting $\psi(x)$ into the
differential equation (\ref{e142}), dividing off the exponential, and collecting
powers of $x$, we obtain a polynomial in $x$ of degree $J-1$. Setting the
coefficients of $x^k$ ($1\leq k\leq J-1$) to $0$ gives a system of $J-1$
simultaneous linear equations for the coefficients $c_k$ ($0\leq k \leq J-2$).
We solve these equations and substitute the values of $c_k$ into the coefficient
of $x^0$. This gives a polynomial $Q_J(E)$ of degree $J$ in the eigenvalue $E$.
The coefficients of this polynomial are functions of the parameters $a$ and $b$.
The first two of these polynomials are
\begin{eqnarray}
Q_1 &=& E -b^2 -a,\nonumber\\
Q_2 &=& E^2 -(2b^2+4a)E+b^4+4ab^2-4b+3a^2.
\label{e144}
\end{eqnarray}
The roots of $Q_J(E)$ are the QES portion of the spectrum of $H$.

The polynomials $Q_J(E)$ simplify dramatically when we substitute
\begin{equation}
E=F+b^2+Ja\qquad{\rm and}\qquad K=4b+a^2.
\label{e145}
\end{equation}
The new polynomials then have the form
\begin{eqnarray}
Q_1 &=& F,\nonumber\\
Q_2 &=& F^2-K,\nonumber\\
Q_3 &=& F^3-4KF-16,\nonumber\\
Q_4 &=& F^4-10KF^2-96F+9K^2,\nonumber\\
Q_5 &=& F^5-20KF^3-336F^2+64K^2F+768K.
\label{e146}
\end{eqnarray}
The roots of these polynomials are all real as long as $K\geq K_{\rm critical}$,
where $K_{\rm critical}$ is a function of $J$. 

Had $\cP\cT$-symmetric quantum mechanics not been discovered, this beautiful
family of quartic QES Hamiltonians would never have been considered because the
quartic term has a negative sign. Lacking the discovery that $\cP\cT$-symmetric
Hamiltonians have positive spectra, the Hamiltonians in (\ref{e141}) would have
been rejected because it would have been assumed that the spectra of such
Hamiltonians would be unbounded below. Quasi-exactly solvable sextic $\cP
\cT$-symmetric Hamiltonians are studied in \cite{MONOU}.

\subsection{Complex Crystals}
\label{ss7-2}

An experimental signal of a complex Hamiltonian might be found in the context of
condensed matter physics. Consider the complex crystal lattice whose potential
is $V(x)=i\sin\,x$. The optical properties of complex crystal lattices were
first studied by Berry and O'Dell, who referred to them as complex diffraction
gratings \cite{MVB}.

\begin{figure*}[t!]
\vspace{2.5in}
\includegraphics{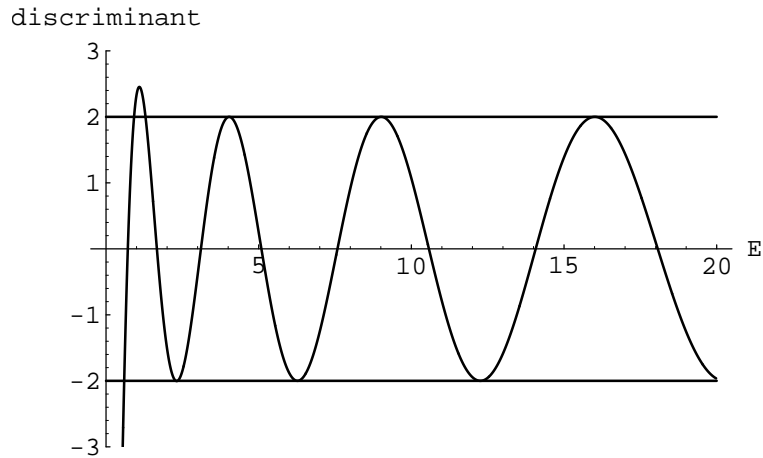}
\caption{Discriminant $\Delta(E)$ plotted as a function of $E$ for the real
periodic potential $V(x)=\sin(x)$. Although it cannot be seen in the figure, all
local maxima lie {\em above} the line $\Delta=2$ and all local minima lie {\em
below} the line $\Delta=-2$. The regions of energy $E$ for which $|\Delta|\leq2$
correspond to allowed energy bands and the regions where $|\Delta|>2$ correspond
to gaps. There are infinitely many gaps and these gaps become exponentially
narrow as $E$ increases. It is clear in this figure that the first maximum lies
above $2$. The first minimum occurs at $E=2.313\,8$, where the discriminant has
the value $-2.003\,878\,7$. The second maximum occurs at $E=4.033\,6$, where the
discriminant is $2.000\,007$. Similar behavior is found for other potentials in
the class $V(x)=\sin^{2N+1}(x)$.}
\label{f24}
\end{figure*}

\begin{figure*}[t!]
\vspace{2.65in}
\includegraphics{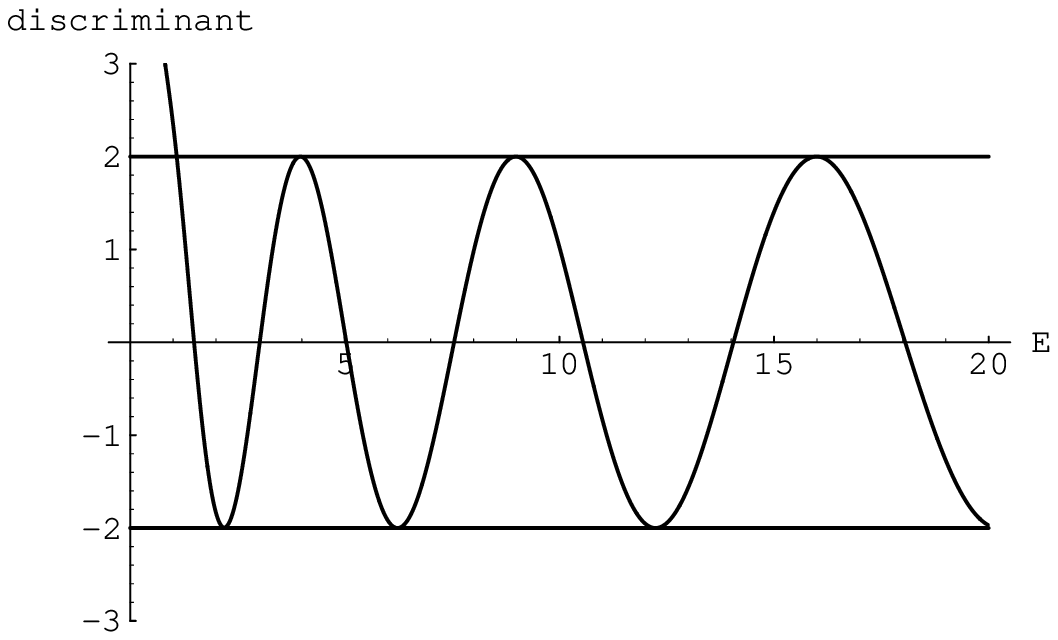}
\caption{Discriminant $\Delta(E)$ plotted as a function of $E$ for the complex
periodic potential $V(x)=i\sin(x)$. For $E>2$ it is not possible to see any
difference between this figure and Fig.~\ref{f24}. However, although it cannot
be seen, all local maxima in this figure lie {\em above} the line $\Delta=2$ and
all local minima lie {\em above} the line $\Delta =-2$. This behavior is
distinctly different from the behavior in Fig.~\ref{f24} exhibited by the real
periodic potential $V(x)=\sin x$. In stark contrast with real periodic
potentials, for the potentials of the form $V(x)=i\sin^{2N+1}(x)$, while the
maxima of the discriminant lie above $+2$, the minima of the discriminant lie
above $-2$. Thus, for these potentials there are no antiperiodic wave functions.
Lengthy and delicate numerical analysis verifies that for the potential $V(x)=i
\sin(x)$ the first three maxima of the discriminant $\Delta(E)$ are located at
$E=3.966\,428\,4$, $E=8.985\,732\,0$, and $E=15.992\,066\,213\,46$. The value of
the discriminant $\Delta(E)$ at these energies is $2.000\,007$, $2.000\,000\,000
\,000\,69$, and $2.000\,000\,000\,000\,000\,000\,04$. The first two minima of
the discriminant are located at $E=2.191\,6$ and $E=6.229\,223$ and at these
energies $\Delta(E)$ has the values $-1.995\,338\,6$ and $-1.999\,999\,995\,27$.
Similar behavior is found for the other complex periodic potentials in the
class.}
\label{f25}
\end{figure*}

While the Hamiltonian $H=\p^2+i\sin\x$ is not Hermitian, it is $\cP\cT$
symmetric and all of its energy bands are {\em real}. At the edge of the bands
the wave function of a particle in such a lattice is bosonic ($2\pi$-periodic),
and unlike the case of ordinary crystal lattices, the wave function is never
fermionic ($4\pi$-periodic). The discriminant for a Hermitian $\sin(x)$
potential is plotted in Fig.~\ref{f24} and the discriminant for a non-Hermitian
$i\sin(x)$ potential is plotted in Fig.~\ref{f25}. The difference between these
two figures is subtle. In Fig.~\ref{f25} the discriminant does not go below $-
2$, and thus there are half as many gaps \cite{A12}. Direct observation of such
a band structure would give unambiguous evidence of a $\cP\cT$-symmetric
Hamiltonian \cite{ZZ}. Complex periodic potentials having more elaborate band
structures have also been found \cite{KS1,KS2,KS3,KS4}.

\subsection{Quantum Brachistochrone}
\label{ss7-3}

We pointed out in Subsec.~\ref{ss6-4} that there is a similarity transformation
that maps a non-Hermitian $\cP\cT$-symmetric Hamiltonian $H$ to a Hermitian
Hamiltonian $h$ [see (\ref{e132})]. The two Hamiltonians, $H$ and $h$, have the 
same eigenvalues, but this does not mean they describe the same physics. To
illustrate the differences between $H$ and $h$, we show how to solve the quantum
brachistochrone problem for $\cP\cT$-symmetric and for Hermitian quantum
mechanics, and we show that the solution to this problem in these two
formulations of quantum mechanics is not the same.

The fancy word {\em brachistochrone} means ``shortest time.'' Thus, the {\em
quantum brachistochrone} problem is defined as follows: Suppose we are given
initial and final quantum states $|\psi_I\rangle$ and $|\psi_F\rangle$ in a
Hilbert space. There exist infinitely many Hamiltonians $H$ under which $|\psi_I
\rangle$ evolves into $|\psi_F\rangle$ in some time $t$:
\begin{equation}
|\psi_F\rangle=e^{-iHt/\hbar}|\psi_I\rangle.
\label{e147}
\end{equation}
The problem is to find the Hamiltonian $H$ that minimizes the evolution time $t$
subject to the constraint that $\omega$, the difference between the largest and
smallest eigenvalues of $H$, is held fixed. The shortest evolution time is
denoted by $\tau$.

In Hermitian quantum mechanics there is an unavoidable lower bound $\tau$ on the
time required to transform one state into another. Thus, the minimum time
required to flip unitarily a spin-up state into a spin-down state of an electron
is an important physical quantity because it gives an upper bound on the
speed of a quantum computer.

In this paper we have shown that Hermitian quantum mechanics can be extended
into the complex domain while retaining the reality of the energy eigenvalues,
the unitarity of time evolution, and the probabilistic interpretation. It
has recently been discovered that within this complex framework a spin-up state
can be transformed {\em arbitrarily quickly} to a spin-down state by a simple
non-Hermitian Hamiltonian \cite{BR1}.

Let us first show how to find the value of $\tau$ for the case of a Hermitian
Hamiltonian: This problem is easy because finding the optimal evolution time
$\tau$ requires only the solution to the simpler problem of finding the optimal
evolution time for the $2\times2$ matrix Hamiltonians acting on the
two-dimensional subspace spanned by $|\psi_I\rangle$ and $|\psi_F\rangle$
\cite{rrr2}. To solve the Hermitian version of the two-dimensional quantum
brachistochrone problem, we choose the basis so that the initial and final
states are given by
\begin{equation}
|\psi_I\rangle=\left( \begin{array}{c} 1\\0\end{array} \right)\quad{\rm and}
\qquad |\psi_F\rangle=\left(\begin{array}{c} a\\b\end{array}\right),
\label{e148}
\end{equation}
where $|a|^2+|b|^2=1$. The most general $2\times2$ Hermitian Hamiltonian is
\begin{equation}
H=\left(\begin{array}{cc} s & r{\rm e}^{-i\theta} \cr r{\rm e}^{i\theta} & u
\end{array}\right),
\label{e149}
\end{equation}
where the parameters $r$, $s$, $u$, and $\theta$ are real. The eigenvalue
constraint is
\begin{equation}
\omega^2=(s-u)^2+4r^2.
\label{e150}
\end{equation}
The Hamiltonian $H$ in (\ref{e149}) can be expressed in terms of the Pauli
matrices
\begin{equation}
\sigma_1=\left( \begin{array}{cc} 0&1\\ 1&0\end{array}\right),\quad
\sigma_2=\left( \begin{array}{cc} 0&-i\\ i&0\end{array}\right),\quad
\sigma_3=\left( \begin{array}{cc} 1&0\\ 0&-1\end{array}\right),
\label{e151}
\end{equation}
as $H=\half(s+u){\bf 1}+\half\omega{\boldsymbol{\sigma}}\!\cdot\!{\bf n}$, where
${\bf n}=\frac{2}{\omega}\left(r\cos\theta,r\sin\theta,\frac{s-u}{2}\right)$ is
a unit vector. The matrix identity
\begin{equation}
\exp(i\phi\,{\boldsymbol{\sigma}}\!\cdot\!{\bf n})=\cos\phi\,{\bf
1} + i \sin\phi\,{\boldsymbol{\sigma}}\!\cdot\!{\bf n}
\label{e152}
\end{equation}
then simplifies the relation $|\psi_F\rangle=e^{-iH\tau/\hbar}|\psi_I\rangle$ to
\begin{equation}
\left(\begin{array}{c} a\\ b\end{array}\right)=e^{-i(s+u)t/(2\hbar)}\left(
\begin{array}{c}\cos\left(\frac{\omega t}{2\hbar}\right)-i\frac{s-u}{\omega}
\sin\left(\frac{\omega t}{2\hbar}\right) \\ {}\\ -i\frac{2r}{\omega}e^{i\theta}
\sin\left(\frac{\omega t}{2\hbar}\right)\end{array}\right).
\label{e153}
\end{equation}

From the second component of (\ref{e153}) we obtain $|b|=\frac{2r}{\omega}\sin
\left(\frac{\omega t}{2\hbar}\right)$, which gives the time required to
transform the initial state: $t=\frac{2\hbar}{\omega}\arcsin\big(\frac{\omega|b|
}{2r}\big)$. To optimize this relation over all $r>0$, we note that (\ref{e150})
implies that the maximum value of $r$ is $\half\omega$ and that at this maximum
$s=u$. The optimal time is thus
\begin{equation}
\tau=\frac{2\hbar}{\omega}\arcsin|b|.
\label{e154}
\end{equation}
Note that if $a=0$ and $b=1$ we have $\tau=2\pi\hbar/\omega$ for the smallest
time required to transform $\left({1\atop0}\right)$ to the orthogonal state
$\left({0\atop1}\right)$.

For general $a$ and $b$, at the optimal time $\tau$ we have $a=e^{i\tau s/\hbar}
\sqrt{1-|b|^2}$ and $b=ie^{i\tau s/\hbar}|b|e^{i\theta}$, which satisfies the
condition $|a|^2+|b|^2=1$ that the norm of the state does not change under
unitary time evolution. The parameters $s$ and $\theta$ are determined by the
phases of $a$ and $b$. We set $a=|a|e^{i{\rm arg}(a)}$ and $b=|b|e^{i{\rm arg}
(b)}$ and find that the optimal Hamiltonian is
\begin{equation}
H=\left(\begin{array}{cc}\frac{\omega{\rm arg}(a)}{2\arcsin|b|} & \frac{\omega}
{4}e^{-i[{\rm arg}(b)-{\rm arg}(a)-\frac{\pi}{2}]}\cr\frac{\omega}{4}e^{i[{\rm
arg}(b)-{\rm arg}(a)-\frac{\pi}{2}]}&\frac{\omega{\rm arg}(a)}{2\arcsin|b|}
\end{array}\right).
\label{e155}
\end{equation}
The overall phase of $|\psi_{\rm F}\rangle$ is not physically relevant, so the
quantity ${\rm arg}(a)$ may be chosen arbitrarily; we may thus assume that it is
$0$. We are free to choose ${\rm arg}(a)$ because there is no absolute energy in
quantum mechanics; one can add a constant to the eigenvalues of the Hamiltonian
without altering the physics. Equivalently, this means that the value of $\tau$
cannot depend on the trace $s+u$ of $H$.

How do we interpret the result for $\tau$ in (\ref{e154})? While this equation
resembles the time-energy uncertainty principle, it is really the much simpler
statement that {\em rate$\times$time$=$distance}. The constraint (\ref{e150}) on
$H$ is a bound on the standard deviation $\Delta H$ of the Hamiltonian, where
$\Delta H$ in a normalized state $|\psi\rangle$ is given by $(\Delta H)^2=
\langle\psi|H^2|\psi\rangle-\langle\psi|H|\psi\rangle^2$. The maximum value of
$\Delta H$ is $\omega/2$. From the Anandan-Aharonov relation~\cite{rrr4}, the
speed of evolution of a quantum state is given by $\Delta H$. The distance
between the initial state $|\psi_I\rangle$ and the final state $|\psi_F\rangle$
is $\delta=2\arccos(|\langle\psi_F|\psi_I\rangle|)$. Thus, the shortest time
$\tau$ for $|\psi_I\rangle$ to evolve to $|\psi_F\rangle=e^{-iH\tau/\hbar}|
\psi_I\rangle$ is bounded below because the speed is bounded above while the
distance is held fixed. The Hamiltonian $H$ that realizes the shortest time
evolution can be understood as follows: The standard deviation $\Delta H$ of
the Hamiltonian in (\ref{e149}) is $r$. Since $\Delta H$ is bounded by $\omega/
2$, to maximize the speed of evolution (and minimize the time of evolution) we
choose $r=\omega /2$.

We now perform the same optimization for the complex $2\times2$ non-Hermitian
$\cP\cT$ Hamiltonian in (\ref{e93}). Following the same procedure used for
Hermitian Hamiltonians, we rewrite $H$ in (\ref{e93}) in the form $H=(r\cos
\theta){\bf 1}+\half\omega{\boldsymbol{\sigma}}\!\cdot\!{\bf n}$, where ${\bf n}
=\frac{2}{\omega}(s,0,ir\sin\theta)$ is a unit vector and the squared difference
between the energy eigenvalues [see (\ref{e95}) is
\begin{equation}
\omega^2=4s^2-4r^2\sin^2\theta.
\label{e156}
\end{equation}
The condition of unbroken $\cP\cT$ symmetry ensures the positivity of
$\omega^2$. This equation emphasizes the key difference between Hermitian and
$\cP\cT$-symmetric Hamiltonians: The corresponding equation (\ref{e150}) for the
Hermitian Hamiltonian has a {\em sum} of squares, while this equation for
$\omega^2$ has a {\em difference} of squares. Thus, Hermitian Hamiltonians
exhibit elliptic behavior, which leads to a nonzero lower bound for the optimal
time $\tau$. The hyperbolic nature of (\ref{e156}) allows $\tau$ to approach
zero because, as we will see, the matrix elements of the $\cP\cT$-symmetric
Hamiltonian can be made large without violating the energy constraint
$E_+-E_-=\omega$.

The $\cP\cT$-symmetric analog of the evolution equation (\ref{e153}) is
\begin{equation}
e^{-iHt/\hbar}\left(\begin{array}{c}1\\ 0\end{array}\right)=\frac{e^{-itr\cos
\theta/\hbar}}{\cos\alpha}\left(\begin{array}{c}\cos\left(\frac{\omega t}{2\hbar
}-\alpha\right) \\ {}\\ -i\sin\left(\frac{\omega t}{2\hbar}\right)\end{array}
\right).
\label{e157}
\end{equation}
We apply this result to the same pair of vectors examined in the Hermitian case:
$|\psi_I\rangle=\left(1\atop0\right)$ and $|\psi_F\rangle=\left(0\atop1\right)$.
(Note that these vectors are not orthogonal with respect to the $\cC\cP\cT$
inner product.) Equation (\ref{e157}) shows that the evolution time to reach $|
\psi_F\rangle$ from $|\psi_I\rangle$ is $t=(2\alpha-\pi)\hbar/\omega$.
Optimizing this result over allowable values for $\alpha$, we find that as
$\alpha$ approaches $\half\pi$ the optimal time $\tau$ tends to zero.

Note that in the limit $\alpha\to\half\pi$ we get $\cos\alpha\to0$. However, in
terms of $\alpha$, the energy constraint (\ref{e156}) becomes $\omega^2=4s^2
\cos^2\alpha$. Since $\omega$ is fixed, in order to have $\alpha$ approach
$\half\pi$ we must require that $s\gg1$. It then follows from the relation $\sin
\alpha=(r/s)\sin\theta$ that $|r|\sim|s|$, so we must also require that $r\gg1$.
Evidently, in order to make $\tau\ll1$, the matrix elements of the $\cP
\cT$-symmetric Hamiltonian (\ref{e93}) must be large.

The result demonstrated here does not violate the uncertainty principle. Indeed,
Hermitian and non-Hermitian $\cP\cT$-symmetric Hamiltonians share the properties
that (i) the evolution time is given by $2\pi\hbar/\omega$, and (ii) $\Delta H
\leq\omega/2$. The key difference is that a pair of states such as $\left({1
\atop0}\right)$ and $\left({0\atop1}\right)$ are orthogonal in a Hermitian
theory, but have separation $\delta=\pi-2|\alpha|$ in the $\cP\cT$-symmetric
theory. This is because the Hilbert space metric of the $\cP\cT$-symmetric
quantum theory {\em depends on the Hamiltonian}. Hence, it is possible to
choose the parameter $\alpha$ to create a wormhole-like effect in the Hilbert
space.

A {\em gedanken} experiment to realize this effect in a laboratory might work as
follows: A Stern-Gerlach filter creates a beam of spin-up electrons. The beam
then passes through a `black box' containing a device governed by a $\cP
\cT$-symmetric Hamiltonian that flips the spins unitarily in a very short time.
The outgoing beam then enters a second Stern-Gerlach device which verifies that
the electrons are now in spin-down states. In effect, the black-box device is
applying a magnetic field in the complex direction $(s,0,ir\sin\theta)$. If the
field strength is sufficiently strong, then spins can be flipped unitarily in
virtually no time because the complex path joining these two states is arbitrary
short without violating the energy constraint. The arbitrarily short alternative
complex pathway from a spin-up state to a spin-down state, as illustrated by
this thought experiment, is reminiscent of the short alternative distance
between two widely separated space-time points as measured through a wormhole in
general relativity.

The results established here provide the possibility of performing experiments
that distinguish between Hermitian and $\cP\cT$-symmetric Hamiltonians. If
practical implementation of complex $\cP\cT$-symmetric Hamiltonians were
feasible, then identifying the optimal unitary transformation would be
particularly important in the design and implementation of fast quantum
communication and computation algorithms. Of course, the wormhole-like effect we
have discussed here can only be realized if it is possible to switch rapidly
between Hermitian and $\cP\cT$-symmetric Hamiltonians by means of similarity
transformations. It is conceivable that so much quantum noise would be generated
that there is a sort of quantum protection mechanism that places a lower bound
on the time required to switch Hilbert spaces. If so, this would limit the
applicability of a Hilbert-space wormhole to improve quantum algorithms.

\subsection{Supersymmetric $\cP\cT$-Symmetric Hamiltonians}
\label{ss7-4}

After the discovery of $\cP\cT$-symmetric Hamiltonians in quantum mechanics, it
was proposed that $\cP\cT$ symmetry might be combined with supersymmetry
\cite{BM7} in the context of quantum field theory. In Ref.~\cite{BM7} it was
shown that one can easily construct two-dimensional supersymmetric quantum field
theories by introducing a $\cP\cT$-symmetric superpotential of the form ${\cal
S}(\phi)=-ig(i\phi)^{1+\epsilon}$. The resulting quantum field theories exhibit
a broken parity symmetry for all $\delta>0$. However, supersymmetry remains
unbroken, which is verified by showing that the ground-state energy density
vanishes and that the fermion-boson mass ratio is unity. Many papers have
subsequently worked on the subject of $\cP\cT$-symmetric supersymmetric quantum
mechanics. (See Ref.~\cite{ZX}.) A particularly interesting paper by Dorey {\em
et al.} examines the connection between supersymmetry and broken $\cP\cT$
symmetry in quantum mechanics \cite{D6}.

\subsection{Other Quantum-Mechanical Applications}
\label{ss7-5}

There are many additional quantum-mechanical applications of non-Hermitian ${\cP
\cT}$-invariant Hamiltonians. In condensed matter physics Hamiltonians rendered
non-Hermitian by an imaginary external field have been introduced to study
delocalization transitions in condensed matter systems such as vortex flux-line
depinning in type-II superconductors \cite{HN}. In this Hatano-Nelson model
there is a critical value of the anisotropy (non-Hermiticity) parameter below
which all eigenvalues are real \cite{RS}. In the theory of Reaction-Diffusion
systems, many models have been constructed for systems described by matrices
that can be non-Hermitian \cite{MH1,MH2} and with the appropriate definition of
the $\cP$ and $\cT$ operators, these systems can be shown to be $\cP\cT$
symmetric. Finally, we mention a recent paper by Hibberd {\em et al.} who found
a transformation that maps a Hamiltonian describing coherent superpositions of
Cooper pairs and condensed molecular bosons to one that is $\cP\cT$-symmetric
\cite{HDL}.

\section{$\cP\cT$-Symmetric Quantum Field Theory}
\label{s8}

Quantum-mechanical theories have only a finite number of degrees of freedom.
Most of the $\cP\cT$-symmetric quantum-mechanical models discussed so far in
this paper have just one degree of freedom; that is, the Hamiltonians for these
theories are constructed from just one pair of dynamical variables, $\x$ and
$\p$. In a quantum field theory the operators $\x(t)$ and $\p(t)$ are replaced
by the quantum fields $\vf({\bf x},t)$ and $\pi({\bf x},t)$, which represent
a continuously infinite number of degrees of freedom, one for each value of the
spatial variable ${\bf x}$. Such theories are vastly more complicated than
quantum-mechanical theories, but constructing quantum field theories that are
non-Hermitian and $\cP\cT$ symmetric is straightforward. For example, the
quantum-field-theoretic Hamiltonians that are analogous to the
quantum-mechanical Hamiltonians in (\ref{e10}) and (\ref{e11}) are
\begin{equation}
\mathcal{H}=\half\pi^2({\bf x},t)+\half[\nabla_{\bf x}\vf({\bf x},t)]^2+
\half\mu^2\vf^2({\bf x},t)+ig\vf^3({\bf x},t)
\label{e158}
\end{equation}
and
\begin{equation}
\mathcal{H}=\half\pi^2({\bf x},t)+\half[\nabla_{\bf x}\vf({\bf x},t)]^2+
\half\mu^2\vf^2({\bf x},t)-{\textstyle\frac{1}{4}}g\vf^4({\bf x},t).
\label{e159}
\end{equation}
As in quantum mechanics, where the operators $\x$ and $\p$ change sign under
parity reflection $\cP$, we assume that the fields in these Hamiltonians are
{\em pseudoscalars} so that they also change sign:
\begin{equation}
\cP\vf({\bf x},t)\cP=-\vf(-{\bf x},t),\qquad\cP\pi({\bf x},t)\cP=-\pi(-{\bf
x},t).
\label{e160}
\end{equation}

Quantum field theories like these that possess $\cP\cT$ symmetry exhibit a rich
variety of behaviors. Cubic field-theory models like that in (\ref{e158}) are of
interest because they arise in the study of the Lee-Yang edge singularity
\cite{A3,A4,A5} and in Reggeon field theory \cite{A7,A8}. In these papers it was
asserted that an $i\vf^3$ field theory is nonunitary. However, by constructing
the $\cC$ operator, we argue in Subsec.~\ref{ss8-1} that this quantum field
theory is, in fact, unitary. We show in Subsec.~\ref{ss8-3} how $\cP\cT$
symmetry eliminates the ghosts in the Lee Model, another cubic quantum field
theory. The field theory described by (\ref{e159}) is striking because it is
asymptotically free, as explained in Subsec.~\ref{ss8-4}. We also examine $\cP
\cT$-symmetric quantum electrodynamics in Subsec.~\ref{ss8-5}, the $\cP
\cT$-symmetric Thirring and Sine-Gordon models in Subsec.~\ref{ss8-6}, and
gravitational and cosmological theories in Subsec.~\ref{ss8-7}. Last, we look
briefly at $\cP\cT$-symmetric classical field theories in Subsec.~\ref{ss8-8}.

\subsection{$i\vf^3$ Quantum Field Theory}
\label{ss8-1}

In courses on quantum field theory, a scalar $g\vf^3$ theory is used as a
pedagogical example of perturbative renormalization even though this model is
not physically realistic because the energy is not bounded below. However, by
calculating $\cC$ perturbatively for the case when the coupling constant $g=i
\epsilon$ is pure imaginary, one obtains a fully acceptable Lorentz invariant
quantum field theory. This calculation shows that it is possible to construct
perturbatively the Hilbert space in which the Hamiltonian for this cubic scalar
field theory in $(D+1)$-dimensional Minkowski space-time is {\em self-adjoint}.
Consequently, such theories have positive spectra and exhibit unitary time
evolution. 

In this subsection we explain how to calculate perturbatively the $\cC$ operator
for the quantum-field-theoretic Hamiltonian in (\ref{e158}) \cite{BBJ2}. We
apply the powerful algebraic techniques explained in Subsecs.~\ref{ss6-1} and
\ref{ss6-2} for the calculation of the $\cC$ operator in quantum mechanics. As
in quantum mechanics, we express $\cC$ in the form $\cC=\exp\left(\epsilon Q_1+
\epsilon^3Q_3+\ldots\right)\cP$, where now $Q_{2n+1}$ ($n=0,\,1,\,2,\,\ldots$)
are real functionals of the fields $\vf_{\bf x}$ and $\pi_{\bf x}$. To find
$Q_n$ for $\mathcal{H}$ in (\ref{e158}) we must solve a system of operator
equations.

We begin by making an {\em ansatz} for $Q_1$ analogous to the {\em ansatz} used
in (\ref{e120}):
\begin{equation}
Q_1=\int\!\!\int\!\!\int d{\bf x}\,d{\bf y}\,d{\bf z}\left(M_{({\bf xyz})}\pi_{
\bf x}\pi_{\bf y}\pi_{\bf z}+N_{{\bf x}({\bf yz})}\vf_{\bf y}\pi_{\bf x}
\vf_{\bf z}\right).
\label{e161}
\end{equation}
In quantum mechanics $M$ and $N$ are constants, but in field theory they are
functions. The notation $M_{({\bf x}{\bf y}{\bf z})}$ indicates that this
function is totally symmetric in its three arguments, and the notation $N_{{\bf
x}({\bf y}{\bf z})}$ indicates that this function is symmetric under the
interchange of the second and third arguments. The unknown functions $M$ and $N$
are form factors; they describe the spatial distribution of three-point
interactions of the field variables in $Q_1$. The nonlocal spatial interaction
of the fields is an intrinsic property of $\cC$. (Note that we have suppressed
the time variable $t$ in the fields and that we use subscripts to indicate the
spatial dependence.)

To determine $M$ and $N$ we substitute $Q_1$ into the first equation in
(\ref{e119}), namely $\left[H_0,Q_1\right]=-2H_1$, which now takes the form
\begin{equation}
\left[\int d{\bf x}\,\pi^2_{\bf x}+\int\!\!\int d{\bf x}\,d{\bf y}\,\vf_{\bf
x}G_{\bf xy}^{-1}\vf_{\bf y},Q_1\right]=-4i\int d{\bf x}\,\vf^3_{\bf x},
\label{e162}
\end{equation}
where the inverse Green's function is given by $G_{{\bf x}{\bf y}}^{-1}\equiv(
\mu^2-\nabla_{\bf x}^2)\delta({\bf x}-{\bf y})$. We obtain the following
coupled system of partial differential equations:
\begin{eqnarray}
&&(\mu^2-\nabla_{\bf x}^2)N_{{\bf x}({\bf y}{\bf z})}
+(\mu^2-\nabla_{\bf y}^2)N_{{\bf y}({\bf x}{\bf z})}
+(\mu^2-\nabla_{\bf z}^2)N_{{\bf z}({\bf x}{\bf y})}\nonumber\\
&&\qquad\qquad\qquad\,=-6\delta({\bf x}-{\bf y})\delta({\bf x}-{\bf z}),
\nonumber\\
&&N_{{\bf x}({\bf y}{\bf z})}+N_{{\bf y}({\bf x}{\bf z})}=
3(\mu^2-\nabla_{\bf z}^2) M_{({\bf w}{\bf x}{\bf y})}.
\label{e163}
\end{eqnarray}

We solve these equations by Fourier transforming to momentum space and get
\begin{equation}
M_{({\bf xyz})}=-\frac{4}{(2\pi)^{2D}}\int\!\!\int d{\bf p}\,
d{\bf q}\frac{e^{i({\bf x}-{\bf y})\cdot{\bf p}+i({\bf x}-{\bf z})\cdot{\bf q}}}
{D({\bf p},{\bf q})},
\label{e164}
\end{equation}
where $D({\bf p},{\bf q})=4[{\bf p}^2{\bf q}^2-({\bf p}\cdot{\bf q})^2]+4\mu^2(
{\bf p}^2+{\bf p}\cdot{\bf q}+{\bf q}^2)+3\mu^4$ is positive, and
\begin{eqnarray}
N_{{\bf x}({\bf yz})}&=&-3\left(\nabla_{\bf y}\cdot\nabla_{\bf
z}+\half\mu^2\right)M_{({\bf xyz})}.
\label{e165}
\end{eqnarray}
For the special case of a $(1+1)$-dimensional quantum field theory the integral
in (\ref{e165}) evaluates to $M_{({\bf x}{\bf y}{\bf z})}=-{\rm K}_0(\mu R)/
\left(\sqrt{3}\pi\mu^2\right)$, where ${\rm K}_0$ is the associated Bessel
function and $R^2=\half[({\bf x}-{\bf y})^2+({\bf y}-{\bf z})^2+({\bf z}-{\bf x}
)^2]$. This completes the calculation of $\cC$ to first order in perturbation
theory.

We mention finally that the $\cC$ operator for this cubic quantum field theory
transforms as a scalar under the action of the homogeneous Lorentz group
\cite{BBCW}. In Ref.~\cite{BBCW} it was argued that because the Hamiltonian
$H_0$ for the unperturbed theory ($g=0$) commutes with the parity operator
$\cP$, the intrinsic parity operator $\cP_{\rm I}$ in the noninteracting theory
transforms as a Lorentz scalar. (The {\em intrinsic} parity operator $\cP_{\rm
I}$ and the parity operator $\cP$ have the same effect on the fields, except
that $\cP_{\rm I}$ does not reverse the sign of the spatial argument of the
field. In quantum mechanics $\cP$ and $\cP_{\rm I}$ are indistinguishable.) When
the coupling constant $g$ is nonzero, the parity symmetry of $H$ is broken and
$\cP_{\rm I}$ is no longer a scalar. However, $\cC$ {\em is} a scalar. Since
$\lim_{g\to0}\cC=\cP_{\rm I}$, one can interpret the $\cC$ operator in quantum
field theory as the complex extension of the intrinsic parity operator when the
imaginary coupling constant is turned on. This means that $\cC$ is
frame-invariant and it shows that the $\cC$ operator plays a truly fundamental
role in non-Hermitian quantum field theory.

\subsection{Other Quantum Field Theories Having Cubic Interactions}
\label{ss8-2}

We can repeat the calculations done in Subsec.~\ref{ss8-1} for cubic quantum
field theories having several interacting scalar fields \cite{BBJ2,BBJ3}. For
example, consider the case of {\em two} scalar fields $\vf_{\bf x}^{(1)}$
and $\vf_{\bf x}^{(2)}$ whose interaction is governed by
\begin{equation}
H=H_0^{(1)}+H_0^{(2)}+i\epsilon\int d{\bf x}\,\big[\vf_{\bf x}^{(1)}\big]^2
\vf_{\bf x}^{(2)},
\label{e166}
\end{equation}
which is the analog of the quantum-mechanical theory described by $H$ in
(\ref{e123}). Here,
\begin{equation}
H_0^{(j)}=\half\int d{\bf x}\,\big[\pi_{\bf x}^{(j)}\big]^2+\half\int\!\!\int
d{\bf x}\,d{\bf y}\,\big[G_{\bf xy}^{(j)}\big]^{-1}\vf_{\bf x}^{(j)}
\vf_{\bf y}^{(j)}.
\label{e167}
\end{equation}
To determine $\cC$ to order $\epsilon$ we introduce the {\em ansatz}
\begin{eqnarray}
Q_1&=&\int\!\!\int\!\!\int d{\bf x}\,d{\bf y}\,d{\bf z}\Big[N_{{\bf x}({\bf yz})
}^{(1)}\left(\pi_{\bf z}^{(1)}\vf_{\bf y}^{(1)}+\vf_{\bf y}^{(1)}\pi_{
\bf z}^{(1)}\right)\vf_{\bf x}^{(2)}\nonumber\\
&&+N_{{\bf x}({\bf yz})}^{(2)}\pi_{\bf x}^{(2)}\vf_{\bf y}^{(1)}\vf_{\bf
z}^{(1)}+M_{{\bf x}({\bf yz})}\pi_{\bf x}^{(2)}\pi_{\bf y}^{(1)}\pi_{\bf z}^{(1)
}\Big],
\label{e168}
\end{eqnarray}
where $M_{{\bf x}({\bf y}{\bf z})}$, $N_{{\bf x}({\bf y}{\bf z})}^{(1)}$, and
$N_{{\bf x}({\bf y}{\bf z})}^{(2)}$ are unknown functions and the parentheses
indicate symmetrization. We get
\begin{eqnarray}
M_{{\bf x}({\bf yz})}&=&- G_m(R_1)\,G_{\mu_2}(R_2),\nonumber\\
N_{{\bf x}({\bf yz})}^{(1)}&=&-\delta(2{\bf x}-{\bf y}-{\bf z})G_m(R_1),
\nonumber\\
N_{{\bf x}({\bf yz})}^{(2)}&=& \half\delta(2{\bf x}-{\bf y}-{\bf z})G_m(R_1)
-\delta({\bf y}-{\bf z})G_{\mu_2}(R_2),
\label{e169}
\end{eqnarray}
where $G_\mu(r)=\frac{r}{\mu}\big(\frac{\mu}{2\pi r}\big)^{D/2}{\rm K}_{-1+D/2}
(\mu r)$ with $r=|{\bf r}|$ is the Green's function in $D$-dimensional space,
$m^2=\mu_1^2-\frac{1}{4}\mu_2^2$, $R_1^2=({\bf y}-{\bf z})^2$, and $R_2^2=
\frac{1}{4}(2{\bf x}-{\bf y}-{\bf z})^2$.

For {\em three} interacting scalar fields whose dynamics is described by
\begin{equation}
H=H_0^{(1)}+H_0^{(2)}+H_0^{(3)}+i\epsilon\int d{\bf x}\,\vf_{\bf x}^{(1)}
\vf_{\bf x}^{(2)}\vf_{\bf x}^{(3)},
\label{e170}
\end{equation}
which is the analog of $H$ in (\ref{e125}), we make the {\em ansatz}
\begin{eqnarray}
Q_1&=&\int\!\!\int\!\!\int d{\bf x}\,d{\bf y}\,d{\bf z}\Big[N_{{\bf x}{\bf y}
{\bf z}}^{(1)}\pi_{\bf x}^{(1)}\vf_{\bf y}^{(2)}\vf_{\bf z}^{(3)}+N_{
{\bf x}{\bf y}{\bf z}}^{(2)}\pi_{\bf x}^{(2)}\vf_{\bf y}^{(3)}\vf_{\bf
z}^{(1)}\nonumber\\
&&\quad +N_{{\bf x}{\bf y}{\bf z}}^{(3)}\pi_{\bf x}^{(3)}\vf_{\bf y}^{(1)}
\vf_{\bf z}^{(2)}+M_{{\bf x}{\bf y}{\bf z}}\pi_{\bf x}^{(1)}\pi_{\bf y}^{(2)
}\pi_{\bf z}^{(3)}\Big].
\label{e171}
\end{eqnarray}

The solutions for the unknown functions are as follows: $M_{\bf xyz}$ is given
by the integral (\ref{e164}) with the more general formula $D({\bf p},{\bf q})=
4[{\bf p}^2{\bf q}^2-({\bf p}\cdot{\bf q})^2]+4[\mu_1^2({\bf q}^2+{\bf p}\cdot{
\bf q})+\mu_2^2({\bf p}^2+{\bf p}\cdot{\bf q})-\mu_3^2{\bf p}\cdot{\bf q}]+\mu^4
$ with $\mu^4=2\mu_1^2\mu_2^2+2\mu_1^2\mu_3^2+2\mu_2^2\mu_3^2-\mu_1^4-\mu_2^4-
\mu_3^4$. The $N$ coefficients are expressed as derivatives acting on $M$:
\begin{eqnarray}
N_{\bf xyz}^{(1)}&=&\left[4\nabla_{\bf y}\cdot\nabla_{\bf z}+2(\mu_2^2
+\mu_3^2-\mu_1^2)\right] M_{\bf xyz},\nonumber\\
N_{\bf xyz}^{(2)}&=&\left[-4\nabla_{\bf y}\cdot\nabla_{\bf z}
-4\nabla_{\bf z}^2+2(\mu_1^2+\mu_3^2-\mu_2^2)\right] M_{\bf xyz},\nonumber\\
N_{\bf xyz}^{(3)}&=&\left[-4\nabla_{\bf y}\cdot\nabla_{\bf z}
-4\nabla_{\bf y}^2+2(\mu_1^2+\mu_2^2-\mu_3^2)\right] M_{\bf xyz}.
\label{e172}
\end{eqnarray}
Once again, the calculation of $\cC$ shows that these cubic field theories are
fully consistent quantum theories.\footnote{In Refs.~\cite{BBCW,LEE} it is shown
that the $\cC$ operator in quantum field theory has the form $\cC=e^Q\cP_{\rm
I}$, where $\cP_{\rm I}$ is the {\em intrinsic} parity reflection operator. The
difference between $\cP$ and $\cP_{\rm I}$ is that $\cP_{\rm I}$ does not
reflect the spatial arguments of the fields. For a cubic interaction Hamiltonian
this distinction is technical. It does not affect the final result for the $Q$
operator in (\ref{e161}), (\ref{e168}), and (\ref{e171}).}

\subsection{The Lee model}
\label{ss8-3}

In this subsection we will show how to use the tools that we have developed to
study non-Hermitian $\cP\cT$-symmetric quantum theories to examine a famous
model quantum field theory known as the {\em Lee model}. In 1954 the Lee model
was proposed as a quantum field theory in which mass, wave function, and charge
renormalization could be performed exactly and in closed form
\cite{BARTON,L1,L2,L3}. We discuss the Lee model here because when the
renormalized coupling constant is taken to be larger than a critical value, the
Hamiltonian becomes non-Hermitian and a (negative-norm) ghost state appears.
The appearance of the ghost state was assumed to be a fundamental defect of the
Lee model. However, we show that the non-Hermitian Lee-model Hamiltonian is
actually $\cP\cT$ symmetric. When the states of this model are examined using
the $\cC$ operator, the ghost state is found to be an ordinary physical state
having positive norm \cite{LEE}.

The idea for studying the Lee model as a non-Hermitian Hamiltonian is due to
Kleefeld, who was the first to point out this transition to $\cP\cT$ symmetry
\cite{KLE}. His work gives a comprehensive history of non-Hermitian
Hamiltonians.

The Lee model has a cubic interaction term that describes the interaction of
three spinless particles called $V$, $N$, and $\theta$. The $V$ and $N$
particles are fermions and behave roughly like nucleons, and the $\theta$
particle is a boson and behaves roughly like a pion. In the model a $V$ may emit
a $\theta$, but when it does so it becomes an $N$: $V\rightarrow N\,+\,\theta$.
Also, an $N$ may absorb a $\theta$, but when it does so it becomes a $V$: $N\,+
\,\theta\rightarrow V$.

The solvability of the Lee model is based on the fact that there is no crossing
symmetry. That is, the $N$ is forbidden to emit an anti-$\theta$ and become a
$V$. Eliminating crossing symmetry makes the Lee model solvable because it
introduces two conservation laws. First, {\em the number of $N$ quanta plus the
number of $V$ quanta is fixed}. Second, {\em the number of $N$ quanta minus the
number of $\theta$ quanta is fixed.} These two highly constraining conservation
laws decompose the Hilbert space of states into an infinite number of
noninteracting sectors. The simplest sector is the vacuum sector. Because of the
conservation laws, there are no vacuum graphs and the bare vacuum is the
physical vacuum. The next two sectors are the one-$\theta$-particle and the
one-$N$-particle sector. These two sectors are also trivial because the two
conservation laws prevent any dynamical processes from occurring there. As a
result, the masses of the $N$ particle and of the $\theta$ particle are not
renormalized; that is, the physical masses of these particles are the same as
their bare masses.

The lowest nontrivial sector is the $V/N\theta$ sector. The physical states in
this sector of the Hilbert space are linear combinations of the bare $V$ and the
bare $N\theta$ states, and these states consist of the one-physical-$V$-particle
state and the physical $N$-$\theta$-scattering states. To find these states one
can look for the poles and cuts of the Green's functions. The renormalization in
this sector is easy to perform. Following the conventional renormalization
procedure, one finds that the mass of the $V$ particle is renormalized; that is,
the mass of the physical $V$ particle is different from its bare mass. In the
Lee model one calculates the unrenormalized coupling constant as a function of
the renormalized coupling constant in closed form. There are many ways to define
the renormalized coupling constant. For example, in an actual scattering
experiment one could define the square of the renormalized coupling constant
$g^2$ as the value of the $N\theta$ scattering amplitude at threshold.

\begin{figure}[b!]\vspace{2.6in}
\includegraphics{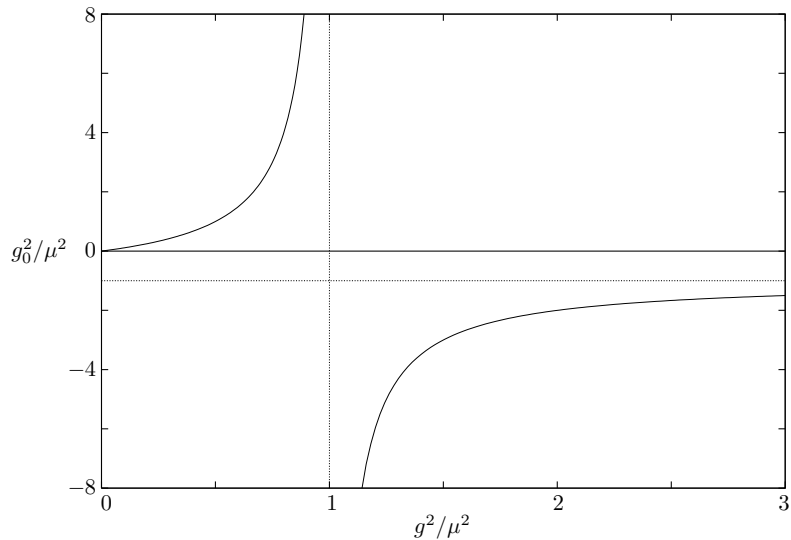}
\caption{Square of the unrenormalized coupling constant, $g_0^2$, plotted as a
function of the square of the renormalized coupling constant $g^2$. Note that
$g^2=0$ when $g_0^2=0$, and as $g^2$ increases from $0$ so does $g_0^2$.
However, as $g^2$ increases past a critical value, $g_0^2$ abruptly becomes
negative. In this regime $g_0$ is imaginary and the Hamiltonian is
non-Hermitian.}
\label{f26}
\end{figure}

The intriguing aspect of the Lee model is the appearance of a ghost state in the
$V/N\theta$ sector. This state appears when one performs coupling-constant
renormalization. Expressing $g_0^2$, the square of the unrenormalized coupling
constant, in terms of $g^2$, the square of the renormalized coupling constant,
one obtains the graph in Fig.~\ref{f1}. In principle, the $g$ is a physical
parameter whose numerical value is determined by a laboratory experiment. If
$g^2$ is measured to be near 0, then from Fig.~\ref{f26} we see that $g_0^2$ is
also small. However, if the experimental value of $g^2$ is larger than this
critical value, then the square of the unrenormalized coupling constant is
negative. In this regime $g_0$ is imaginary and the Hamiltonian is
non-Hermitian. Moreover, a new state appears in the $V/N\theta$ sector, and
because its norm is negative, the state is called a {\em ghost}. Ghost states
are unacceptable in quantum theory because their presence signals a violation of
unitarity and makes a probabilistic interpretation impossible.

There have been many unsuccessful attempts to make sense of the Lee model as a
physical quantum theory in the ghost regime \cite{BARTON,L2,L3}. However, the
methods of $\cP\cT$-symmetric quantum theory enable us to give a physical
interpretation for the $V/N\theta$ sector of the Lee model when $g_0$ becomes
imaginary and $H$ becomes non-Hermitian. The Lee model is a cubic interaction
and we have already shown in Subsecs.~\ref{ss8-1} and \ref{ss8-2} how to make
sense of a Hamiltonian in which there is a cubic interaction multiplied by an
imaginary coupling constant. The procedure is to calculate the $\cC$ operator
and to use it to define a new inner product when the Hamiltonian is
non-Hermitian.

For simplicity, we focus here on the {\em quantum-mechanical} Lee model; the
results for the field-theoretic Lee model in Ref.~\cite{LEE} are qualitatively
identical. The Hamiltonian for the quantum-mechanical Lee model is
\begin{equation}
H=H_0+g_0 H_1=m_{V_0}V^\dag V+m_N N^\dag N+m_\theta a^\dag a
+\left(V^\dag Na+a^\dag N^\dag V\right).
\label{e173}
\end{equation}
The bare states are the eigenstates of $H_0$ and the physical states are the
eigenstates of the full Hamiltonian $H$. The mass parameters $m_N$ and $m_
\theta$ represent the {\em physical} masses of the one-$N$-particle and
one-$\theta$-particle states because these states do not undergo mass
renormalization. However, $m_{V_0}$ is the {\em bare} mass of the $V$ particle.

We treat the $V$, $N$, and $\theta$ particles as pseudoscalars. To understand
why, recall that in quantum mechanics the position operator $x=(a+a^\dag)/\sqrt{
2}$ and the momentum operator $p=i(a^\dag -a)/\sqrt{2}$ both change sign under
parity reflection $\cP$:
\begin{equation}
\cP x\cP=-x,\quad\cP p\cP=-p.
\label{e174}
\end{equation}
Thus, $\cP V\cP=-V$, $\cP N\cP=-N$, $\cP a\cP=-a$, $\cP V^\dag\cP=-V^\dag$,
$\cP N^\dag\cP=-N^\dag$, $\cP a^\dag\cP=-a^\dag$. Under time reversal $\cT$, $p$
and $i$ change sign but $x$ does not:
\begin{equation}
\cT p\cT=-p,\quad\cT i\cT=-i,\quad \cT x\cT=x.
\label{e175}
\end{equation}
Thus, $\cT V\cT=V$, $\cT N\cT=N$, $\cT a\cT=a$, $\cT V^\dag\cT=V^\dag$, $\cT N^
\dag\cT=N^\dag $, $\cT a^\dag\cT=a^\dag$.

When the bare coupling constant $g_0$ is real, $H$ in (\ref{e173}) is Hermitian:
$H^\dag=H$. When $g_0$ is imaginary, $g_0=i\lambda_0\quad(\lambda_0~{\rm real}
)$, $H$ is not Hermitian, but by virtue of the above transformation properties,
$H$ is $\cP\cT$-symmetric: $H^{\cP\cT}=H$.

We assume first that $g_0$ is real so that $H$ is Hermitian and we examine the
simplest nontrivial sector of the quantum-mechanical Lee model; namely, the
$V/N\theta$ sector. We look for the eigenstates of the Hamiltonian $H$ in the
form of linear combinations of the bare one-$V$-particle and the bare
one-$N$-one-$\theta$-particle states. There are two eigenfunctions. We interpret
the eigenfunction corresponding to the lower-energy eigenvalue as the physical
one-$V$-particle state, and we interpret the eigenfunction corresponding with
the higher-energy eigenvalue as the physical one-$N$-one-$\theta$-particle
state. (In the field-theoretic Lee model this higher-energy state corresponds to
the continuum of physical $N$-$\theta$ scattering states.) Thus, we make the
{\em ansatz}
\begin{equation}
|V\rangle=c_{11}|1,0,0\rangle+c_{12}|0,1,1\rangle,\qquad
|N\theta\rangle=c_{21}|1,0,0\rangle+c_{22}|0,1,1\rangle,
\label{e176}
\end{equation}
and demand that these states be eigenstates of $H$ with eigenvalues $m_V$ (the
renormalized $V$-particle mass) and $E_{N\theta}$. The eigenvalue problem
reduces to a pair of elementary algebraic equations:
\begin{equation}
c_{11}m_{V_0}+c_{12}g_0=c_{11}m_V,\qquad
c_{21}g_0+c_{22}\left(m_N+m_\theta\right)=c_{22}E_{N\theta}.
\label{e177}
\end{equation}
The solutions to (\ref{e177}) are
\begin{eqnarray}
m_V&=&\frac{1}{2}\left(m_N+m_\theta+m_{V_0}-\sqrt{\mu_0^2+4g_0^2}\right),
\nonumber\\
E_{N\theta}&=&\frac{1}{2}\left(m_N+m_\theta+m_{V_0}+\sqrt{\mu_0^2+4g_0^2}
\right),
\label{e178}
\end{eqnarray}
where $\mu_0\equiv m_N+m_\theta-m_{V_0}$. Notice that $m_V$, the mass of the
physical $V$ particle, is different from $m_{V_0}$, the mass of the bare $V$
particle, because the $V$ particle undergoes mass renormalization.

Next, we perform wave-function renormalization. Following Barton \cite{BARTON}
we define the wave-function renormalization constant $Z_V$ by $\sqrt{Z_V}=
\langle 0|V|V\rangle$. This gives
\begin{equation}
Z_V^{-1}=\frac{1}{2}g_0^{-2}\sqrt{\mu_0^2+4g_0^2}\left(\sqrt{\mu_0^2+4g_0^2}
-\mu_0\right).
\label{e179}
\end{equation}

Finally, we perform coupling-constant renormalization. Again, following Barton
we note that $\sqrt{Z_V}$ is the ratio between the renormalized coupling
constant $g$ and the bare coupling constant $g_0$ \cite{BARTON}. Thus, $g^2/
g_0^2=Z_V$. Elementary algebra gives the bare coupling constant in terms of the
renormalized mass and coupling constant:
\begin{equation}
g_0^2=g^2/\left(1-g^2/\mu^2\right),
\label{e180}
\end{equation}
where $\mu$ is defined as $\mu\equiv m_N+m_\theta-m_V$. We cannot freely choose
$g$ because the value of $g$ is, in principle, taken from experimental data.
Once $g$ has been determined experimentally, we can use (\ref{e21}) to determine
$g_0$. The relation in (\ref{e21}) is plotted in Fig.~\ref{f26}. This figure
reveals a surprising property of the Lee model: If $g$ is larger than the
critical value $\mu$, then the square of $g_0$ is negative and $g_0$ is
imaginary.

As $g$ approaches its critical value from below, the two energy eigenvalues in
(\ref{e178}) vary accordingly. The energy eigenvalues are the two zeros of the
secular determinant $f(E)$ obtained from applying Cramer's rule to (\ref{e177}).
As $g$ (and $g_0$) increase, the energy of the physical $N\theta$ state
increases. The energy of the $N\theta$ state becomes infinite as $g$ reaches its
critical value. As $g$ increases past its critical value, the upper energy
eigenvalue goes around the bend; it abruptly jumps from being large and positive
to being large and negative. Then, as $g$ continues to increase, this energy
eigenvalue approaches the energy of the physical $V$ particle from below.

When $g$ increases past its critical value, the Hamiltonian $H$ in (\ref{e173})
becomes non-Hermitian, but its eigenvalues in the $V/N\theta$ sector remain
real. (The eigenvalues remain real because $H$ becomes $\cP\cT$ symmetric. All
cubic $\cP\cT$-symmetric Hamiltonians that we have studied have been shown to
have real spectra.) However, in the $\cP\cT$-symmetric regime it is no longer
appropriate to interpret the lower eigenvalue as the energy of the physical $N
\theta$ state. Rather, it is the energy of a new kind of state $|G\rangle$
called a {\em ghost}. As is shown in Refs.~\cite{L2,L3,BARTON}, the Hermitian
norm of this state is {\em negative}.

A physical interpretation of the ghost state emerges easily when we use the
procedure developed in Ref.~\cite{LEE}. We begin by verifying that in the $\cP
\cT$-symmetric regime, where $g_0$ is imaginary, the states of the Hamiltonian
are eigenstates of the $\cP\cT$ operator, and we then choose the multiplicative
phases of these states so that their $\cP\cT$ eigenvalues are unity:
\begin{equation}
\cP\cT|G\rangle=|G\rangle,\qquad\cP\cT|V\rangle=|V\rangle.
\label{e181}
\end{equation}
It is then straightforward to verify that the $\cP\cT$ norm of the $V$ state is
positive, while the $\cP\cT$ norm of the ghost state is negative.

We now follow the procedures described in Sec.~\ref{s6} to calculate $\cC$. We
express the $\cC$ operator as an exponential of a function $Q$ multiplying the
parity operator: $\cC=\exp\left[Q(V^\dag,V;N^\dag,N;a^\dag,a)\right]\cP$. We
then impose the operator equations $\cC^2={\bf 1}$, $[\cC,\cP\cT]=0$, and $[\cC,
H]=0$. The condition $\cC^2={\bf 1}$ gives
\begin{equation}
Q(V^\dag,V;N^\dag,N;a^\dag,a)=-Q(-V^\dag,-V;-N^\dag,-N;-a^\dag,-a).
\label{e182}
\end{equation}
Thus, $Q(V^\dag,V;N^\dag,N;a^\dag,a)$ is an odd function in total powers of
$V^\dag$, $V$, $N^\dag$, $N$, $a^\dag$, and $a$. Next, we impose the condition
$[\cC,\cP\cT]=0$ and obtain
\begin{equation}
Q(V^\dag,V;N^\dag,N;a^\dag,a)=Q^*(-V^\dag,-V;-N^\dag,-N;-a^\dag,-a),
\label{e183}
\end{equation}
where $*$ denotes complex conjugation.

Last, we impose the condition that $\cC$ commutes with $H$, which requires that
\begin{equation}
\left[e^Q, H_0\right]= g_0\left[e^Q,H_1\right]_+.
\label{e184}
\end{equation}
Although in Subsecs.~\ref{ss8-1} and \ref{ss8-2} we were only able to find the
$\cC$ operator to leading order in perturbation theory, for the Lee model one
can calculate $\cC$ exactly and in closed form. To do so, we seek a solution for
$Q$ as a formal Taylor series in powers of $g_0$:
\begin{equation}
Q=\sum_{n=0}^{\infty}g_0^{2 n+1}Q_{2n+1}.
\label{e185}
\end{equation}
Only odd powers of $g_0$ appear in this series, and $Q_{2n+1}$ are all
anti-Hermitian: $Q_{2n+1}^\dag=-Q_{2n+1}$. From (\ref{e185}) we get
\begin{eqnarray}
Q_{2n+1}=(-1)^n\frac{2^{2n+1}}{(2n+1)\mu_0^{2n+1}}\left(V^\dag Nan_\theta^n
-n_\theta^na^\dag N^\dag V\right),
\label{e186}
\end{eqnarray}
where $n_\theta=a^\dag a$ is the number operator for $\theta$-particle quanta.

We then sum over all $Q_{2n+1}$ and obtain the {\em exact} result that
\begin{equation}
Q=V^\dag Na\frac{1}{\sqrt{n_\theta}}{\rm arctan}\left(\frac{2g_0\sqrt{n_\theta}}
{\mu_0}\right)-\frac{1}{\sqrt{n_\theta}}{\rm arctan}\left(\frac{2g_0
\sqrt{n_\theta}}{\mu_0}\right) a^\dag N^\dag V.
\label{e187}
\end{equation}
We exponentiate this result to obtain the $\cC$ operator. The exponential of $Q$
simplifies considerably because we are treating the $V$ and $N$ particles as
fermions and therefore we can use the identity $n_{V,N}^2=n_{V,N}$. The {\em
exact} result for $e^Q$ is
\begin{eqnarray}
e^Q&=&\left[1-\frac{2g_0\sqrt{n_\theta}}{\sqrt{\mu_0^2+4g_0^2n_\theta}}a^\dag
N^\dag V+\frac{\mu_0n_N\left(1-n_V\right)}{\sqrt{\mu_0^2+4g_0^2n_\theta}}+\frac{
\mu_0n_V\left(1-n_N\right)}{\sqrt{\mu_0^2+4g_0^2\left(n_\theta+1\right)}}\right.
\nonumber\\
&&\left.+V^\dag Na\frac{2g_0\sqrt{n_\theta}}{\sqrt{\mu_0^2+4g_0^
2n_\theta}}- n_V-n_N+n_Vn_N \right].
\label{e188}
\end{eqnarray}

We can also express the parity operator $\cP$ in terms of number operators:
\begin{equation}
\cP=e^{i\pi\left(n_V+n_N+n_\theta\right)}=\left(1-2n_V\right)\left(1-2n_N
\right)e^{i\pi n_\theta}.
\label{e189}
\end{equation}
Combining $e^Q$ and $\cP$, we obtain the exact expression for $\cC$:
\begin{eqnarray}
\cC&=&\left[1-\frac{2g_0\sqrt{n_\theta}}{\sqrt{\mu_0^2+4g_0^2n_\theta}}a^\dag
N^\dag V +\frac{\mu_0n_N\left(1-n_V\right)}{\sqrt{\mu_0^
2+4g_0^2n_\theta}}+\frac{\mu_0n_V\left(1-n_N\right)}{\sqrt{\mu_0^2+4g_0^2\left(
n_\theta+1\right)}}\right.\nonumber\\
&&\!\!\!\!\!\!\!\!\!\!\!\!\!
\left.+V^\dag Na\frac{2g_0\sqrt{n_\theta}}{\sqrt{\mu_0^2+4g_0^2n_\theta}
}- n_V-n_N+n_Vn_N\right]\!\!\left(1-2n_V\right)\!\left(1-2n_N\right)
e^{i\pi n_\theta}.
\label{e190}
\end{eqnarray}

Using this $\cC$ operator to calculate the $\cC\cP\cT$ norm of the $V$ state and
of the ghost state, we find that these norms are both positive. Furthermore, the
the time evolution is unitary. This establishes that with the proper definition
of the inner product the quantum-mechanical Lee model is a physically acceptable
and consistent quantum theory, even in the ghost regime where the unrenormalized
coupling constant is imaginary. The procedure of redefining the inner product to
show that the ghost state is a physical state is a powerful technique that has
been used by Curtright {\em et al.} and by Ivanov and Smilga for more advanced
problems \cite{CM,SMIL}.

\subsection{The Higgs Sector of the Standard Model of Particle Physics}
\label{ss8-4}

The distinguishing features of the $-g\vf^4$ quantum field theory in 
(\ref{e159}) are that its spectrum is real and bounded below, it is
perturbatively renormalizable, it has a dimensionless coupling constant in
four-dimensional space-time, and it is {\em asymptotically free} \cite{AF}.
The property of asymptotic freedom was established many years ago by Symanzik
\cite{Sy}, as has been emphasized in a recent paper by Kleefeld \cite{KKK}. As
Kleefeld explains, a $+g\vf^4$ theory in four-dimensional space-time is trivial
because it is not asymptotically free. However, Symanzik proposed a
``precarious'' theory with a negative quartic coupling constant as a candidate
for an asymptotically free theory of strong interactions. Symanzik used the term
``precarious'' because the negative sign of the coupling constant suggests that
this theory is energetically unstable. However, as we have argued in this paper,
imposing $\cP\cT$-symmetric boundary conditions (in this case on the
functional-integral representation of the quantum field theory) gives a spectrum
that is bounded below. Thus, Symansik's proposal of a nontrivial theory is
resurrected.

The $-g\vf^4$ quantum field theory has another remarkable property. Although the
theory seems to be parity invariant, the $\cP\cT$-symmetric boundary conditions
violate parity invariance, as explained in Subsec.~\ref{ss2-7}. Hence, the
one-point Green's function (the expectation value of the field $\vf$) does not
vanish. (Techniques for calculating this expectation value are explained in
\cite{BMY}.) Thus, a nonzero vacuum expectation value can be achieved without
having to have spontaneous symmetry breaking because parity symmetry is {\em
permanently} broken. These properties suggest that a $-g\vf^4$ quantum field
theory might be useful in describing the Higgs sector of the standard model.

Perhaps the Higgs particle state is a consequence of the field-theoretic parity
anomaly in the same way that the quantum-mechanical parity anomaly described in
Subsecs.~\ref{ss2-7} and \ref{ss2-8} gives rise to bound states. Recent research
in this area has been done by Jones {\em et al.}, who studied transformations of
functional integrals \cite{JMR}, and by Meisinger and Ogilvie, who worked on
large-$N$ approximations and matrix models \cite{MO}.

\subsection{$\cP\cT$-Symmetric Quantum Electrodynamics}
\label{ss8-5}

If the unrenormalized electric charge $e$ in the Hamiltonian for quantum
electrodynamics were imaginary, then the Hamiltonian would be non-Hermitian.
However, if one specifies that the potential $A^\mu$ in this theory transforms
as an {\em axial} vector instead of a vector, then the Hamiltonian becomes
$\cP\cT$ symmetric \cite{QED2}. Specifically, we assume that the four-vector
potential and the electromagnetic fields transform under $\cP$ like
\begin{equation}
\cP:\quad {\bf E\to E},\quad {\bf B\to-B}, \quad {\bf A\to A}, \quad A^0\to-A^0.
\label{e191}
\end{equation}
Under time reversal, the transformations are assumed to be conventional:
\begin{equation}
\cT:\quad {\bf E\to E},\quad{\bf B\to-B},\quad{\bf A\to-A},\quad A^0\to A^0.
\label{e192}
\end{equation}
The Lagrangian of the theory possesses an imaginary coupling constant in order
that it be invariant under the product of these two symmetries:
\begin{equation}
\mathcal{L}=-\textstyle{\frac{1}{4}}F^{\mu\nu}F_{\mu\nu}+\half\psi^\dagger\gamma
^0\gamma^\mu\frac1i\partial_\mu\psi+\half m\psi^\dagger\gamma^0\psi+ie\psi^\dag
\gamma^0\gamma^\mu\psi A_\mu.
\label{e193}
\end{equation}
The corresponding Hamiltonian density is then
\begin{equation}
\mathcal{H}=\half(E^2+B^2)+\psi^\dagger\left[\gamma^0\gamma^k\left(-i\nabla_k+ie
A_k\right)+m\gamma^0\right]\psi.
\label{e194}
\end{equation}
The Lorentz transformation properties of the fermions are unchanged from the
usual ones. Thus, the electric current appearing in the Lagrangian and
Hamiltonian densities, $j^\mu=\psi^\dagger\gamma^0\gamma^\mu\psi$, transforms
conventionally under both $\cP$ and $\cT$:
\begin{equation}
\cP j^\mu({\bf x},t)\cP=\left(\begin{array}{c}j^0\\-{\bf j}\end{array}\right)
(-{\bf x},t),\qquad \cT j^\mu({\bf x},t)\cT=\left(\begin{array}{c} j^0\\-{\bf j}
\end{array}\right)({\bf x},-t).
\label{e195}
\end{equation}

Because its interaction is cubic, this non-Hermitian theory of
``electrodynamics'' is the analog of the spinless $i\vf^3$ quantum field theory
discussed in Subsec.~\ref{ss8-1}. ${\cP\cT}$-symmetric electrodynamics is
especially interesting because it is an asymptotically free theory (unlike
ordinary electrodynamics) and because the sign of the Casimir force is the
opposite of that in ordinary electrodynamics \cite{QED2,QED1}. This theory is
remarkable because finiteness conditions enable it to determine its own coupling
constant \cite{QED1}. 

The $\cC$ operator for $\cP\cT$-symmetric quantum electrodynamics has been
constructed perturbatively to first order in $e$ \cite{QED3}. This construction
is too technical to describe here, but it demonstrates that non-Hermitian
quantum electrodynamics is a viable and consistent unitary quantum field theory.
$\cP\cT$-symmetric quantum electrodynamics is more interesting than an $i\phi^3$
quantum field theory because it possesses many of the features of conventional
quantum electrodynamics, including Abelian gauge invariance. The only
asymptotically free quantum field theories described by Hermitian Hamiltonians
are those that possess a {\em non-Abelian} gauge invariance; $\cP\cT$ symmetry
allows for new kinds of asymptotically free theories, such as the $-\vf^4$
theory discussed in Subsec.~\ref{ss8-4}, that do not possess a non-Abelian gauge
invariance.

\subsection{Dual $\cP\cT$-Symmetric Quantum Field Theories}
\label{ss8-6}

Until now we have focused on bosonic $\cP\cT$-symmetric Hamiltonians, but it is
just as easy to construct fermionic $\cP\cT$-symmetric Hamiltonians. We look
first at free theories. The Lagrangian density for a conventional Hermitian free
fermion field theory is
\begin{equation}
{\cal L}({\bf x},t)=\bar\psi({\bf x},t)(i\pslash-m)\psi({\bf x},t)
\label{e196}
\end{equation}
and the corresponding Hamiltonian density is
\begin{equation}
{\cal H}({\bf x},t)=\bar\psi({\bf x},t)(-i\delslash+m)\psi({\bf x},t),
\label{e197}
\end{equation}
where $\bar\psi({\bf x},t)=\psi^\dagger({\bf x},t)\gamma_0$.

In $(1+1)$-dimensional space-time we adopt the following conventions:
$\gamma_0=\sigma_1$ and $\gamma_1=i\sigma_2$, where the Pauli $\sigma$ matrices
are given in (\ref{e151}). With these definitions, we have $\gamma_0^2=1$ and
$\gamma_1^2=-1$. We also define $\gamma_5=\gamma_0\gamma_1=\sigma_3$, so that
$\gamma_5^2=1$. The parity-reflection operator $\cP$ has the effect
\begin{equation}
\cP\psi(x,t)\cP=\gamma_0\psi(-x,t),\qquad\cP\bar\psi(x,t)\cP=\bar\psi(-x,t)
\gamma_0.
\label{e198}
\end{equation}
The effect of the time-reversal operator $\cT$,
\begin{equation}
\cT\psi(x,t)\cT=\gamma_0\psi(x,-t),\qquad\cT\bar\psi(x,t)\cT=\bar\psi(x,-t)
\gamma_0,
\label{e199}
\end{equation}
is similar to that of $\cP$, except that $\cT$ is antilinear and takes the
complex conjugate of complex numbers. From these definitions the Hamiltonian $H=
\int dx\,{\cal H}(x,t)$, where $\cal H$ is given in (\ref{e197}), is Hermitian:
$H=H^\dag$. Also, $H$ is separately invariant under parity reflection and under
time reversal: $\cP H\cP=H$ and $\cT H\cT=H$.

We can construct a non-Hermitian fermionic Hamiltonian by adding a
$\gamma_5$-dependent mass term to the Hamiltonian density in (\ref{e197}):
\begin{equation}
{\cal H}(x,t)=\bar\psi(x,t)(-i\delslash+m_1+m_2\gamma_5)\psi(x,t)\quad(m_2~{\rm
real}).
\label{e200}
\end{equation}
The Hamiltonian $H=\int dx\,{\cal H}(x,t)$ associated with this Hamiltonian
density is not Hermitian because the $m_2$ term changes sign under Hermitian
conjugation. This sign change occurs because $\gamma_0$ and $\gamma_5$
anticommute. Also, $H$ is not invariant under $\cP$ or under $\cT$ separately
because the $m_2$ term changes sign under each of these reflections. However,
$H$ {\em is} invariant under combined $\cal P$ and $\cal T$ reflection. Thus,
$H$ is $\cP\cT$-symmetric.

To see whether the $\cP\cT$ symmetry of $H$ is broken or unbroken, we must check
to see whether the spectrum of $H$ is real. We do so by solving the field
equations. The field equation associated with $\cal H$ in (\ref{e200}) is
\begin{equation}
\left(i\pslash-m_1-m_2\gamma_5\right)\psi(x,t)=0.
\label{e201}
\end{equation}
If we iterate this equation and use $\pslash^2=\partial^2$, we obtain
\begin{equation}
\left(\partial^2+\mu^2\right)\psi(x,t)=0,
\label{e202}
\end{equation}
which is the two-dimensional Klein-Gordon equation with $\mu^2=m_1^2-m_2^2$. The
physical mass that propagates under this equation is real when the inequality
\begin{equation}
m_1^2\geq m_2^2
\label{e203}
\end{equation}
is satisfied. This condition defines the two-dimensional parametric region of
{\em unbroken} $\cP\cT$ symmetry. When (\ref{e203}) is not satisfied, $\cP\cT$
symmetry is {\em broken}. At the boundary between the regions of broken and
unbroken $\cP\cT$ symmetry (the line $m_2=0$), the Hamiltonian is Hermitian.
[The same is true in quantum mechanics. Recall that for the Hamiltonian in
(\ref{e12}) the region of broken (unbroken) $\cP\cT$ symmetry is $\epsilon<0$
($\epsilon>0$). At the boundary $\epsilon=0$ of these two regions the
Hamiltonian is Hermitian.]

The $\cC$ operator associated with the $\cP\cT$-symmetric Hamiltonian density
$\cal H$ in (\ref{e200}) is given by $\cC=e^Q\cP$, where \cite{BJR}
\begin{eqnarray}
Q&=&-\tanh^{-1}\!\ve\!\int dx\,\bar\psi(x,t)\gamma_1\psi(x,t)\nonumber\\
&=&-\tanh^{-1}\!\ve\!\int dx\,\psi^\dag(x,t)\gamma_5\psi(x,t).
\label{e204}
\end{eqnarray}
The inverse hyperbolic tangent in this equation requires that $|\ve|\leq1$, or
equivalently that $m_1^2\geq m_2^2$, which corresponds to the unbroken region of
$\cP\cT$ symmetry. We use (\ref{e204}) to construct the equivalent Hermitian
Hamiltonian $h$ as in (\ref{e132}):
\begin{eqnarray}
h&=&\exp\left[\half\tanh^{-1}\!\ve\int\!dx\,\psi^\dag(x,t)\gamma_5\psi(x,t)
\right]H\nonumber\\
&&\qquad\times\exp\left[-\half\tanh^{-1}\!\ve\int\!dx\,\psi^\dag(x,t)\gamma_5
\psi(x,t)\right].
\label{e205}
\end{eqnarray}

The commutation relations $[\gamma_5,\gamma_0]=-2\gamma_1$ and $[\gamma_5,
\gamma_1]=-2\gamma_0$ simplify $h$ in (\ref{e205}):
\begin{equation}
h=\int dx\,\bar\psi(x,t)(-i\delslash+\mu)\psi(x,t),
\label{e206}
\end{equation}
where $\mu^2=m^2(1-\ve^2)=m_1^2-m_2^2$, in agreement with (\ref{e202}).
Replacing $H$ by $h$ changes the $\gamma_5$-dependent mass term $m\bar\psi(1+\ve
\gamma_5)\psi$ to a conventional fermion mass term $\mu\bar\psi\psi$. Thus, the
non-Hermitian $\cP\cT$-symmetric Hamiltonian density in (\ref{e200}) is
equivalent to the Hermitian Hamiltonian density in (\ref{e197}) with $m$
replaced by $\mu$.

If we introduce a four-point fermion interaction term in (\ref{e196}), we obtain
the Lagrangian density for the massive Thirring model in $(1+1)$ dimensions:
\begin{equation}
{\cal L}=\bar\psi(i\pslash-m)\psi+\half g(\bar\psi\gamma^\mu\psi)(\bar\psi
\gamma_\mu\psi),
\label{e207}
\end{equation}
whose corresponding Hamiltonian density is
\begin{equation}
{\cal H}=\bar\psi(-i\delslash+m)\psi-\half g(\bar\psi\gamma^\mu\psi)(\bar\psi
\gamma_\mu\psi),
\label{e208}
\end{equation}
We can then construct the $\cP\cT$-symmetric Thirring model
\begin{equation}
{\cal H}=\bar\psi(-i\delslash+m+\ve m\gamma_5)\psi-\half g(\bar\psi\gamma^\mu
\psi)(\bar\psi\gamma_\mu\psi)
\label{e209}
\end{equation}
by introducing a $\gamma_5$-dependent mass term. The additional term is
non-Hermitian but $\cP\cT$-symmetric because it is odd under both parity
reflection and time reversal. Remarkably, the $Q$ operator for the interacting
case $g\neq0$ is {\em identical} to the $Q$ operator for the case $g=0$ because
in $(1+1)$-dimensional space the interaction term $(\bar\psi\gamma^\mu\psi)(\bar
\psi\gamma_\mu\psi)$ commutes with the $Q$ in (\ref{e204}) \cite{BJR}. Thus, the
non-Hermitian $\cal PT$-symmetric Hamiltonian density in (\ref{e209}) is
equivalent to the Hermitian Hamiltonian density in (\ref{e208}) with the mass
$m$ replaced by $\mu$, where $\mu^2=m^2(1-\ve^2)=m_1^2-m_2^2$.

The same holds true for the $(3+1)$-dimensional interacting Thirring model by
virtue of the commutation relation $[\gamma_5,\gamma_0\gamma_\mu]=0$, but
because this higher-dimensional field theory is nonrenormalizable, the $Q$
operator may only have a formal significance.

In $(1+1)$ dimensions the massive Thirring Model (\ref{e207}) is {\em dual} to
the $(1+1)$-dimensional Sine-Gordon model \cite{DUAL}, whose Lagrangian density
is
\begin{equation}
{\cal L}=\half(\p_\mu\vf)^2+m^2\lambda^{-2}(\cos\lambda\vf-1),
\label{e210}
\end{equation}
whose corresponding Hamiltonian density is
\begin{equation}
{\cal H}=\half\pi^2+\half(\nabla\vf)^2+m^2\lambda^{-2}(1-\cos\lambda\vf),
\label{e211}
\end{equation}
where $\pi(x,t)=\partial_0\vf(x,t)$ and where in $(1+1)$-dimensional space
$\nabla\vf(x,t)$ is just $\p_1\vf(x,t)$. The duality between the Thirring model
and the Sine-Gordon model is expressed as an algebraic relationship between the
coupling constants $g$ and $\lambda$:
\begin{equation}
\frac{\lambda^2}{4\pi}=\frac{1}{1-g/\pi}.
\label{e212}
\end{equation}
Note that the free fermion theory ($g=0$) is equivalent to the Sine-Gordon model
with the special value for the coupling constant $\lambda^2=4\pi$.

The $\cP\cT$-symmetric extension (\ref{e209}) of the modified Thirring model is,
by the same analysis, dual to a modified Sine-Gordon model with the Hamiltonian
density
\begin{equation}
{\cal H}=\half\pi^2+\half(\nabla\vf)^2+m^2\lambda^{-2}(1-\cos\lambda\vf-i
\ve\sin\lambda\vf),
\label{e213}
\end{equation}
which is $\cP\cT$-symmetric and not Hermitian. The $Q$ operator for this
Hamiltonian is
\begin{equation}
Q=\frac{2}{\lambda}\tanh^{-1}\!\ve\!\int\!dx\,\pi(x,t).
\label{e214}
\end{equation}
Thus, the equivalent Hermitian Hamiltonian $h$ is given by
\begin{equation}
h=\exp\left[-\frac{1}{\lambda}\tanh^{-1}\!\ve\!\int\!dx\,\pi(x,t)\right]H
\exp\left[\frac{1}{\lambda}\tanh^{-1}\!\ve\!\int\!dx\,\pi(x,t)\right].
\label{e215}
\end{equation}
Note that the operation that transforms $H$ to $h$ has the same effect as
shifting the boson field $\vf$ by an imaginary constant:
\begin{equation}
\vf\to\vf+\frac{i}{\lambda}\tanh^{-1}\!\ve.
\label{e216}
\end{equation}
Under this transformation the interaction term $m^2\lambda^{-2}(1-\cos\lambda\vf
-i\ve\sin\lambda\vf)$ in (\ref{e213}) becomes $-m^2\lambda^{-2}(1-\ve^2)\cos
\lambda\vf$, apart from an additive constant. Hence, $h$ is the Hamiltonian for
the Hermitian Sine-Gordon model, but with mass $\mu$ given by $\mu^2=m^2(1-
\ve^2)=m_1^2-m_2^2$. This change in the mass is the same as for the Thirring
model. Being Hermitian, $h$ is even in the parameter $\ve$ that breaks the
Hermiticity of $H$.

The idea of generating a non-Hermitian but $\cP\cT$-symmetric Hamiltonian from a
Hermitian Hamiltonian by shifting the field operator as in (\ref{e216}), first
introduced in the context of quantum mechanics in Ref.~\cite{ZNOJ}, suggests a
way to construct solvable fermionic $\cP\cT$-invariant models whenever there is
a boson-fermion duality.

\subsection{$\cP\cT$-Symmetric Gravitational and Cosmological Theories}
\label{ss8-7}

In Subsec.~\ref{ss8-5} we showed that to construct a $\cP\cT$-symmetric model of
quantum electrodynamics, one need only replace the electric charge $e$ by $ie$
and then replace the vector potential $A^\mu$ by an axial-vector potential. The 
result is a non-Hermitian but $\cP\cT$-symmetric Hamiltonian. An interesting 
classical aspect of this model is that the sign of the Coluomb force is
reversed, so that like charges feel an attractive force and unlike charges
feel a repulsive force.

One can use the same idea to construct a $\cP\cT$-symmetric model of massless
spin-2 particles (gravitons). One simply replaces the gravitational coupling
constant $G$ by $iG$ and then requires the two-component tensor field to behave
like an axial tensor under parity reflection. The result is a non-Hermitian $\cP
\cT$-symmetric Hamiltonian, which at the classical level describes a {\em
repulsive} gravitational force. It would be interesting to investigate the
possible connections between such a model and the notion of dark energy and the
recent observations that the expansion of the universe is accelerating.

The connection discussed in Subsec.~\ref{ss2-6} between reflectionless
potentials and $\cal PT$ symmetry may find application in quantum cosmology.
There has been much attention given to anti-de Sitter cosmologies \cite{ADS} and
de Sitter cosmologies \cite{DS1,DS2}. In the AdS description the universe
propagates reflectionlessly in the presence of a wrong-sign potential ($-x^6$,
for example). In the dS case the usual Hermitian quantum mechanics must be
abandoned and be replaced by a non-Hermitian one in which there are
`meta-observables'. The non-Hermitian inner product that is used in the dS case
is based on the $\cC\cP\cT$ theorem in the same way that the $\cC\cP\cT$ inner
product is used in $\cal PT$-symmetric quantum theory \cite{BBJ1}. The
condition of reflectionless, which is equivalent to the requirement of $\cP\cT$
symmetry, is what allows the wrong-sign potential to have a positive spectrum.
Calculating the lowest energy level in this potential would be equivalent to
determining the cosmological constant \cite{Moffat}.

\subsection{Classical Field Theory}
\label{ss8-8}

The procedure for constructing $\cP\cT$-symmetric Hamiltonians is to begin with
a Hamiltonian that is both Hermitian and $\cP\cT$ symmetric, and then to
introduce a parameter $\epsilon$ that extends the Hamiltonian into the complex
domain while maintaining its $\cP\cT$ symmetry. This is the procedure that was
used to construct the new kinds Hamiltonians in (\ref{e12}). We can follow the
same procedure for classical nonlinear wave equations because many wave
equations are $\cP\cT$ symmetric.

As an example, consider the Korteweg-de Vries (KdV) equation
\begin{equation}
u_t+uu_x+u_{xxx}=0.
\label{e217}
\end{equation}
To demonstrate that this equation is $\cP\cT$ symmetric, we define a classical
parity reflection $\cP$ to be the replacement $x\to-x$, and since $u=u(x,t)$ is
a velocity, the sign of $u$ also changes under $\cP$: $u\to-u$. We define a
classical time reversal $\cT$ to be the replacement $t\to-t$, and again, since
$u$ is a velocity, the sign of $u$ also changes under $\cT$: $u\to-u$. Following
the quantum-mechanical formalism, we also require that $i\to-i$ under time
reversal. Note that the KdV equation is not symmetric under $\cP$ or $\cT$
separately, but that it {\it is} symmetric under combined $\cP\cT$ reflection.
The KdV equation is a special case of the Camassa-Holm equation \cite{r8}, which
is also $\cP\cT$ symmetric. Other nonlinear wave equations such as the
generalized KdV equation $u_t+u^ku_x+u_{xxx}=0$, the Sine-Gordon equation $u_{tt
}-u_{xx}+g\sin u=0$, and the Boussinesq equation are $\cP\cT$ symmetric as well.

The observation that there are many nonlinear wave equations possessing $\cP\cT$
symmetry suggests that one can generate rich and interesting families of new
complex nonlinear $\cP\cT$-symmetric wave equations by following the same
procedure that was used in quantum mechanics and one can try to discover which
properties (conservation laws, solitons, integrability, stochastic behavior) of
the original wave equations are preserved and which are lost. One possible
procedure for generating new nonlinear equations from the KdV equation is to
introduce the real parameter $\epsilon$ as follows:
\begin{equation}
u_t-iu(iu_x)^\epsilon+u_{xxx}=0.
\label{e218}
\end{equation}
Various members of this family of equations have been studied in
Ref.~\cite{KDV}. Of course, there are other ways to extend the KdV equation into
the complex domain while preserving $\cP\cT$ symmetry; see, for example 
Ref.~\cite{Fring}.

\section{Final Remarks}
\label{s9}

In this paper we have shown how to extend physical theories into the complex
domain. The complex domain is huge compared with the real domain, and therefore
there are many exciting new theories to explore. The obvious potential problem
with extending a real theory into complex space is that one may lose some of the
characteristics that a valid physical theory must possess. Thus, it is necessary
that this complex extension be tightly constrained. We have shown that the
essential physical axioms of a quantum theory are maintained if the complex
extension is done in such a way as to preserve $\cP\cT$ symmetry.

The complex theories that we have constructed are often far more elaborate and
diverse than theories that are restricted to the real domain. Upon entering the
complex world we have found a gold mine of new physical theories that have
strange and fascinating properties. We have just begun to study the vast new
panorama that has opened up and we can hardly begin to guess what new kinds of
phenomena have yet to be discovered.

\vspace{0.5cm}

\begin{footnotesize}
\noindent As an Ulam Scholar, CMB receives financial support from the Center for
Nonlinear Studies at the Los Alamos National Laboratory and he is supported by a
grant from the U.S. Department of Energy.
\end{footnotesize}

\end{document}